\documentclass[twocolumn,tighten]{aastex631}
\usepackage{comment}
\usepackage{url}
\usepackage{booktabs}
\usepackage{color}

\received{xxx}
\revised{yyy}
\accepted{zzz}

\submitjournal{ApJ}

\shorttitle{Multi-wavelength properties of MAXI J1820+070}
\shortauthors{Echibur\'u-Trujillo et al.}
\graphicspath{{./}{figures/}}
\begin{document}

\title{Chasing the break: Tracing the full evolution of a black hole X-ray binary jet with multi-wavelength spectral modeling}

%%%%%%%%%%%%%%%%%%%%%%%%%%%%%%%%% AUTHORS %%%%%%%%%%%%%%%%%%%%%%%%%%%%%%%%%%
\author[0000-0001-8436-1847]{Constanza Echibur\'u-Trujillo}
\affiliation{Department of Astrophysical and Planetary Sciences, JILA, Duane Physics Bldg., 2000 Colorado Ave., University of Colorado, Boulder, CO 80309, USA}
\affiliation{Department of Physics, McGill University, 3600 rue University, Montreal, Québec, H3A 2T8, Canada}
\affiliation{Trottier Space Institute, 3550 Rue University, Montréal, Québec, H3A 2A7, Canada}
\correspondingauthor{C. Echibur\'u-Trujillo}
\email{constanza.echiburu@colorado.edu}

\author[0000-0003-3906-4354]{Alexandra J. Tetarenko}\thanks{Former NASA Einstein Fellow}
\affiliation{Department of Physics and Astronomy, University of Lethbridge, Lethbridge, Alberta, T1K 3M4, Canada}
\affiliation{Department of Physics and Astronomy, Texas Tech University, Lubbock, TX 79409-1051, USA}
\affiliation{East Asian Observatory,
660 N. A`oh\={o}k\={u} Place, University
Park, Hilo, Hawaii 96720, USA}

\author[0000-0001-6803-2138]{Daryl Haggard}
\affiliation{Department of Physics, McGill University, 3600 rue University, Montreal, Québec, H3A 2T8, Canada}
\affiliation{Trottier Space Institute, 3550 Rue University, Montréal, Québec, H3A 2A7, Canada}

\author[0000-0002-7930-2276]{Thomas D. Russell}
\affiliation{INAF, Istituto di Astrofisica Spaziale e Fisica Cosmica, Via U. La Malfa 153, I-90146 Palermo, Italy}

\author[0000-0002-9677-1533]{Karri I. I. Koljonen}
\affiliation{Institutt for Fysikk, Norwegian University of Science and Technology, Høgskloreringen 5, Trondheim 7491, Norway}
\affiliation{Finnish Centre for Astronomy with ESO (FINCA), University of Turku, Väisäläntie 20, FI-21500 Piikkiö, Finland}
\affiliation{Aalto University, Metsahovi Radio Observatory, PO Box 13000, FI-00076 Aalto, Finland}

\author[0000-0003-2506-6041]{Arash Bahramian}
\affiliation{International Centre for Radio Astronomy Research, Curtin University, Bentley, WA 6102, Australia}

\author[0000-0002-1742-2125]{Jingyi Wang}
\affiliation{MIT Kavli Institute for Astrophysics and Space Research, MIT, 70 Vassar Street, Cambridge, MA 02139, USA}

\author{Michael Bremer}
\affiliation{Institut de Radioastronomie Millimétrique, 300 Rue de la Piscine, 38406 Saint Martin d'Hères, France}

\author[0000-0002-7735-5796]{Joe Bright}
\affiliation{Astrophysics, Department of Physics, University of Oxford, Keble Road, Oxford OX1 3RH, UK}

\author{Piergiorgio Casella}
\affiliation{INAF - Osservatorio Astronomico di Roma, Via Frascati 33, I-00078 Monteporzio Catone, Italy}

\author[0000-0002-3500-631X]{David M. Russell}
\affiliation{Center for Astrophysics and Space Science (CASS), New York University Abu Dhabi, PO Box 129188, Abu Dhabi, UAE}

\author{Diego Altamirano}
\affiliation{School of Physics and Astronomy, University of Southampton, Southampton, SO17 1BJ, UK}

\author[0000-0003-1285-4057]{M. Cristina Baglio}
\affiliation{INAF-Osservatorio Astronomico di Brera, Via Bianchi 46, I-23807 Merate (LC), Italy}

\author{Tomaso Belloni}\thanks{Deceased.}
\affiliation{INAF-Osservatorio Astronomico di Brera, Via Bianchi 46, I-23807 Merate (LC), Italy}

\author[0000-0002-4767-9925]{Chiara Ceccobello}
\affiliation{Department of Space, Earth and Environment, Chalmers University of Technology, Onsala Space Observatory, SE-439 92 Onsala, Sweden}

\author{Stephane Corbel}
\affiliation{Université Paris-Saclay, Université Paris-Cité, CEA, CNRS, AIM, 91191, Gif-surYvette, France}

\author{Maria Diaz Trigo}
\affiliation{ESO, Karl-Schwarzschild-Strasse 2, D-85748 Garching bei München, Germany}

\author[0000-0003-1897-6872]{Dipankar Maitra}
\affiliation{Department of Physics and Astronomy, Wheaton College, Norton, MA 02766, USA}

\author[0000-0003-1541-7557]{Aldrin Gabuya}
\affiliation{Al Sadeem Observatory, Al Wathba South, Abu Dhabi, UAE}
\affiliation{Rizal Technological University, Mandaluyong City, Philippines}

\author[0000-0001-5802-6041]{Elena Gallo}
\affiliation{Department of Astronomy, University of Michigan, 1085 South University Avenue, Ann Arbor, MI 48109, USA}

\author{Sebastian Heinz}
\affiliation{Department of Astronomy, University of Wisconsin Madison, 475 N. Charter Street, Madison, WI 53706, USA}

\author[0000-0001-8371-2713]{Jeroen Homan}
\affiliation{Eureka Scientific, Inc., 2452 Delmer Street, Oakland, CA 94602, USA}

\author[0000-0003-0172-0854]{Erin Kara}
\affiliation{MIT Kavli Institute for Astrophysics and Space Research, MIT, 70 Vassar Street, Cambridge, MA 02139, USA}

\author{Elmar K\"ording}
\affiliation{Department of Astrophysics/IMAPP, Radboud University,
P.O. Box 9010, 6500 GL Nijmegen, The Netherlands.}

\author{Fraser Lewis}
\affiliation{Faulkes Telescope Project, School of Physics and Astronomy, Cardiff University, The Parade, Cardiff, CF24 3AA Wales, UK}
\affiliation{Astrophysics Research Institute, Liverpool John Moores University, 146 Brownlow Hill, Liverpool L3 5RF, UK}

\author[0000-0002-2235-3347]{Matteo Lucchini}
\affiliation{MIT Kavli Institute for Astrophysics and Space Research, MIT, 70 Vassar Street, Cambridge, MA 02139, USA}
\affiliation{Anton Pannekoek Institute for Astronomy, University of Amsterdam, Science Park 904, Amsterdam NL-1098 XH, the Netherlands}
\affiliation{SRON Netherlands Institute for Space Research, Niels Bohrweg 4, 2333CA Leiden, The Netherlands}

\author[0000-0001-9564-0876]{Sera Markoff}
\affiliation{Anton Pannekoek Institute for Astronomy, University of Amsterdam, Science Park 904, Amsterdam NL-1098 XH, the Netherlands}
\affiliation{Gravitation Astroparticle Physics Amsterdam (GRAPPA) Institute, University of Amsterdam, Science Park NL-904, Amsterdam 1098 XH, the Netherlands}

\author{Simone Migliari}
\affiliation{Aurora Technology, Calle Principe de Vergara, 211, 1-B, 28002 Madrid, Spain}

\author[0000-0003-3124-2814]{James C. A. Miller-Jones}
\affiliation{International Centre for Radio Astronomy Research, Curtin University, Bentley, WA 6102, Australia}

\author[0000-0002-4151-4468]{Jerome Rodriguez}
\affiliation{Université Paris-Saclay, Université Paris-Cité, CEA, CNRS, AIM, 91191, Gif-surYvette, France}

\author{Payaswini Saikia}
\affiliation{Center for Astrophysics and Space Science (CASS), New York University Abu Dhabi, PO Box 129188, Abu Dhabi, UAE}

\author[0000-0003-0167-0981]{Craig L. Sarazin}
\affiliation{Department of Astronomy, University of Virginia, 530 McCormick Road, Charlottesville, VA 22904-4325, USA}

\author[0000-0003-1331-5442]{Tariq Shahbaz}
\affiliation{Instituto de Astrof\'\i{}sica de Canarias (IAC), E-38205 La Laguna,  Tenerife, Spain}
\affiliation{Departamento de  Astrof\'\i{}sica, Universidad de La Laguna (ULL),  E-38206 La Laguna, Tenerife, Spain}

\author{Gregory Sivakoff}
\affiliation{Department of Physics, University of Alberta, CCIS 4-181, Edmonton AB T6G 2E1, Canada}

\author[0000-0002-4622-796X]{Roberto Soria}
\affiliation{INAF, Osservatorio Astrofisico di Torino, Strada Osservatorio 20, I-10025 Pino Torinese, Italy}
\affiliation{College of Astronomy and Space Sciences, University of the Chinese Academy of Sciences, Beijing 100049, China}
\affiliation{Sydney Institute for Astronomy, School of Physics A28, The University of Sydney, Sydney, NSW 2006, Australia}

\author[0000-0003-1033-1340]{Vincenzo Testa}
\affiliation{INAF - Osservatorio Astronomico di Roma, Via Frascati 33, I-00078 Monteporzio Catone, Italy}

\author[0000-0003-2636-6644]{Bailey E. Tetarenko}
\affiliation{Department of Physics, McGill University, 3600 rue University, Montreal, Québec, H3A 2T8, Canada}
\affiliation{Trottier Space Institute, 3550 Rue University, Montréal, Québec, H3A 2A7, Canada}

\author{Valeriu Tudose}
\affiliation{Institute for Space Sciences, Atomistilor 409, PO Box MG-23, 077125 Bucharest-Magurele, Romani}

%%%%%%%%%%%%%%%%%%%%%%%%%%%%%%%%% ABSTRACT %%%%%%%%%%%%%%%%%%%%%%%%%%%%%%%%%%
\begin{abstract}
Black hole X-ray binaries (BH XRBs) are ideal targets to study the connection between accretion inflow and jet outflow. 
Here we present quasi-simultaneous, multi-wavelength observations of the Galactic black hole system MAXI J1820+070, throughout its 2018--2019 outburst. Our data set includes coverage from the radio through X-ray bands from 17 different instruments/telescopes, and encompasses 19 epochs over a 7 month time period, resulting in one of the most well-sampled multi-wavelength data sets of a BH XRB outburst to date. With our data, we compile and model the broad-band spectra of this source using a phenomenological model that includes emission from the jet, companion star, and accretion flow. This modeling allows us to track the evolution of the spectral break in the jet spectrum, a key observable that samples the jet launching region. We find that the spectral break location changes over at least $\approx3$ orders of magnitude in electromagnetic frequency over this period.
Using these spectral break measurements, we link the full cycle of jet behavior, including the rising, quenching, and re-ignition, to the changing accretion flow properties as the source evolves through its different accretion states. Our analyses show a consistent jet behavior with other sources in similar phases of their outbursts, reinforcing that the jet quenching and recovery may be a global feature of BH XRB systems in outburst. Our results also provide valuable evidence supporting a close connection between the geometry of the inner accretion flow and the base of the jet.

\end{abstract}


%%%%%%%%%%%%%%%%%%%%%%%%%%%%%%%%% INTRO %%%%%%%%%%%%%%%%%%%%%%%%%%%%%%%%%%
\section{Introduction} 
\label{sec:Intro}

Black hole X-ray binaries (BH XRBs) consist of a stellar-mass BH accreting material from a companion star. Since this accreted material carries angular momentum, it forms an accretion disk around the BH, where some of the material can be transported away from the disk in the form of a relativistic plasma jet \citep{2006csxs.book..381F}. The physical processes involved in jet launching are still a matter of debate, as are the composition of the jet material, and the amount of energy carried away from the system. However, the launching mechanism is thought to be connected to the accretion process, suggesting a close relationship between emission properties of the disk and the jet in these systems \citep[e.g.,][]{2004MNRAS.355.1105F,2012MNRAS.421..468M,2014MNRAS.439.1390R,2015ApJ...814..139K,2020MNRAS.498.5772R,2021MNRAS.505.3393W}. Thanks to their close distances of $\sim$kpc \citep[see][and references therein]{2004MNRAS.354..355J,2016ApJS..222...15T}, and because they present variability on timescales ranging from hours to a few days, BH XRBs are ideal systems to track changes in the accretion inflow and jet outflow in real time as the sources evolve through different accretion states, and therefore, provide insight into the disk-jet connection.

The different accretion states observed in BH XRBs during a typical outburst are marked by changes in the structure of the accretion flow \citep{2005Ap&SS.300..107H,2011BASI...39..409B}. At low mass accretion rates, the system is in the hard state, where the inner accretion flow is hot, optically thin, and geometrically thick (although still debated, this is known as the corona). The hard state is associated with the presence of a compact jet, a continuous and highly-collimated outflow with opening angles $<10^{\circ}$ \citep{2006MNRAS.367.1432M} and Lorentz factors 1.3--3.5 \citep{2019ApJ...887...21S} detected in radio bands. As the accretion rate increases, the system moves from the hard state into the soft state, a transition state known as the hard intermediate state (HIMS). During this process, jets are observed to take the form of discrete clouds of plasma (known as jet ejecta), while the emission from the compact jet begins to switch off. With increasing accretion rates, the system is settled in the soft state, where most of the emission can be characterized by an optically thick, geometrically thin disk that extends down to the innermost stable circular orbit \citep[ISCO,][]{1973A&A....24..337S}, and locally emits a thermal blackbody spectrum. The compact jet emission in the soft state is completely quenched \citep[e.g.,][]{2019ApJ...883..198R,2021MNRAS.504..444C}. As the mass accretion rate decreases again, the system begins its transition back to the hard state, through the soft intermediate state (SIMS), where the compact radio jet emission is observed to recover.

The changing structure of inflows and outflows across accretion states manifests observationally as changes in the broad-band (radio through X-rays) emission spectrum of BH XRBs. For this reason, multi-wavelength observing campaigns play an important role in understanding the evolution of these sources throughout an outburst. 
During the rise phase of an outburst, when BH XRBs are found in the hard state, the jet component dominates the lower electromagnetic frequency broad-band emission. This jet emission is characterized by a flat to slightly inverted optically thick spectrum ($f_\nu\propto\nu^\alpha$, where $\alpha \sim 0$), extending from radio to sub-mm frequencies and above \citep{2002ApJ...573L..35C,2010MNRAS.404L..21C,2015ApJ...805...30T}. The jet spectrum transitions from optically thick to optically thin emission (with $\alpha\sim-0.6$), which is observed as a spectral break at $\nu_{\mathrm{b}} \sim 10^{11-14}$ Hz,  \citep[sub-mm/infrared frequencies,][]{2013MNRAS.429..815R}. Thus, the compact jet component is typically modeled as a broken power-law. Throughout this paper we refer to this spectral break, which results from synchrotron self-absorption, as the \textit{jet} spectral break. There can also be a synchrotron cooling break at higher frequencies \citep[e.g.][]{2014MNRAS.439.1390R}, resulting from the highest energy electrons that radiate faster than the dynamical time scale of the system. At higher electromagnetic frequencies (optical to X-ray bands), the emission originates mainly from the accretion flow, and it is well described by an irradiated disk \citep{2008MNRAS.388..753G,2009MNRAS.392.1106G,2010LNP...794...17G}. 
In this model, the thermal disk provides seed photons that are intercepted by hot electrons in the inner flow (the corona). This interaction, known as inverse Compton scattering, results in a hard power-law spectrum with a high energy cutoff in the range 20-100 keV, and photon indices $\Gamma \lesssim 2$. Some of these Comptonized photons can illuminate the disk, producing an iron emission line and Compton reflection component \citep[e.g.,][]{2005A&A...430..761M}.

As the source evolves into the intermediate states, the jet spectral break, initially located around the IR region, is observed to move towards lower electromagnetic frequencies \citep[towards the radio wavebands;][]{2013MNRAS.436.2625V,2014MNRAS.439.1390R,2020MNRAS.498.5772R}. The jet spectral evolution appears to be correlated with the quenching of the compact jet \citep{2020MNRAS.498.5772R}. Discrete jet ejecta can also become detectable at this stage of an outburst \citep{2004ApJ...617.1272C,2004MNRAS.355.1105F}, but tend to display much brighter flux densities than the compact jet \citep[e.g.,][]{2017MNRAS.469.3141T}. Some studies suggest that the breakup of the compact jet and the launching of ejections may be related to a change in the speed of the jet flow, leading to internal shocks when faster moving plasma catches up with slower moving plasma \citep{2010MNRAS.401..394J,2013MNRAS.429L..20M,2014MNRAS.443..299M}. Alternatively, the ejecta may result from the ejection of the corona \citep{2003ApJ...597.1023V,2003ApJ...595.1032R,2008ApJ...675.1449R}. In this scenario, the compact jet quenching might be related to the jet acceleration zone becoming disconnected from the system, and its propagation away from the source could explain the emergence of ejecta \citep{2020MNRAS.498.5772R}. It has been suggested that these jet knots can be produced towards the end of the hard state in the rise of an outburst, during which time the corona may contract and become less vertically extended \citep[e.g.,][]{2019Natur.565..198K}. Once in the HIMS \citep{2004MNRAS.355.1105F,2009MNRAS.396.1370F}, observations indicate that coronal height increases, possibly representing material being ejected \citep[e.g.,][]{2022ApJ...930...18W}. In the simplest scenario, the corona extends above the disk, and is responsible for the hard X-ray emission, some of which is intercepted and reprocessed by the disk, producing the soft X-ray component. If the distance between the corona and the disk increases, the hard X-ray photons would reach the observer before the soft ones, causing a delay known as soft reverberation lag. This delay is then a consequence of the changing disk-corona geometry and light travel times. Thanks to the high time resolution of the Neutron star Interior Composition Explorer (\textit{NICER}) X-ray Timing instrument, and its low energy coverage at good effective area, such measurements have been possible \citep{2021ApJ...910L...3W,2021A&A...654A..14D}, providing the aforementioned insights into the corona-jet connection.

Once the source enters the softer states (SIMS and soft state), the broad-band spectrum is dominated by the thermal disk, with a softer X-ray spectrum ($\Gamma \gtrsim 2$), and any radio emission detected is attributed to the remnants of the jet ejecta, or their collisions with the local interstellar medium \citep[ISM, e.g.,][]{2002Sci...298..196C,2004ApJ...617.1272C,2019ApJ...883..198R,2020NatAs...4..697B,2021MNRAS.504..444C}. Towards the outburst decay the source returns to the hard state (going through the SIMS and HIMS in reverse), but this time with lower luminosities \citep{2007A&ARv..15....1D,2003A&A...409..697M}. Over the soft to hard state transition at the end of the outburst, the compact jet is observed to re-ignite, first in the radio and then in the optical/IR bands \citep[e.g.,][]{2012MNRAS.421..468M,2013ApJ...779...95K,2013MNRAS.431L.107C,2014MNRAS.439.1390R,2020MNRAS.495..182R}, where the spectral break is observed to move in the opposite direction to the forward transition, i.e., from lower to higher frequencies \citep{2014MNRAS.439.1390R}.

The location of the jet spectral break and the flux density at that electromagnetic frequency are key pieces of information needed in understanding the jet launching mechanism and energetics \citep{2003MNRAS.343L..59H,2011A&A...529A...3C,2014MNRAS.438..959P,2018MNRAS.473.4417C,2021MNRAS.501.5910L}, since this break traces the jet base region where the particles are first accelerated  \citep{2001A&A...372L..25M,2005ApJ...635.1203M,2010LNP...794..143M,2017SSRv..207....5R}. For instance, accurate measurements of the spectral break can provide constraints on the cross-sectional radius and magnetic field strength at the base of the jet \citep{1979rpa..book.....R,2009ApJ...703L..63C,2011A&A...529A...3C,2011ApJ...740L..13G}, although this is dependent on simple one-zone models. 
Tracking the spectral break location and its connection to changes in the accretion flow (probed through X-ray emission) require multi-wavelength coverage of the broad-band spectrum throughout different stages of an outburst. Observations exist for only a handful of systems so far: MAXI J1836--194 \citep{2013ApJ...768L..35R,2014MNRAS.439.1390R}, V404 Cygni \citep{2019MNRAS.482.2950T}, and MAXI J1535--571 \citep{2020MNRAS.498.5772R,Baglio18}. However, all of these previous works have only probed a portion of the jet evolution cycle during outburst. In this work, we present a multi-wavelength data set during the 2018/2019 outburst of the BH XRB MAXI J1820+070, which has allowed us to track the broad-band spectrum throughout the full outburst cycle, sampling the establishment, quenching, and re-ignition of the compact jet for the first time. 

\subsection{MAXI J1820+070}
The Galactic BH XRB MAXI J1820+070 (ASSASN--18ey, hereafter J1820) was first detected in the optical band with the All-Sky Automated Survey for SuperNovae \citep[ASAS-SN;][]{2014ApJ...788...48S,2017PASP..129j4502K}, on 2018 March 6 \citep[MJD 58184.079861;][]{2018ApJ...867L...9T}. 
Later, it was detected in X-rays with the Monitor of All-sky X-ray Image \citep[MAXI;][]{2009PASJ...61..999M} Gas Slit Camera \citep[GCS;][]{2011PASJ...63S.623M} on 2018 March 11 \citep{2018ATel11399....1K}. The system was first identified as a likely BH XRB in outburst by \citealt{baglio2018ATel11418}. The nature of the compact object was dynamically confirmed by \citealt{2019ApJ...882L..21T}, and later refined in \citealt{2020ApJ...893L..37T}, to be a BH with $M_{\mathrm{BH}}=8.48_{-0.72}^{+0.79} \, M_{\odot}$. The system also hosts a type K3-5 companion with an orbital period of 16.5 hours. Thanks to its high X-ray flux $\sim 3.99\times 10^{-8}$ erg cm$^{-2}$ s$^{-1}$ ($\sim4$ Crab in $20-50$ keV, \citealt{2019ApJ...870...92R}), a close distance ($2.96\pm0.33$\,kpc, \citealt{2020MNRAS.493L..81A}) and a low Galactic extinction ($N_H=1.5\times 10^{21}$ cm$^{-2}$, \citealt{2018ATel11423....1U}), the source was an excellent candidate for an extended multi-wavelength campaign during its outburst.

J1820 remained in the hard state until a rapid softening of the X-ray spectrum on 2018 July 5 indicated it was entering the soft state \citep{2018ATel11820....1H}. During this state transition, the broad-band emission was dominated by a thermal disk from optical to soft X-rays, while the radio to infrared flux decreased, suggesting a quenching of the compact jet \citep{2018ATel11833....1C,2018ATel11831....1T}. Additionally, strong radio flares were detected \citep{2018ATel11827....1B}, consistent with the launching of jet ejecta \citep{2020NatAs...4..697B,2021MNRAS.505.3393W}. In late September, the X-ray spectrum exhibited spectral hardening \citep{2018ATel12068....1H,2018ATel12064....1M}, suggesting that the source started its return to the hard state. In the following months the outburst continued to decay, reaching quiescence in 2019 February \citep{2019ATel12534....1R}. Since then, J1820 has shown little activity, with re-brightening episodes in 2019 \citep{2019ATel12567....1U,2019ATel12573....1B,2019ATel12577....1W,2019ATel13014....1H,2019ATel13025....1X,2019ATel13041....1B}, 2020 \citep{2020ATel13502....1A,2020ATel13530....1S}, 2021 \citep{2021ATel14492....1B}, and a possible re-activation of the compact jet in 2022 \citep{2022ATel15277....1C}, but has not entered a full outburst again with state changes. The latest reports indicate that the source continues to fade into quiescence \citep{2023ATel16192....1B,2023ATel16200....1H}, and no other multi-wavelength observations have since been reported.

Broad-band spectral analyses have been performed from J1820's outburst in 2018. However, these analyses only sampled isolated epochs of the outburst, focusing mainly on the hard state \citep[e.g.,][on April 12]{2021ApJ...910...21R}, and/or a limited region of the electromagnetic spectrum \citep[e.g.,][]{2018ApJ...868...54S,2019MNRAS.487.5946B,2019ApJ...874..183S,2020MNRAS.498.5873C,2021A&A...656A..63M,2022MNRAS.514.6102P,2022MNRAS.514.3894O,2023A&A...669A..65C}. In this work, we characterize the broad-band spectrum (from radio to X-ray bands) of J1820 over the course of its full 2018 outburst. We place particular attention to the evolution of the spectral parameters of the jet, and their connection to the accretion flow parameters.

This paper is organized as follows: In Section \ref{sec:ObsAndData} we describe the observations and reduction of each data set utilized in this work. The details of the spectral modeling are presented in Section \ref{sec:Results}, together with the best-fit broad-band spectrum of each observational epoch. In Section \ref{sec:discussion} we discuss the evolution of the spectral parameters, focusing on those connecting the accretion flow to the jet. To date, this connection remains uncertain. We also compare J1820's evolution to the observed behavior of other BH XRBs in similar phases of their outbursts. Our analyses are complementary to the significant work that has been made in studying the time-domain properties of J1820 \cite[e.g.,][]{2019Natur.565..198K,2019MNRAS.490L..62P,2020ApJ...896...33W,2021MNRAS.505.3452P,2021MNRAS.504.3862T,2021ApJ...909L...9Z}, some of which we discuss in Section \ref{sub:connecting}. Finally, we summarize our conclusions and findings in Section \ref{sec:Conclusion}.

\begin{figure*}
    \centering
        \includegraphics[width=0.9\textwidth]{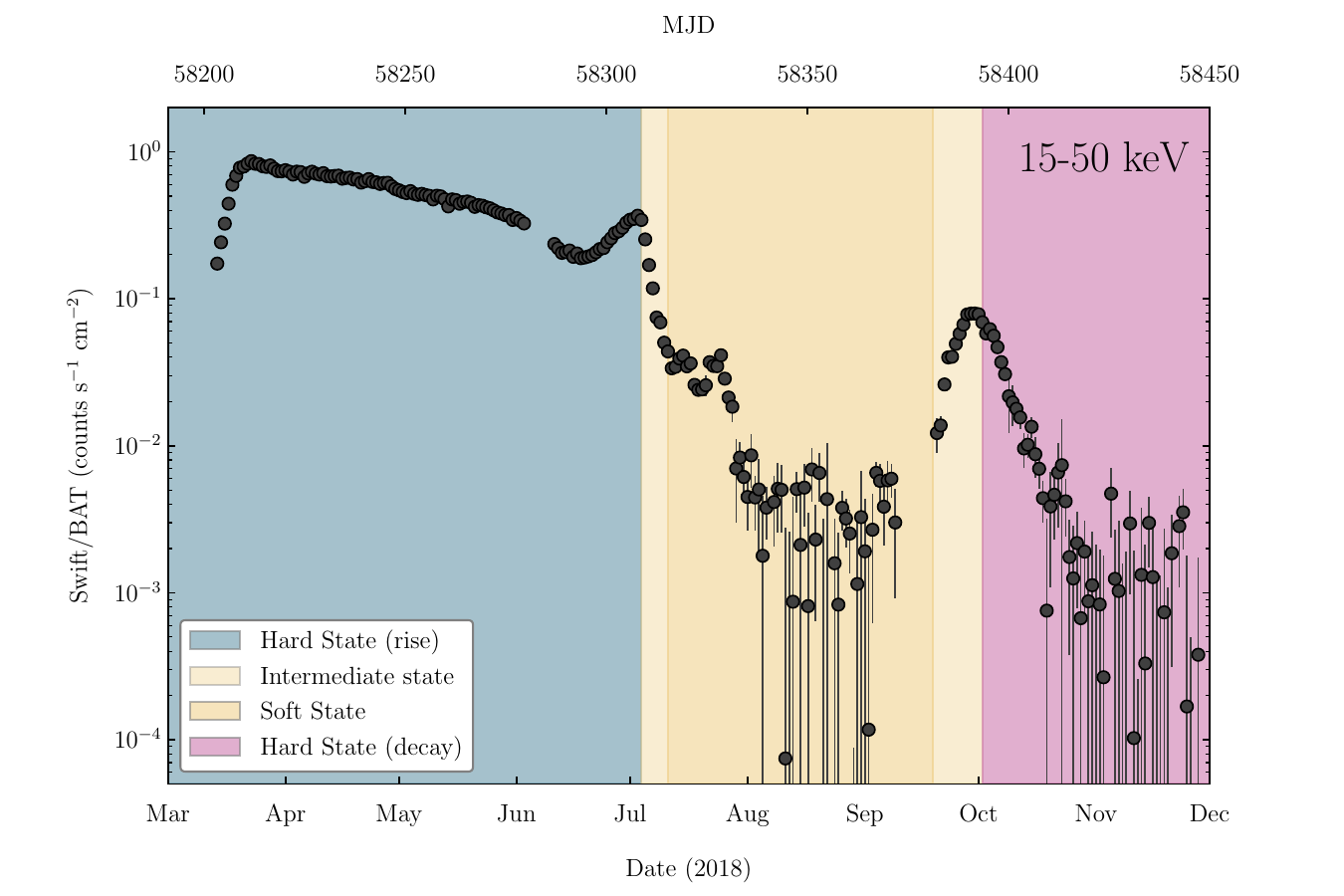}
    \caption{MAXI J1820+070 \textit{Swift}/BAT daily light curve in the 15--50 keV energy range throughout its 2018/2019 outburst. Data were obtained from the ``BAT Transient Monitor'' \citep{2013ApJS..209...14K}. The background shading represents the accretion states identified in \citealt{2019ApJ...874..183S}: rising hard state (blue), intermediate and soft states (yellow), and declining hard state (pink).}
    \label{fig:lightcurve}
\end{figure*}

%%%%%%%%%%%%%%%%%%%%%%%%%%%%%%%%% OBS AND DATA %%%%%%%%%%%%%%%%%%%%%%%%%%%%%%%%%%
\section{Observations and Data Analysis} \label{sec:ObsAndData}

\subsection{Radio/(sub)-millimetre}

\subsubsection{VLA}
J1820 was observed with the National Radio Astronomy Observatory's (NRAO) Karl G. Jansky Very Large Array (VLA; Project Code: 18A--470) on 2018 April 12 (observing 6 hrs on source). During these observations, the array was in the A configuration, and split into 3 sub-arrays, observing at the C (4--8 GHz), X (8--12 GHz), or K (18--26 GHz) bands. The correlator was set up in 8-bit mode and was comprised of 2 base-bands, with 8 spectral windows of 64 2-MHz channels each, giving a total bandwidth of 1.024 GHz per base-band. This VLA data was calibrated and imaged within the Common Astronomy Software Application package (\textsc{casa} v5.4; \citealt{mc07}). Flux densities of the source were measured by fitting a point source in the image plane (using the \texttt{imfit} task), and all flux density measurements are provided in Table \ref{tab:radio_fluxes}. Further details on the observations and calibration of these data are provided in \citealt{atet21}. Refer to Section \ref{subsubsec:literature} for additional VLA data from the literature.

\subsubsection{ALMA}
The Atacama Large Millimetre/Sub-Millimetre Array (ALMA) observed J1820 (Project Code: 2017.1.01103.T) between 2018 April 12 and July 06 (observing up to 5 hrs on source per epoch). During our observations, the 12-m array was in its Cycle 5 C3 configuration, with 46 antennas, observing in Band 7 (central frequency of 343.5 GHz).  The ALMA correlator was set up to yield 4$\times$2 GHz wide base-bands. These ALMA data were reduced and imaged within \textsc{casa}. Flux densities of the source were measured by fitting a point source in the image plane (using the \texttt{imfit} task), and all flux density measurements are provided in Table \ref{tab:radio_fluxes}. Details on the observations and calibration process of these data can be found in \citealt{atet21}.

\subsubsection{JCMT SCUBA-2}
The James Clerk Maxwell Telescope (JCMT; Project Code: M18BP025) observed J1820 between 2018 October 22 and November 14, in the 850$\mu$m (350 GHz) and 450$\mu$m (666 GHz) bands. The observations consisted of a series of $\sim$30 min scans on target with the SCUBA-2 detector \citep{chap13,holl13}. To perform absolute flux calibration, observations of the calibrator CRL2688 were used to derive a flux conversion factor \citep{demp13,mairs_jcmt}. The daisy configuration was used to produce 3-arcmin maps of the target source region. During the observations, the weather band was Grade 2 or 3 at the telescope, with a 225 GHz opacity of 0.05--0.12. Data were reduced in the \textsc{starlink} software package using both standard procedures outlined in the SCUBA-2 cookbook\footnote{\url{http://starlink.eao.hawaii.edu/devdocs/sc21.htx/sc21.html}} and SCUBA-2 Quickguide\footnote{\url{https://www.eaobservatory.org/jcmt/instrumentation/continuum/scuba-2/ data-reduction/reducing-scuba2-data}}. We note that the source was only detected at 850$\mu$m, and all flux density measurements are provided in Table \ref{tab:radio_fluxes}.

\subsubsection{IRAM NOEMA}
J1820 was observed with the Institute de Radioastronomie Millim\'etrique’s Northern Extended Millimetre Array (IRAM NOEMA) under the project codes W17BN and W17BM in 2018. W17BN was observed first, and recorded data in three different bands: W17BN001 at 97.5 GHz was observed on 2018 March 16, W17BN002 at 140.0 GHz on 2018 March 20, and finally W17BN003 at 230.0 GHz also on 2018 March 20. At that time the interferometer was in extended configuration, observations were performed respectively with 9, 8 and 8 antennas. Under W17BM the source was flux-monitored at 140.0 GHz on 2018 May 10, May 18 and May 21 in compact 8, 6, and 8 antenna configurations, respectively. For amplitude and phase calibration we used the quasars B1827+062 and B1749+096, and as flux calibrator the carbon star MWC349. The PolyFiX correlator was used in broadband mode, providing a bandpass of 7.744 GHz in dual linear polarisation in both upper and lower sideband with 2 MHz resolution. The spectral bandpass was calibrated on different strong quasars, e.g., 3C279, 3C273, 3C84, B2013+370 and B1749+096.
Calibration of the NOEMA data was performed in the dedicated CLIC program that is part of the gildas4 software package {\sc gildas}\footnote{\url{https://www.iram.fr/IRAMFR/GILDAS}} using
standard procedures. All data were then exported to {\sc casa}\footnote{To convert a NOEMA data set for use in {\sc casa}, we followed the procedures outlined at \url{https://www.iram.fr/IRAMFR/ARC/documents/filler/casa-gildas.pdf}.} for imaging (using natural weighting to maximize sensitivity).  Flux densities of the source were measured by fitting a point source in the image plane (using the \texttt{imfit} task). These measurements are presented in Table \ref{tab:radio_fluxes}.

\subsubsection{SMA}
The Sub-millimeter Array (SMA; Project Codes: 2017B-S010 and 2018A-S011) observed J1820 between 2018 April 12 and September 29 (observing up to 7 hours on source per epoch). All of our observations utilized the SWARM correlator in dual receiver mode, tuned to central frequencies of 224/230 for RxA/RxB, with 7 or 8 antennas observing in the array. This setup yields two 8 GHz side-bands per receiver, giving a total bandwidth of 32 GHz. The SMA data were converted to \textsc{casa} MS format using custom scripts provided by SMA\footnote{\url{https://lweb.cfa.harvard.edu/rtdc/SMAdata/process/casa/convertcasa/}}. Then all flagging, calibration, and imaging (using natural weighting to maximize sensitivity) of the data were performed within \textsc{casa} using procedures outlined in the \textsc{casa} Guides for SMA data reduction\footnote{\url{https://www.cfa.harvard.edu/sma/casa}.}. We used 3C454.3, 3C279, and 3C345 as bandpass calibrators, J1743$+$038 and J1751+096 as phase calibrators, and Neptune, Titan, and Callisto as flux calibrators\footnote{The SMA calibrator list can be found at \url{http://sma1.sma.hawaii.edu/callist/callist.html.}}. Flux densities of the source were measured by fitting a point source in the image plane (using the \texttt{imfit} task), and all flux density measurements are provided in Table \ref{tab:radio_fluxes}.

\subsubsection{AMI-LA}
J1820 was also observed with the Arcminute Microkelvin Imager Large Array (AMI-LA; \citealt{zwart2008, hickish2018}) during the 2018--2019 outburst. Observations were carried out at a central frequency of 15.5 GHz with 4096 channels spanning the range 13--18 GHz. The raw data from the correlator were binned into $8\times0.626$ GHz channels to produce `quick-look' data which were reduced by the custom software \textsc{reduce\_dc}, which includes flagging of hardware errors and radio frequency interference and performs flux, bandpass, and complex gain calibration (using 3C286 and J1824$+$1044). Additional flagging and imaging were performed using \textsc{casa}. The flux density of MAXI J1820$+$070 was measured using the \textsc{casa} task \textsc{imfit}. Here we use a sub-set of the AMI-LA observations taken during this outburst, which were quasi-simultaneous with our other multi-wavelength measurements. We note that the AMI-LA resolution is not sufficient to be able to resolve discrete jet ejections or distinguish the core compact jet from the ejections. Therefore, the AMI-LA flux density may be a combination of the compact jet and the jet ejections. While this is not an issue for the hard state observations, it may impact transition/soft state observations. All flux density measurements can be found in Table \ref{tab:radio_fluxes}.

\subsubsection{Additional data from the literature} \label{subsubsec:literature}
We include additional long-wavelength data found in the literature from the Low Frequency ARray (LOFAR; \citealt{2018ATel11887....1B}), the Karoo Array Telescope (MeerKAT; \citealt{2020NatAs...4..697B}), the Multi-Element Radio Linked Interferometer Network (eMERLIN; \citealt{2020NatAs...4..697B}), the VLA (Project Code: 18A--277; \citealt{2021ApJ...907...34S}), and the Very Long Baseline Array (VLBA, Project Code: BM467; \citealt{2020MNRAS.493L..81A}) facilities. The details of all these data sets are presented in Table \ref{tab:radio_fluxes}.

\subsection{Infrared/Optical/UV}
\subsubsection{VLT VISIR}

Mid-IR observations of the field of J1820 were made with the Very Large Telescope (VLT) on 14 dates from 2018 April to October, under the programs 0101.D-0634 and 0102.D-0514 (PI: D. Russell). The VLT Imager and Spectrometer for the mid-Infrared \citep[VISIR;][]{LagageVISIR} instrument on the VLT was used in small-field imaging mode (the pixel scale was 45 mas pixel$^{-1}$). Five filters ($M$, $J8.9$, $B10.7$, $B11.7$ and $Q1$) were used on different dates, with central wavelengths of 
4.67, 8.70, 10.64, 11.51 and 17.65 $\mu$m, respectively
\citep[see also][]{russell2018ATel11533}. For each observation, the integration time on source was composed of a number of nodding cycles, with chopping and nodding between source and sky. The total observing time was typically almost twice the integration time.

Observations of standard stars were made on the same nights as the target, in the same filters. All data (target and standard stars) were reduced using the VISIR pipeline in the \emph{gasgano} environment. Raw images from the chop/nod cycle were recombined. Photometry was performed on the combined images using the \texttt{phot} task in \textsc{iraf}, with an aperture large enough that small seeing variations did not affect the fraction of flux in the aperture. For some standard star observations, ESO provided pipeline-reduced images and counts/flux ratio values. Our counts/flux ratio values calculated separately agree with those of ESO's pipeline to a level of 0.4--2.4 per cent, for those standards, in all filters, with no apparent change with differing seeing. The standards were used to estimate the counts/flux ratio needed to convert count rates to flux densities. Some standard star observations were rejected due to a nearby star overlapping with the PSF of the standard (this was the case for standards HD 075691 and HD 000787), or because they were observed during twilight.

Since we have so many standard star observations, we investigate the variations in the counts/flux ratio throughout the whole observing period. The overall long-term and night-to-night stability of the photometric calibration of VISIR is known to be good \citep{Dobrzycka12}. From the tests we carried out, under photometric conditions the sky transparency variations are less than a few percent \citep[similar results were found in][]{Baglio18}. However, under poorer conditions, airmass and visibility affect some of our observations. For our observations taken in 2018, we find that the counts/flux ratio for the standards changed by $>10\%$ (and sometimes by much larger amounts) if either (a) the conditions were poor (thin or thick cloud, or high wind), or (b) the airmass was greater than 1.4. For the remaining observations (clear conditions, no strong wind, airmass $\leq 1.4$), we find that the counts/flux ratio agreed on all dates, for all standards, to a level of $\pm 6\%$, $\pm 4\%$, $\pm 4\%$, $\pm 3\%$ and $\pm 2\%$ compared to the mean value for the $M$, $J8.9$, $B10.7$, $B11.7$ and $Q1$ filters, respectively. These low level variations in the conversion factor in the standards from night to night are likely due to intrinsic differences between the conversion factors derived for different standard stars, or possible background variations due to the water vapour content of the atmosphere above Paranal. There also appears to be no trend between the counts/flux ratio and time, or with airmass for airmass values $\leq 1.4$. For standards taken at higher airmass (1.4--1.9) under clear conditions, we find that the counts/flux ratio differed by up to 8 per cent from the mean value.

For the target observations, to convert counts to flux densities in mJy, we adopt the mean value of the counts/flux ratio for each filter (for all standards taken under good conditions as explained above), for all target observations taken in clear conditions with airmass $\leq 1.4$. The error on the flux incorporates the error on the photometry (due to the S/N of the target) and the standard deviation of the counts/flux ratio (a systematic error) combined in quadrature. For observations of the target taken in poor conditions (thin or thick cloud, high wind, or airmass $>1.4$) we individually evaluate these carefully. In particular, data on 2018 April 21 were observed under thick cloud and airmass 1.70--1.77. Nevertheless, enough standards were observed before and after the target, in each filter, that we are able to assess the variations of the counts/flux ratio due to clouds, and assess that the ratio of the airmass of the target and standards could cause a factor of $1.69 \pm 0.69$ uncertainty in cloud cover for $M$-band and similar uncertainties for $J8.9$, $B10.7$ and $B11.7$ filters. We incorporate these uncertainties into the error calculations for the data taken on this date. Additionally, on 2018 May 11 high winds affected the counts/flux ratio; for this date we calibrate the target observations using the standards taken on the same date only, using the variations of the counts/flux ratio before and after the target observations to estimate the error contribution. The results are shown in Table \ref{tab:iropt_fluxes}.

\subsubsection{VLT X-Shooter}

We undertook an observing campaign of the 2018 outburst of MAXI J1820$+$070 using the X-Shooter instrument at the ESO’s Very Large Telescope (VLT, Project Code: 0101.D-0356; \citealt{vernet11}). The nine observing epochs spanned different accretion states throughout the outburst (hard, soft, intermediate). Depending on the source intensity, each observation consisted of either five, four, or two exposures arranged in AB pairs alternating between the source and sky positions with a nod throw length of five arcseconds and a jitter box of one arcsecond. The exposure time for each observation was $\sim$45 minutes in aggregate. We use the i'-band filter for the slit acquisition and correcting the normalization of the X-Shooter spectra due to slit losses. We use slit widths of 1.3, 1.3, and 1.2 arcseconds for UV, VIS, and NIR arms, respectively, and the detector readout mode was selected to be 100k/1pt/hg/1x2. 
We reduce the X-Shooter data with ESO pipeline v3.5.0 in \textsc{esoreflex} \citep{freudling13}. The telluric absorption is corrected using \textsc{molecfit} \citep{smette15,kausch15}. Since this work focuses only on the continuum emission, and due to the high spectral resolution of X-Shooter, we bin the data to a few tens of data points, where the representative values correspond to the mean and standard deviation in each bin. 
The resulting data sets are presented in Table \ref{tab:xshooter_fluxes}.

\subsubsection{LCO and Al Sadeem Observatory}

We monitored J1820 during its 2018 outburst extensively with the Las Cumbres Observatory (LCO) optical network of robotic telescopes \citep[e.g.][]{baglio2018ATel11418,baglio2018ATel12128,russell2018ATel11533,2019AN....340..278R}. This is part of an on-going monitoring campaign of $\sim$50 low-mass X-ray binaries coordinated by the Faulkes Telescope Project \citep{Lewis2008,Lewis2018}. The monitoring of J1820 includes data taken at the 1-meter LCO telescopes at Siding Spring Observatory (Australia), Cerro Tololo Inter-American Observatory (Chile), McDonald Observatory (Texas), and the South African Astronomical Observatory (SAAO; South Africa), as well as the 2 m Faulkes Telescopes at Haleakala Observatory (Maui, Hawai`i, USA) and Siding Spring Observatory (Australia). Images were taken in the SDSS $g^{\prime}$, $r^{\prime}$, $i^{\prime}$ and PanSTARRS $Y$-band filters (spanning 477--1004 nm central wavelengths). Here, data are included that were acquired within $\sim1$ d of the {VISIR} observations.

The data are initially processed using the LCO Banzai pipeline \citep{banzai}. Photometry is performed on the reduced data using the real-time data analysis pipeline, \textsc{XB-NEWS}) \citep[see][]{2019AN....340..278R,Pirbhoy2020,2020MNRAS.498.3429G}. The \textsc{XB-NEWS} pipeline downloads new images of targets of interest from the LCO archive along with their associated calibration data, performs several quality control steps to ensure that only good quality images are analysed, and computes an astrometric solution for each image using Gaia DR2 positions\footnote{\url{https://www.cosmos.esa.int/web/gaia/dr2}}. Aperture photometry is then performed on all the stars in each image, solving for zero-point calibrations between epochs \citep{Bramich2012}, and flux calibrating the photometry using the ATLAS All-Sky Stellar Reference Catalog \citep[ATLAS-REFCAT2;][]{Tonry2018}. The pipeline also performs multi-aperture photometry \citep[azimuthally-averaged PSF profile fitting photometry,][]{Stetson1990} for point sources.
We detect the source with high significance throughout the outburst; during the epochs of interest the magnitude varied from $g^{\prime} = 12.1$, $i^{\prime} = 12.4$, to $g^{\prime} = 14.6$, $i^{\prime} = 14.2$.

We also monitored the source extensively with the Al Sadeem Observatory\footnote{\url{http://alsadeemastronomy.ae/}} \citep[see also][]{russell2018ATel11533,2019AN....340..278R,baglio2018ATel12128,baglio2019ATel12596,baglio2021ATel14582}. The observatory is located in Al Wathba South, outside the city of Abu Dhabi in the United Arab Emirates. A Meade LX850 16-inch (41-cm) telescope was used, using a SBIG STT-8300 camera with Baader LRGB CCD filters (blue, green and red filters with similar central wavelengths to $g^{\prime}$, $V$ and $R$-bands). Bias and flat field images were combined, and the science images were reduced using these images. Photometry is then performed on the science images, using the \texttt{PHOT} task in \textsc{iraf}. Several stars from the APASS catalogue \citep{2012JAVSO..40..430H} in the field were used for flux calibration. For $R$-band, we derive $R$ magnitudes of the field stars from the APASS $g^{\prime}$, $r^{\prime}$ and $V$ magnitudes, adopting the conversions of \cite*{2006A&A...460..339J}.

\subsubsection{REM}
REM \citep[Rapid Eye Mount,][]{2004SPIE.5492.1613C} is a 60-cm robotic telescope, located at the ESO-La Silla Observatory, and is equipped with the optical camera, ROS2 \citep{2014SPIE.9147E..6XM}, and the IR camera, REMIR \citep{2003SPIE.4841..627V}. The two cameras observe simultaneously in the same field of view ($\sim 10^{\prime}\times 10^{\prime}$) thanks to a dichroic placed before the telescope focal plane. 

ROS2 observed J1820 simultaneously in its four filters (the Sloan/SDSS $g^{\prime}$, $r^{\prime}$, $i^{\prime}$, and $z^{\prime}$), and with REMIR we cycled through the $J$, $H$, and $K$ filters (Project Code: 37025); see Table \ref{tab:iropt_fluxes} for all REM measurements.
All the observations, as well as the preliminary reduction and calibration procedures are carried out in a fully automated way by the robotic system with the pipeline \textsc{aqua} (Automatic QUick Analysis; \citealt{2004SPIE.5496..729T}). The resulting products are pre-processed images and initial catalogues. Both the ROS2 and REMIR frames are astrometrically calibrated.

REMIR acquires a series of 30-second long frames by rotating a filter wedge along the optical path, thus obtaining five displaced images which are then combined together by median filtering. The resulting ``empty sky" image is subtracted from each original frame. The five (sky-subtracted and flat-fielded) frames are then registered and summed, obtaining the final science image. Through this process, a final exposure of 150 seconds is reached for each of the filters.
These resulting final images were reduced and analyzed with the PSF--fitting photometry package, \textsc{daophot} \citep{1987PASP...99..191S}, and calibrated against the 2MASS catalog.

ROS2 has a multi-channel system that splits the light in four different beams feeding four quadrants of a 2k $\times$ 2k CCD equipped with 
four filters; $g^{\prime}$, $r^{\prime}$, $i^{\prime}$, and $z^{\prime}$. The four images are thus acquired simultaneously, with an exposure of 180 seconds, and then reduced and calibrated using standard procedures with bias and flat-field frames obtained at twilight or during the daytime. The field-of-view of ROS2 is
approx. 9.1 $\times$ 9.1 sq. arcmin. with a pixel scale of 0.58 arcsec.
The ROS2 calibration was performed via secondary standards in the field on objects having SDSS or PANSTARRS magnitudes.
Photometric standards are also taken on every candidate-photometric night, and a calibration relation is derived with zero points, color term,
and atmospheric extinction term. This service is provided by the observatory as another data product.

\subsubsection{AAVSO}
We include additional optical data from the American Association of Variable Star Observers (AAVSO) international data base \citep{AAVSO_CITE}, in the $B$ (0.44 $\mu$m), $V$ (0.55 $\mu$m), and $I$ (0.80 $\mu$m) bands. Here we collect the photometry available within 1--2 days before and after the main date of our broad-band spectra. For the dates with several entries, we average the measurements to get a single magnitude representative of each date with errors corresponding to the standard deviation. The average value is then converted into a flux density. A summary of the results can be found in Table \ref{tab:iropt_fluxes}.

\subsubsection{Swift/UVOT}

The UV/optical instrument UVOT onboard \textit{Swift} observed J1820 simultaneously with the X-ray instrument XRT through a good portion of its 2018 outburst. The exposures span across 7 months, and in most cases include data from the six filters available from the UV ($UW1$, $UM2$, $UW2$) and optical ($V$, $B$, $U$). To analyze UVOT observations we use the \textsc{heasoft} software v6.25\footnote{\url{https://heasarc.gsfc.nasa.gov/docs/software/heasoft/}} and followed the guidelines provided in the \textit{UVOT Data Analysis Guide}\footnote{\url{https://www.swift.ac.uk/analysis/uvot/}}.

We first run the task \texttt{uvotdetect} on the images to obtain the centroid position of the source. To match the UVOT calibration, we select circular regions of 5$\arcsec$ centered on the \texttt{uvotdetect} position to define the source extraction region. For the background we use a circular aperture of 20$\arcsec$, chosen near the source and ensuring that the regions are not contaminated by nearby sources. Aperture photometry is performed using the task \texttt{uvotsource} to extract counts from those regions.
%Coincidence loss corrections
Due to the counting nature of CCD detectors, the UVOT instrument suffers from coincidence loss \citep{2000MNRAS.312...83F}, a similar phenomena to X-ray pile-up. Coincidence loss occurs when multiple photons arrive at the same pixel within one read-out frame of the detector. Since only one photon is recorded instead of two (or more), the true photon flux is underestimated. Furthermore, in the case of bright sources, the UVOT Point Spread Function (PSF) is highly distorted. Because this effect is more likely to occur at high rates, we first analyze the count rate of the source and background regions in all our exposures. Following the analysis of \citealt{2010MNRAS.406.1687B}, we discard all observations with source count rate higher than $\sim$40 counts s$^{-1}$, and background count rate higher than 0.01 counts s$^{-1}$. The observations that are simultaneous to XRT (see below) are detailed in Table \ref{tab:UVOTdata_sim}, while the non-simultaneous are summarized in Table \ref{tab:UVOTdata_nonsim}. Once we have the final exposures corrected for this effect, we obtain flux densities for all of them. These are summarized in Table \ref{tab:uv_fluxes}. The reported fluxes include 1$\sigma$ statistical errors, and the systematic uncertainty that arises from the shape of the instrument’s PSF. As a result, we have a total of 40 UVOT observations simultaneous to XRT, totalling 130 flux density measurements.

\subsection{X-ray}
\label{subsec:xray}

\subsubsection{Swift/XRT}
The 20 X-ray exposures simultaneous with the UV are presented in Table \ref{tab:XRTdata}, all of which were taken in the Window Timing Mode (WT). We first run the \textsc{heasoft} task \texttt{xrtpipeline} to build the standard data products with the latest calibration applied. On each event file, we select a circular region with a 30 pixel radius for both the source and background regions. Using \texttt{xselect} we extract count rates and build the source and background spectra, filtering grade 0 events to reduce the effect of pile-up. However, in WT mode pile-up becomes important for intensities of $\sim$100 counts s$^{-1}$ and above. Therefore, given the count rate of some of the observations, further pile-up analysis is required. 

%Pile-up analysis
To determine the level of pile-up we follow the \textit{Swift}/XRT analysis threads\footnote{\url{https://www.swift.ac.uk/analysis/xrt/pileup.php.}} for WT mode, specifically the spectral distortion method (also described in \citealt{2006A&A...456..917R}). The analyses are performed using the X-ray Spectral Fitting Package (\textsc{xspec} v12.10.1; \citealt{1996ASPC..101...17A}).

The overall effect of pile-up is to distort the shape of the spectrum, because multiple soft energy photons are stored as a single high energy photon. As a result, we find an excess of high energy photons and the power-law X-ray spectrum hardens, i.e., the photon index decreases. To mitigate this problem we select an annular source region, in which the inner circle corresponds to an exclusion region. This means that counts within the inner circle are not considered in the count extraction. The size of the inner circle can be varied to increase or decrease the count rate within the extraction region.
Thus, each size defines a new source region from which a spectrum will be extracted. By fitting a power-law to the X-ray spectrum, it is possible to study the behaviour of the photon index (slope) parameter when the extraction region changes, which allows us to determine the level of pile-up. When increasing the size of the inner circular region no longer impacts the value of the photon index, the region affected by pile-up has effectively been excluded. Once the final pile-up-corrected source region is determined, we modify the size of the background regions to match the size of the new source region (as this is required for WT data), and re-extract the source and background X-ray spectra. We then build the ancillary response files (ARFs) through the \texttt{xrtmkarf} task, which shows the corresponding redistribution matrix file (RMF) as well. We finalize the process using \texttt{grppha} of \textsc{ftools}\footnote{\url{http://heasarc.gsfc.nasa.gov/ftools}} \citep{black95} to assign bad channels (0-29) and group the spectra to a minimum of 20 counts per bin. This number of counts allows us to use $\chi^2$ statistics by ensuring Gaussian errors in each bin. In addition to the bad channels, we decided to ignore all channels below 1.0 keV, due to uncertainties in the low energy regime of WT data\footnote{\url{https://www.swift.ac.uk/analysis/xrt/digest\_cal.php.}} and to further prevent the effects of pile-up, as well as channels above 10 keV.

\subsection{Broad-band spectra}

To construct our broad-band spectra, each epoch is defined on the basis of the radio/sub-mm observations, where we collect OIR, UV, and X-ray data within $\pm2$ days of these data. In this way, the data are grouped in 19 representative epochs of the outburst. The accretion states of J1820 during its 2018 outburst are characterized by \citealt{2019ApJ...874..183S}, using observations from MAXI/GSC and \textit{Swift}/BAT, and are shown in Table~\ref{tab:hid}.

%%%%%%%%%%%%%%%%%%%%%%%%%%%%%%%%% RESULTS %%%%%%%%%%%%%%%%%%%%%%%%%%%%%%%%%%
\section{Results} \label{sec:Results}

In this section, we outline the phenomenological model applied to describe the 19 broad-band spectra of J1820, as well as the fitting methodology and the results of our broad-band spectral modeling.

%%%%%%%%%%%%%%%%%%%%%%%%%%%%%%%%% SPEC MODELING %%%%%%%%%%%%%%%%%%%%%%%%%%%%%%%%%%
\subsection{Spectral Modeling} \label{sec:SpecModel}

\begin{figure*}[h]
    \centering
        \includegraphics[width=0.51\textwidth]{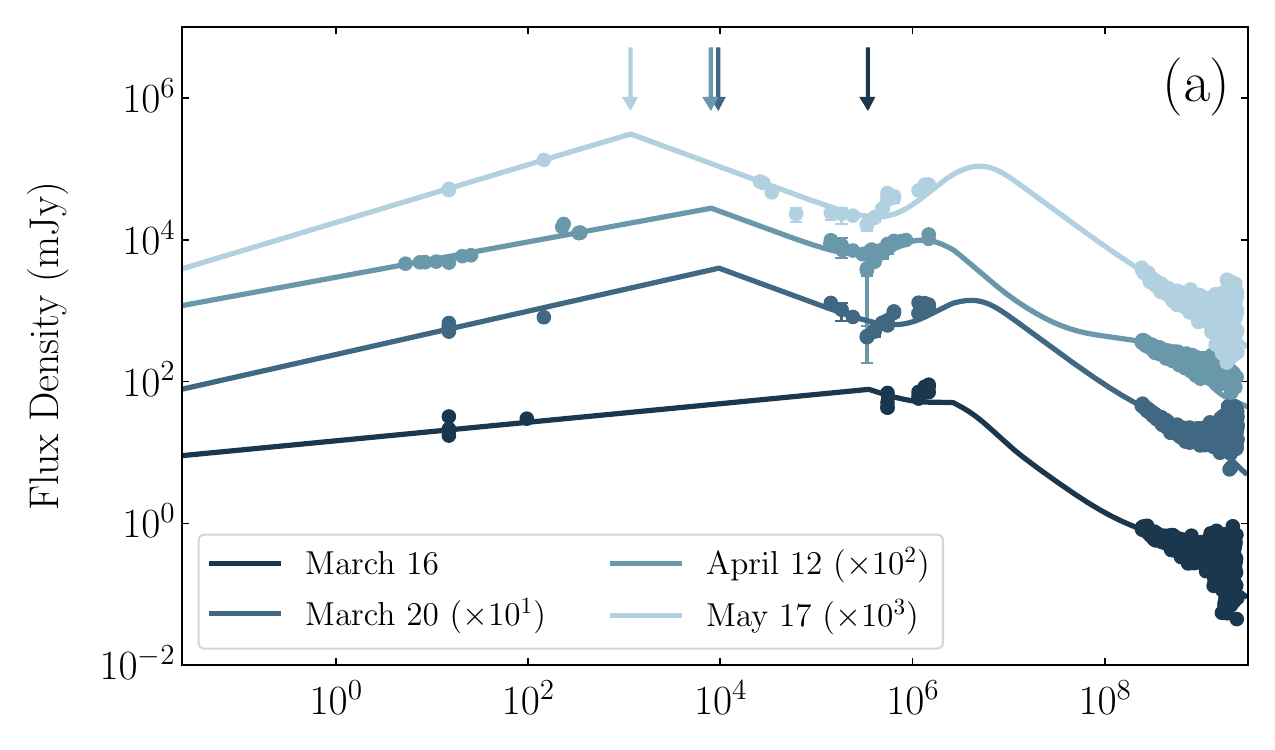}
        \includegraphics[width=0.48\textwidth]{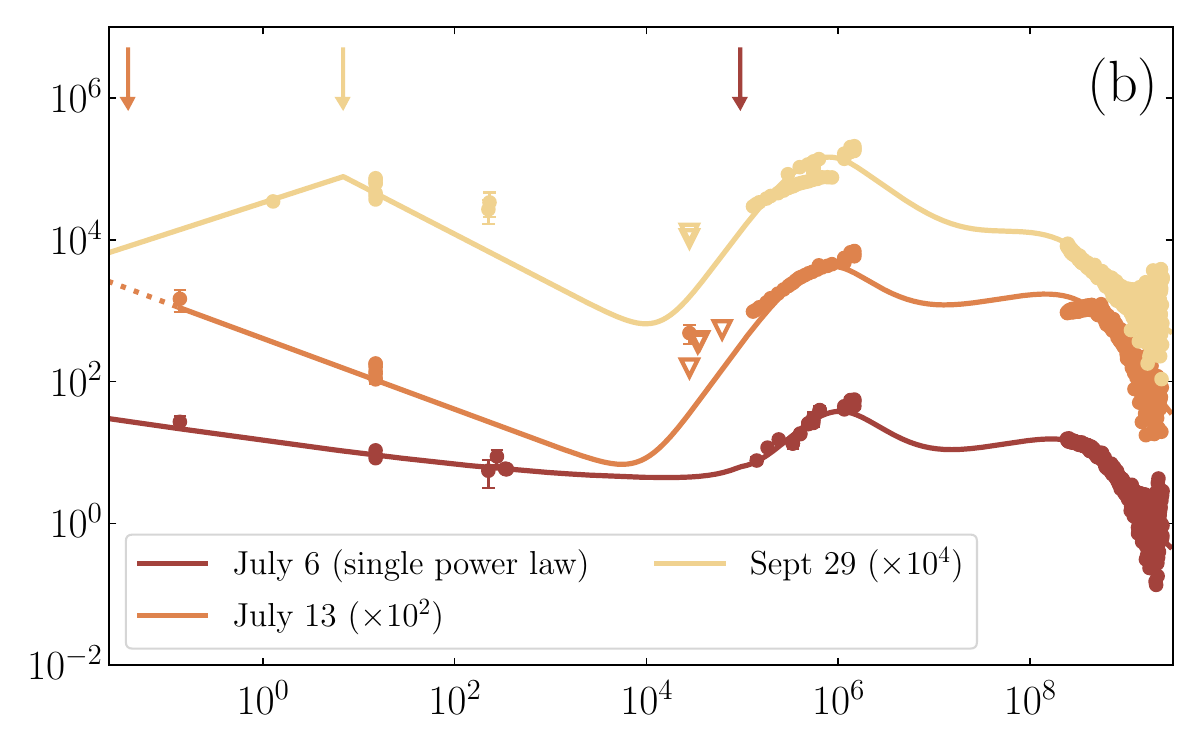}
        \includegraphics[width=0.51\textwidth]{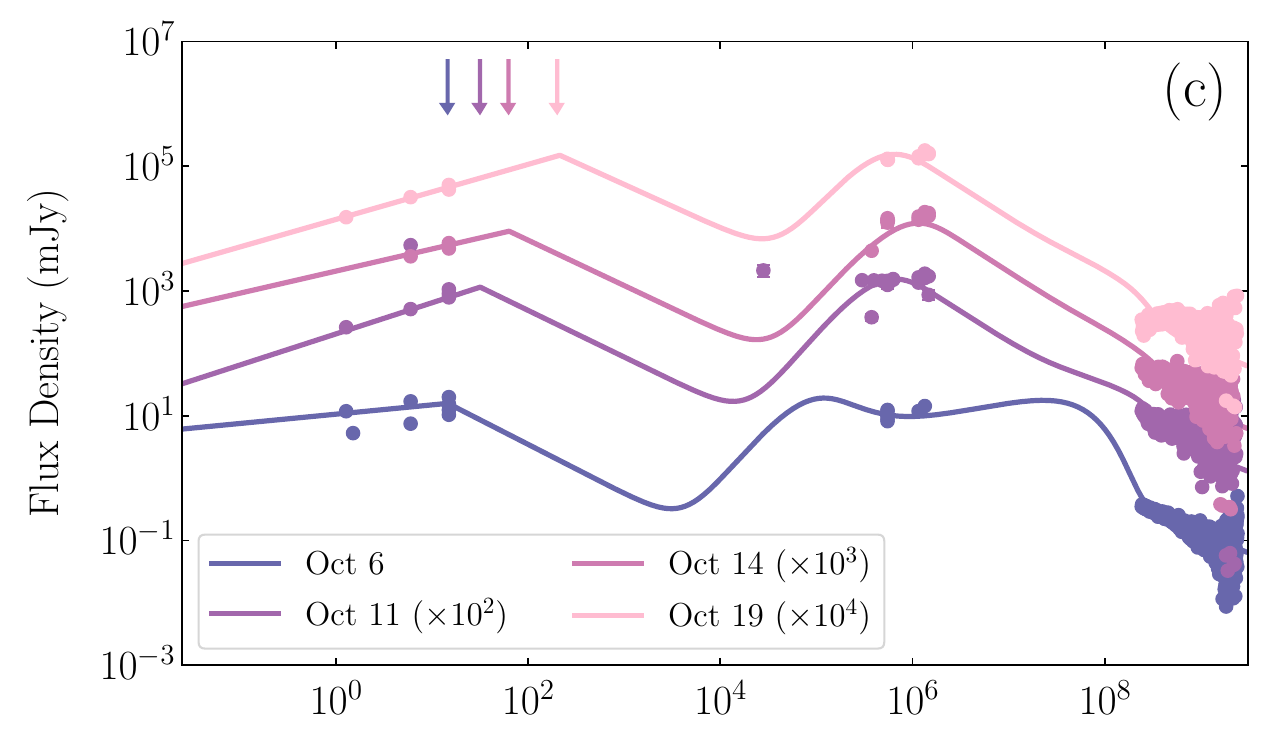}
        \includegraphics[width=0.48\textwidth]{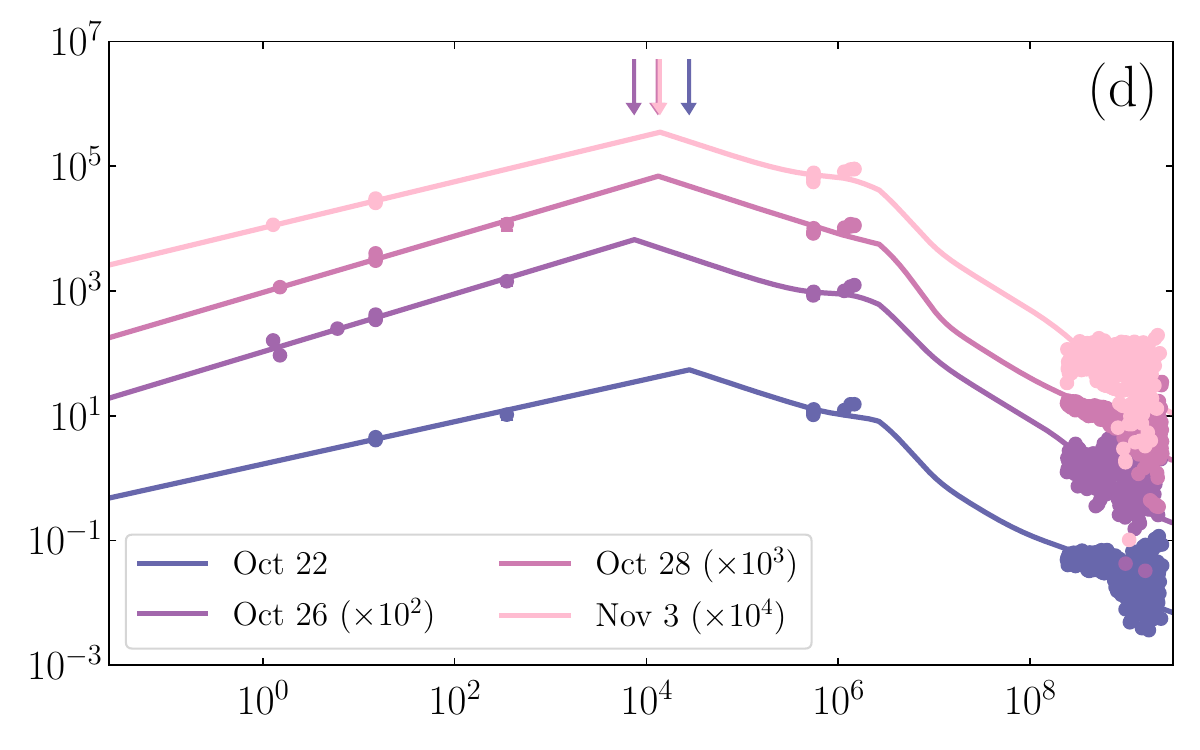}
        \includegraphics[width=0.49\textwidth]{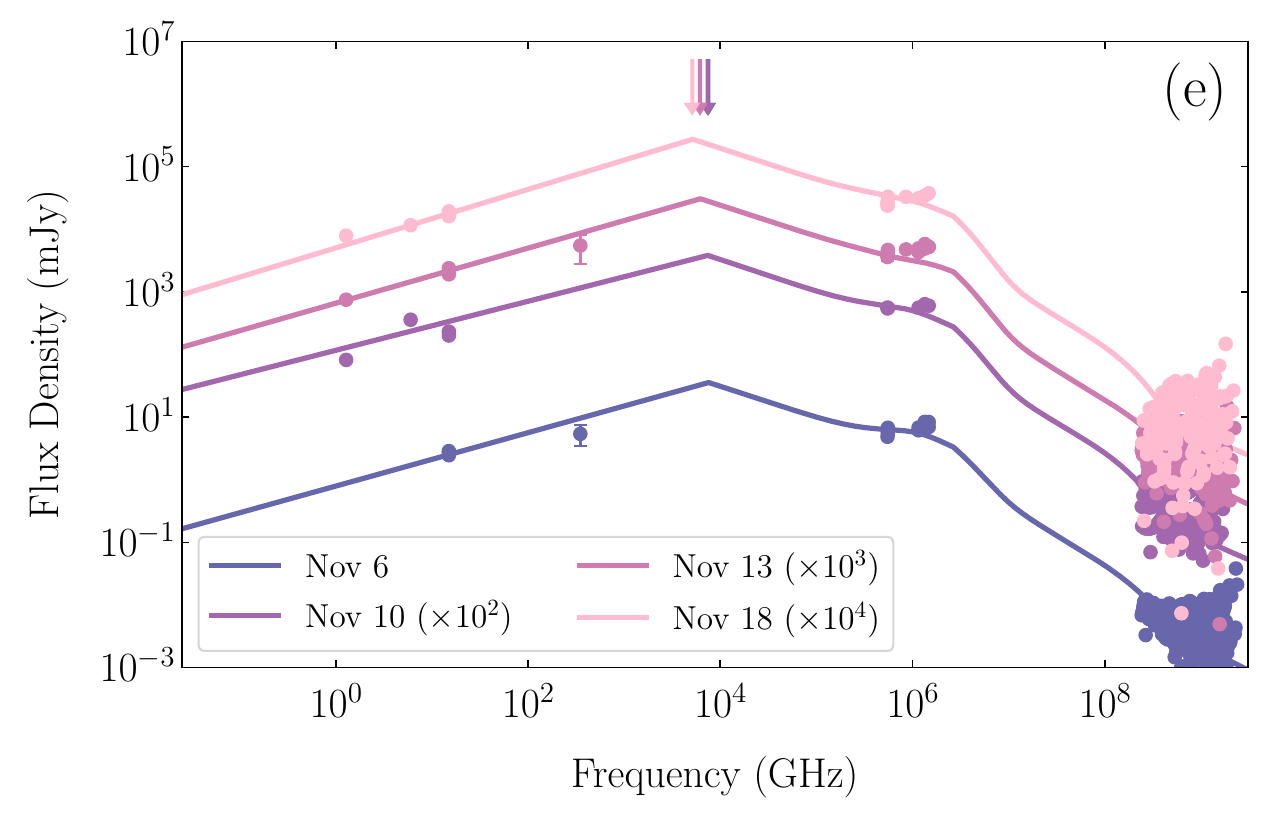}
    \caption{Broad-band spectral evolution of J1820 over the course of its 2018/2019 outburst. In the {\it panels} of each spectrum, the points represent the data and the solid lines represent the best-fit model. We show the fit residuals in Figure \ref{fig:allseds_res}. Colors indicate different epochs/accretion states and arrows mark the position of the spectral break for each individual epoch of the same color. The best-fit models and data points are scaled for better visualization, as identified in the legends (increasing with time). The optical, UV and X-ray data are corrected for reddening and absorption. {\it Panel a} displays the broad-band spectral models corresponding to the rising hard state (color code blue); {\it Panel b} shows the spectral models of the intermediate (July 6 and September 29) and soft (July 13) states (color code yellow); {\it Panels c, d} and {\it e} show the spectral models corresponding to the declining hard state (color code pink). Note that the color code for the accretion states are matched in Figures \ref{fig:lightcurve}, \ref{fig:par_evol}, \ref{fig:BandR} and \ref{fig:qpo_lag}. The position of the spectral break on July 6 must be interpreted carefully since the jet contribution is modeled with a combination of emission from jet ejecta, which dominates at lower electromagnetic frequencies, and a fainter compact jet component (see Section \ref{subsec:July6}). Similarly, on July 13, we do not have sufficient data at low frequencies to constrain the position of the spectral break. Thus, the arrow corresponds to an upper limit, and the dotted line is an extension of the power-law above the break, reflecting our inability to predict its shape. On July 13 and September 29, the lower triangles represent upper limits on the VISIR data for these epochs, which are not included in the fit. We clearly observe different broad-band spectral shapes of J1820 throughout the outburst.}
    \label{fig:allseds}
\end{figure*}

We use \textsc{xspec} to model the broad-band spectrum across multiple epochs. To model each epoch (containing data from radio to X-rays) in the same model phase space in \textsc{xspec}, we first create spectrum files for all the data. The spectrum files corresponding to radio/sub-mm, IR and optical data are created using the tool \texttt{flx2xsp} available through \textsc{ftools}. For the UV, the UVOT software employs its own routine \texttt{uvot2pha}, which allows us to create spectral files from UVOT images. For the X-ray, we directly use the instrument response and spectrum files obtained from the procedure described in Section \ref{subsec:xray}.

\begin{deluxetable}{lll}
\tablecaption{Accretion states for all the observed epochs of J1820 during its 2018/2019 outburst.\label{tab:hid}}
\tablewidth{0pt}
\tablecolumns{3}
\tablehead{\colhead{Spectral state\tablenotemark{\footnotesize{a}}} & \colhead{Color code} & \colhead{Observed Epochs (2018)}}
\def\arraystretch{1.2}
\startdata
Rising Hard & light/dark & March 16, 20 \\
& blue & April 12, May 17\\[0.18cm]
Intermediate & red/yellow & July 6, September 29  \\[0.18cm]
Soft & orange & July 13 \\[0.18cm]
Declining Hard & pink/purple & October 6, 11, 14, 19, 22,\\
& & 26, 28  \\
& & November 3, 6, 10, 13, 18\\
\enddata
\tablenotetext{a}{We use the accretion states defined in \citealt{2019ApJ...874..183S} for this work.}
\end{deluxetable}

The multi-component phenomenological model (see Figure \ref{fig:model_components}) we employ to describe the broad-band spectra consists of: (1) a broken power-law (\texttt{bknpower}\footnote{The photon index parameters ($\Gamma_{1,2}$) in the \texttt{bknpower} part of the total model can be mapped to the more traditional spectral index formalism for the jet spectrum via $\Gamma=1-\alpha$.} in \textsc{xspec}), representing the synchrotron emission from the compact jet in the radio to IR bands, with a high energy cutoff (\texttt{highecut} in \textsc{xspec}) to prevent the synchrotron emission from extending unbroken into the X-ray bands, since the cooling break is expected to lie somewhere below X-rays; (2) a blackbody (\texttt{bbodyrad} in \textsc{xspec}), representing the emission from the stellar companion in the optical band; (3) an irradiated disk (\texttt{diskir} in \textsc{xspec}), representing the accretion flow emission in the optical to X-ray bands \citep[which combines the \texttt{diskbb} and \texttt{thcomp} models,][]{2008MNRAS.388..753G,2009MNRAS.392.1106G}. Additionally, absorption due to the presence of gas and dust in the interstellar medium is modeled with \texttt{redden} acting on IR/optical/UV bands, \citep{1989ApJ...345..245C}, and \texttt{tbabs} \citep{2000ApJ...542..914W} acting on the X-ray bands. The full phenomenological model in \textsc{xspec} formalism is: \texttt{\small{redden*tbabs(highecut*bknpower+bbodyrad+diskir)}}.

In our model, we have 19 total parameters\footnote{Note there are 4 additional parameters defining the extra single power-law component (implemented through \texttt{pegpwrlw} in \textsc{xspec}, which is defined by photon index, normalization, and lower/upper energy limits) added to the total model in the July 6 epoch. This extra component is used to model the emission from the jet ejecta; see \S\ref{subsec:July6} for details.}, where up to 11 of these parameters are fixed to known or expected values. In particular, the absorption parameters $E(B-V)=0.18$ \citep{2018ApJ...867L...9T} and $N_{\mathrm{H}}=1.5 \times 10^{21}$ $\mathrm{cm}^{-2}$ \citep{2018ATel11423....1U} are fixed to their known values in the direction of the source. The \texttt{highecut} energy and e-folding energy were both fixed at 0.01 keV.
The \texttt{bbodyrad} model parameters (surface temperature and normalization) are also fixed based on the known companion star spectral type \citep[K3-5][]{2019ApJ...882L..21T} and the known distance to J1820 \citep{2020MNRAS.493L..81A}.
Parameters from the irradiated disk portion of the model that are fixed across all epochs to typical BH XRB values from the literature \citep[e.g.,][]{2008MNRAS.388..753G} include: the temperature of the corona $T_{\mathrm{e}}=100$ keV, radius of the illuminated disk $R_\mathrm{{irr}}/R_{\mathrm{in}}=1.2$ (where $R_{\mathrm{in}}$ is the disk inner radius), the fraction of luminosity in the Compton tail that is thermalized in the inner disk $\mathrm{f_{in}}=0.1$, and the radius of the outer disk $\log(R_{\mathrm{out}}/R_{\mathrm{in}})=4.5$. Lastly, in some epochs we need to fix the spectral index of the optically thin piece of the \texttt{bknpower} jet emission model to standard values ($\alpha_{\rm thin}=-0.5$), as we do not have enough data to accurately constrain this parameter.

The broad-band spectra are fit individually with the \textsc{xspec} implementation of the Markov Chain Monte Carlo algorithm (MCMC, where \textsc{xspec} uses the Goodman-Weare algorithm; \citealt{2010CAMCS...5...65G}). To initialize the parameters for each MCMC run, we manually explore the parameter space for our best sampled epoch (April 12) to determine a reasonable initial guess for the MCMC algorithm. For each MCMC run, we standarized the number of walkers to 14 per free parameter and run the chains for $10^6$ steps, with a burn-in corresponding to 30\% of the chain length, since this was sufficient for the chains to converge.
The convergence of the parameters is assessed with the Geweke convergence criteria \citep{Geweke92evaluatingthe} that is output by \texttt{chain} in \textsc{xspec}, as well as visually inspecting the chains of each parameter and the posterior distributions (see, for example, Figure \ref{fig:April12cornerplot}). The best-fit parameters reported correspond to the median of the posterior distribution and the uncertainties represent the 68\% confidence interval,  i.e., the 16\% and 84\% quantiles of the posterior distribution. 

The priors used for each model parameter are outlined in Table~\ref{tab:priors}. The majority of our chosen priors are based on typical values observed for BH XRBs. For example, to describe the shape of the broken power-law jet emission spectrum, typical photon index ranges are $\Gamma_1=0.5-1$, and $\Gamma_2=1.5-1.8$, and thus the resulting priors used are uniform distributions covering these ranges. However, for the photon index of the X-ray spectrum ($\Gamma$), we use measurements reported in previous work of J1820 in the literature; \citealt{2021NatCo..12.1025Y} for the rising hard state, \citealt{2019ApJ...874..183S} for the intermediate/soft states, and \citealt{2021ApJ...907...34S} for the declining hard state, where priors are also uniform. In some of our epochs the jet spectral break priors are adjusted according to the available data. For instance, on July 13 the spectral break cannot be higher than $\sim0.14$ GHz (our first radio data point). Similarly, for October 6--19, the sparse data in the radio-IR region prevents us from accurately constraining the jet spectral break. Thus, we use as priors the last radio data point at $\sim15$ GHz and $\sim10^5$ GHz (start of optical band). The best-fit models are shown Figure~\ref{fig:allseds} and the best-fit parameters are listed in Table~\ref{tab:results} (refer to Table~\ref{tab:snu} for flux densities at spectral break).

\renewcommand\tabcolsep{3pt}
\begin{deluxetable}{cccc}
\tablecaption{Priors used in the MCMC simulations. \label{tab:priors}}
\tablewidth{0pt}
\tablecolumns{4}
\tablehead{\colhead{Parameter} & \colhead{Model} & \colhead{Minimum value} & \colhead{Maximum value}}
\def\arraystretch{1.2}
\startdata
$\Gamma_1$ & \texttt{bknpower} & 0.5 & 1.0  \\
BreakE (keV)\tablenotemark{\footnotesize{a}}& \texttt{bknpower} & 0. & $10^{6}$ \\
$\Gamma_2$ & \texttt{bknpower} & 1.5 & 1.8 \\
BPL Norm & \texttt{bknpower} & 0 & $10^{24}$ \\
$kT_{\mathrm{disk}}$ (keV)& \texttt{diskir} & 0.01 & 5\\
$L_{\mathrm{c}}/L_{\mathrm{d}}$ & \texttt{diskir} & 0 & 10 \\
$f_{\mathrm{out}}$ & \texttt{diskir} & 0 & 0.1 \\
Disk Norm & \texttt{diskir} & 0 & $10^{24}$\\
$\Gamma$\tablenotemark{\footnotesize{b}} & \texttt{diskir} & \dots & \dots \\
\enddata
\tablenotetext{a}{For some epochs the priors on the energy break depend on the data available to constrain this parameter. See text for details.}
\tablenotetext{b}{Priors for the X-ray photon index parameter, $\Gamma$, are taken from previous measurements in the literature. See \S\ref{sec:SpecModel} for details.}
\end{deluxetable}
 \renewcommand\tabcolsep{6pt}

 \renewcommand\tabcolsep{3.5pt}
\begin{deluxetable*}{cccccccccc}
\centerwidetable
\tabletypesize{\scriptsize}
\tablecaption{Best-fit parameters of the multi-component phenomenological model obtained from our MCMC runs. The jet component is modeled with \texttt{bknpower}, where $\alpha_{\mathrm{thick}}$ represents the spectral index of the optically thick synchrotron emission, $\alpha_{\mathrm{thin}}$ is the spectral index of the optically thin synchrotron emission, and BPL Norm is the normalization at 1 keV. 
The accretion flow is modeled with \texttt{diskir}, where $kT_{\mathrm{disk}}$ is the innermost temperature of the unilluminated disk, $\Gamma$ is the photon index of the Comptonized X-ray emission, $L_{\mathrm{c}}/L_{\mathrm{d}}$ is the ratio of the luminosity in the Comptonized emission ($L_{\mathrm{c}}$) to the disk intrinsic luminosity ($L_{\mathrm{d}}$), f$_{\mathrm{out}}$ is the fraction of flux intercepted by the outer disk, and Disk Norm is the disk normalization. The best-fit values are obtained from the median of the posterior distributions output from the MCMC runs, while the uncertainties are the 16\% and 84\% quantiles.\label{tab:results}}
\tablewidth{0pt}
\tablecolumns{10}
\tablehead{& \multicolumn{4}{c}{\texttt{bknpower}} & \multicolumn{5}{c}{\texttt{diskir}}\\
\cmidrule(lr){2-5}
\cmidrule(lr){6-10}\\
\colhead{Date} & \colhead{$\alpha_{\mathrm{thick}}$} & \colhead{$\nu_{\mathrm{b}}$} & \colhead{$\alpha_{\mathrm{thin}}$} & \colhead{BPL Norm} & \colhead{$kT_{\mathrm{disk}}$} & \colhead{$\Gamma$} & \colhead{$L_{\mathrm{c}}/L_{\mathrm{d}}$} & \colhead{f$_{\mathrm{out}}$} & \colhead{Disk Norm} \\
(2018) & & (Hz) & &  ($\times 10^3$)  & (keV) &  &  & ($\times 10^{-2}$) & ($\times 10^{2}$) \\}
\def\arraystretch{1.5}
\startdata
% rising hard state
March 16 & $0.130 \pm 0.002$ & $\leq 3.45 \times 10^{14}$ & (-0.5) & $0.30^{+0.01}_{-0.01}$ & $0.79 \pm 0.02$ & $ \ge 1.49$ & $9.97^{+0.02}_{-0.05}$ & $2.04 \pm 0.09$ & $2.94^{+0.29}_{-0.27}$ \\
March 20 & $0.31^{+0.01}_{-0.02}$ & $9.54^{+2.06}_{-1.11} \times 10^{12}$ & (-0.5) & $14.98^{+4.}_{-4.71}$ & $0.91 \pm 0.01$ & $\ge 1.55$ & $9.99^{+0.01}_{-0.03}$ & $3.46^{+0.07}_{-0.06}$ & $7.02^{+0.29}_{-0.27}$ \\
April 12 & $0.2513 \pm 0.0004$ & $7.98^{+0.06}_{-0.06} \times 10^{12}$ & (-0.5) & $6.16 \pm 0.03$ & $0.423^{+0.002}_{-0.002}$ & $\ge 1.6$ & $9.98^{+0.02}_{-0.04}$ & $0.663 \pm 0.006$ & $144.81^{+2.39}_{-2.38}$ \\
May 17 & $0.408 \pm 0.003$ & $1.16^{+0.03}_{-0.03} \times 10^{12}$ & (-0.5) & $76.51^{+3.87}_{-3.75}$ & $0.88^{+0.01}_{-0.01}$ & $1.660^{+0.001}_{-0.001}$ & $9.93^{+0.05}_{-0.10}$ & $6.45 \pm 0.14$ & $4.04 \pm 0.12$ \\
\midrule % intermediate/soft
July 6 & $0.12 \pm 0.07$ & $\leq 1.53 \times 10^{14}$ & (-0.7) & $\leq 0.03$ & $0.69 \pm 0.01$ & $\ge 2.17$ & $0.66 \pm 0.03$ & $0.193^{+0.004}_{-0.004}$ & $164.72^{+6.25}_{-5.88}$ \\
July 13 & $0.45^{+0.04}_{-0.11}$ & $\leq 1.40 \times 10^{8}$ & (-0.5) & $\leq 6752 $ & $0.542 \pm 0.001$ & $\ge 2.60$ & $0.78 \pm 0.01$ & $0.244 \pm 0.001$ & $361.38^{+1.94}_{-1.90}$ \\
Sep 29 & $0.44^{+0.04}_{-0.10}$ & $6.88^{+1.60}_{-0.49} \times 10^{9}$ & (-0.5) & $\leq 61$ & $0.297 \pm 0.001$ & $1.94^{+0.03}_{-0.02}$ & $2.94 \pm 0.11$ & $0.82^{+0.03}_{-0.02}$ & $141.92^{+0.85}_{-0.83}$ \\
\midrule % declining hard state
Oct 6 & $0.152 \pm 0.002$ & $\ge 1.46 \times 10^{10}$ & $\ge -0.80$ & $0.32^{+0.01}_{-0.01}$ & $0.101 \pm 0.001$ & $\ge 1.68$ & $0.98^{+0.05}_{-0.04}$ & $0.05 \pm 0.01$ & $62810.96^{+3289.45}_{-3110.23}$ \\
Oct 11 & $\leq 0.50$ & $\ge 2.06\times 10^{10}$ & $\ge -0.79$ & $53.19^{+0.40}_{-0.77}$ & $0.153 \pm 0.001$ & $\ge 1.63$ & $9.95^{+0.04}_{-0.08}$ & $2.22 \pm 0.03$ & $258.82^{+3.95}_{-3.98}$ \\
Oct 14 & $0.355 \pm 0.007$ & $\ge 2.80 \times 10^{10}$ & $\ge -0.78$ & $3.30^{+0.42}_{-0.37}$ & $0.199 \pm 0.002$ & $\ge 1.62$ & $9.96^{+0.02}_{-0.05}$ & $6.19^{+0.16}_{-0.15}$ & $40.05^{+1.45}_{-1.43}$ \\
Oct 19 & $0.44 \pm 0.01$ & $\ge 6.35 \times 10^{10}$ & $\ge -0.77$ & $12.24^{+2.81}_{-2.34}$ & $0.123 \pm 0.002$ & $\ge 1.67$ & $9.93^{+0.06}_{-0.11}$ & $4.78^{+0.15}_{-0.14}$ & $263.59^{+26.24}_{-23.30}$ \\
Oct 22 & $0.34 \pm 0.03$ & $2.79^{+0.86}_{-0.66} \times 10^{13}$ & $\leq -0.51$ & $\leq 3.4$ & $0.47 \pm 0.02$ & $\ge 1.58$ & $9.86^{+0.11}_{-0.22}$ & $2.46^{+0.26}_{-0.24}$ & $1.47^{+0.22}_{-0.20}$ \\
Oct 26 & $0.464 \pm 0.009$ & $7.46^{+0.91}_{-0.84} \times 10^{12}$ & $\leq -0.5$ & $13.82^{+2.17}_{-1.87}$ & $0.22 \pm 0.02$ & $\ge 1.64$ & $9.67^{+0.25}_{-0.50}$ & $8.14^{+0.85}_{-0.79}$ & $7.74^{+3.83}_{-2.41}$ \\
Oct 28 & $0.45 \pm 0.01$ & $1.32^{+0.12}_{-0.11} \times 10^{13}$ & $\leq -0.5$ & $9.92^{+2.12}_{-1.74}$ & $0.60^{+0.03}_{-0.02}$ & $\ge 1.66$ & $9.72^{+0.21}_{-0.42}$ & $4.32^{+0.27}_{-0.25}$ & $0.13 \pm 0.02$ \\
Nov 3 & $0.37^{+0.02}_{-0.01}$ & $1.38^{+0.21}_{-0.19} \times 10^{13}$ & $\leq -0.5$ & $2.20^{+0.65}_{-0.47}$ & $0.19 \pm 0.01$ & $\ge 1.75$ & $9.72^{+0.21}_{-0.39}$ & $\ge 9.47$ & $7.58^{+3.42}_{- 2.18}$ \\
Nov 6 & $0.42^{+0.04}_{-0.03}$ & $7.52^{+1.80}_{-1.61} \times 10^{12}$ & $\leq -0.5$ & $5.01^{+4.36}_{-1.82}$ & $0.15 \pm 0.01$ & $1.78 \pm 0.02$ & $9.71^{+0.21}_{-0.41}$ & $\ge 9.71$ & $14.27^{+4.91}_{-3.79}$ \\
Nov 10 & $0.39 \pm 0.01$ & $7.39^{+1.61}_{-1.59} \times 10^{12}$ & $\leq -0.5$ & $3.77^{+0.99}_{-0.75}$ & $0.14^{+0.02}_{-0.01}$ & $1.70^{+0.09}_{-0.06}$ & $8.45^{+1.12}_{-1.83}$ & $\ge 8.74$ & $15.66^{+14.11}_{-8.09}$ \\
Nov 13 & $0.44 \pm 0.01$ & $6.17^{+0.70}_{-0.60} \times 10^{12}$ & $\leq -0.5$ & $5.41^{+1.40}_{-1.11}$ & $0.22 \pm 0.01$ & $\leq 1.77$ & $9.69^{+0.23}_{-0.48}$ & $\ge 9.74$ & $1.49^{+0.33}_{-0.28}$ \\
Nov 18 & $0.47 \pm 0.2$ & $5.13^{+0.83}_{-0.70} \times 10^{12}$ & $\leq -0.5$ & $7.05^{+2.34}_{-1.99}$ & $0.15 \pm 0.01$ & $1.66^{+0.06}_{-0.04}$ & $9.06^{+0.69}_{-1.33}$ & $\ge 9.36$ & $5.82^{+2.98}_{-2.34}$ \\
\enddata
\label{tab:results}
\end{deluxetable*}
 \renewcommand\tabcolsep{6pt}

\begin{figure*}[t]
    \centering
    \includegraphics[width=0.5\textwidth]{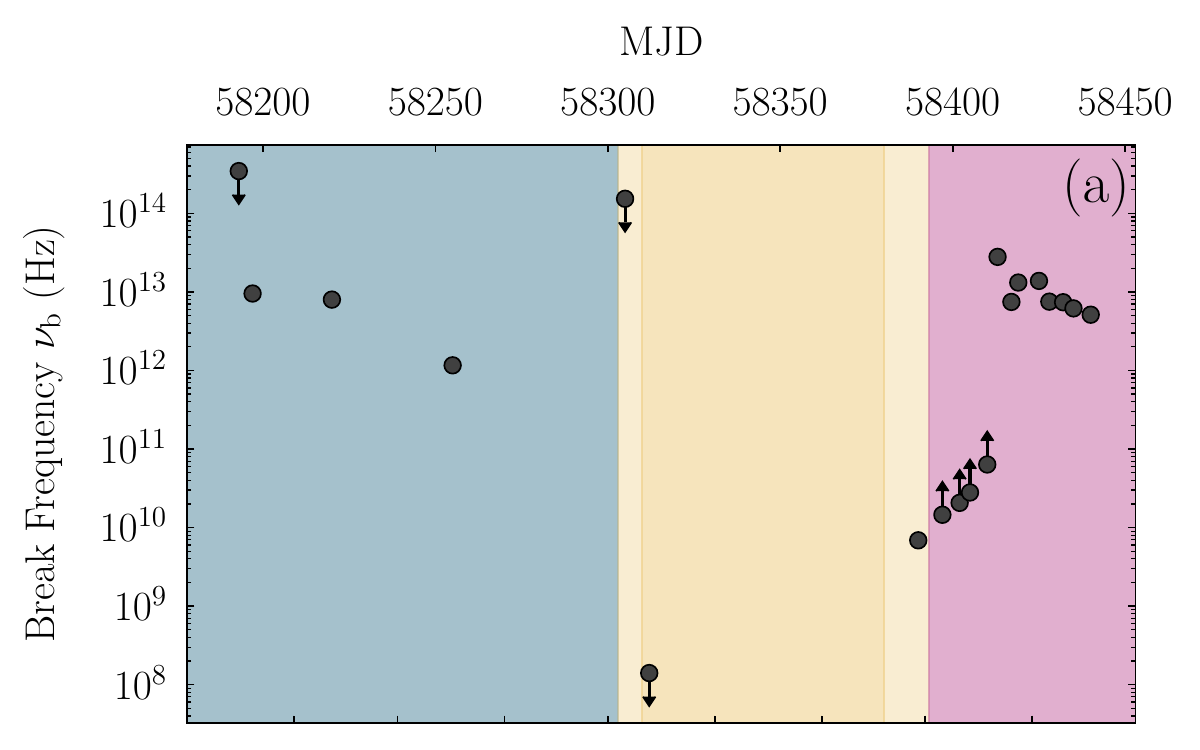}
    \includegraphics[width=0.49\textwidth]{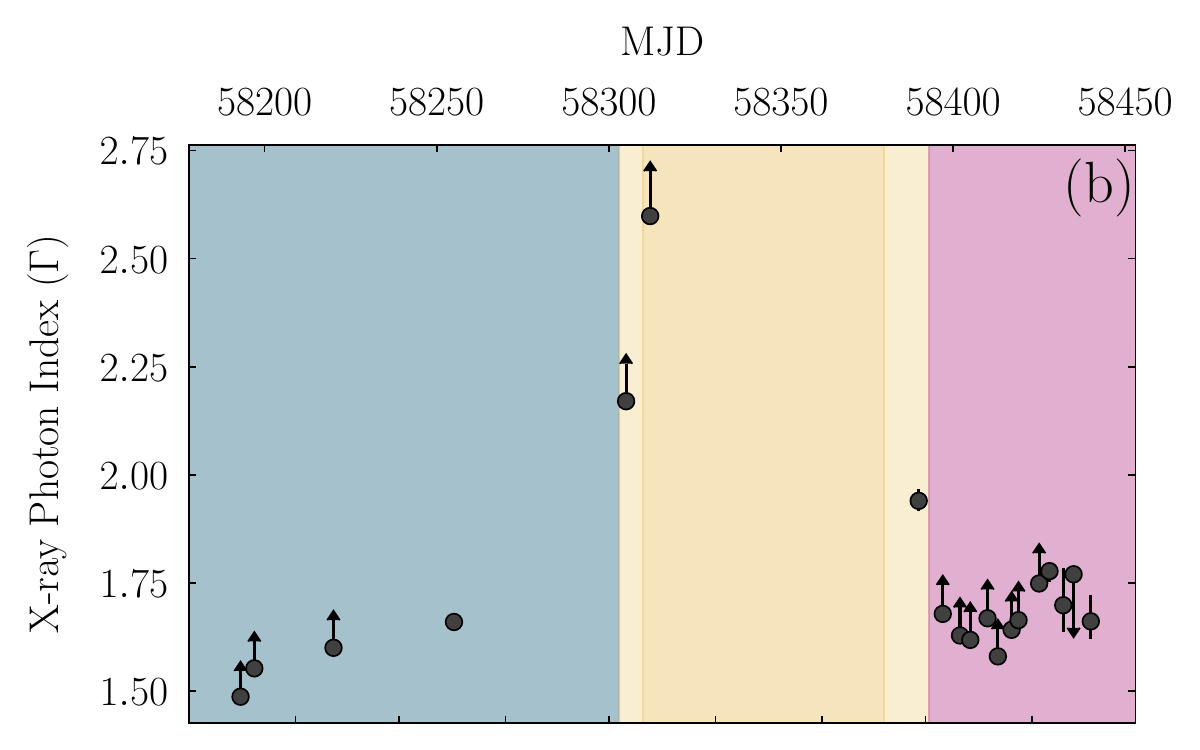}
    \includegraphics[width=0.49\textwidth]{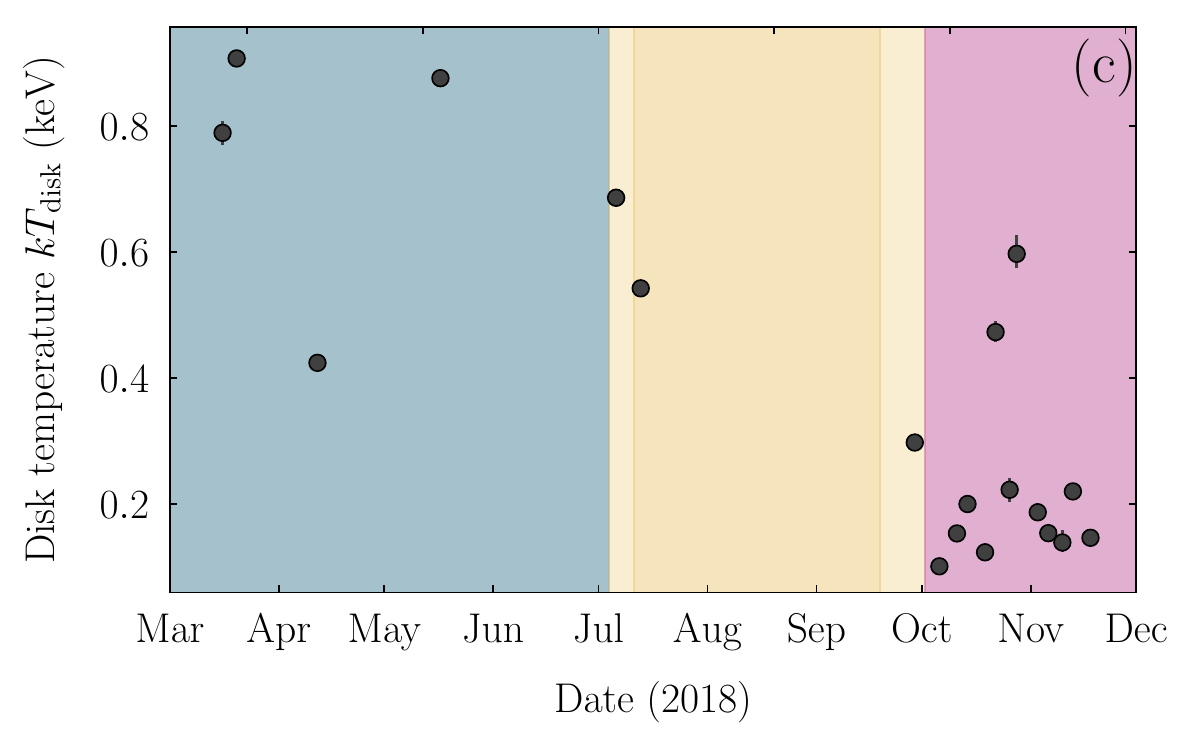}
    \includegraphics[width=0.49\textwidth]{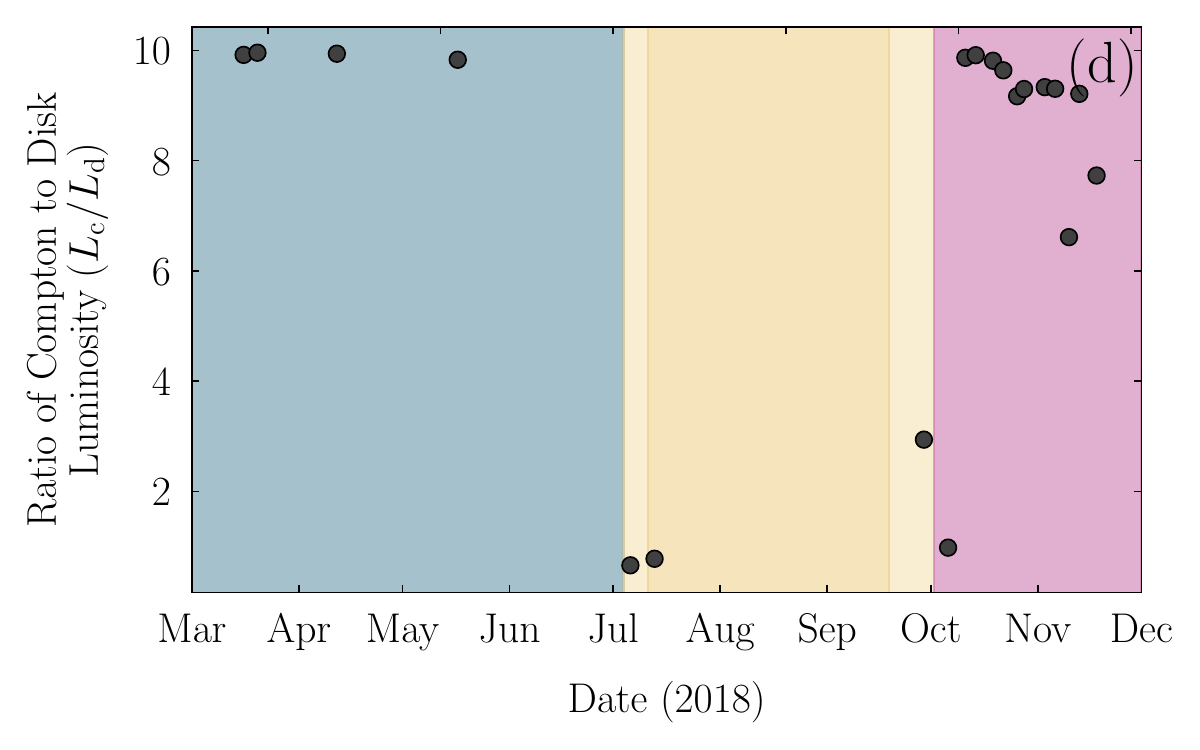}
    \caption{Evolution of the broad-band spectral parameters in J1820 throughout its 2018/2019 outburst; the jet spectral break ($\nu_{\mathrm{b}}$, {\it panel a}), X-ray power-law photon index ($\Gamma$, {\it panel b}), disk temperature ($kT_{\mathrm{disk}}$, {\it panel c}) and the ratio of the luminosity in the Comptonized emission to the disk intrinsic luminosity ($L_{\mathrm{c}}/L_{\mathrm{d}}$, {\it panel d}). Error bars are included in all data points, although in some cases these are smaller than the marker size. Arrows represent upper/lower limits. For $L_{\mathrm{c}}/L_{\mathrm{d}}$, we note that some values are consistent with the hard limit of the parameter ($L_{\mathrm{c}}/L_{\mathrm{d}}=10$), but we omit the arrows for visualization. The background shading in all panels is matched to Figure~\ref{fig:lightcurve}, and represents the accretion states identified in \citealt{2019ApJ...874..183S}: rising hard state (blue), intermediate and soft states (yellow), and declining hard state (pink). Note that the value of $\nu_{\mathrm{b}}$ on July 6 must be interpreted carefully (see Section \ref{subsec:July6}), as the spectrum is dominated by jet ejecta emission, preventing us from constraining the compact jet parameters accurately. All parameters show a distinct evolutionary pattern as the source progresses throughout the different accretion states of the outburst, and we observe an opposite trend in the evolution of the break frequency ($\nu_b$) and the X-ray photon index ($\Gamma$).}
    \label{fig:par_evol}
\end{figure*}

%%%%%%%%%% Transition states
\subsection{Analysis of spectra in transition states}

Among the broad-band spectra analyzed in this work, special care is taken during the transition (July 6) and soft state (July 13) epochs. During these dates, the rapid evolution of the system produced complexities in the spectral modeling, due to the presence of an extra jet ejecta component \citep[July 6,][]{2021MNRAS.505.3393W}, or due to flux variability (July 13). Here we discuss the details of the spectral analyses in both of these epochs.

%%%%%%%%%% July 6
\subsubsection{July 6} \label{subsec:July6}

The hard to soft accretion state transition is associated with the launching of discrete jet ejections \citep{2020ApJ...891L..29H}. On 2018 July 6, discrete jet ejections were resolved by VLBI imaging \citep{2020NatAs...4..697B,2021MNRAS.505.3393W} in J1820, which motivates fitting an alternative phenomenological model for this epoch. In particular, jet ejections on average produce a steep, optically thin radio to mm spectrum. Thus, on this date, we model the emission of jet ejections by including an additional single power-law in addition to the broken power-law representing the compact jet, since both may be present in the transition state. This component is modeled with \texttt{pegpwrlw} with a photon index fixed at 1.2. We keep the cutoff of both models as indicated in Section \ref{sec:SpecModel}. For this epoch, we extend the prior of the break frequency of the compact jet to be within the frequency range in which the jet ejecta do not dominate, i.e., above (sub)-mm. 
Thus, the spectral break position reported in Table \ref{tab:results} for this epoch corresponds to an upper limit (the higher frequency typically observed in BH XRBs), to reflect our data limitations rather than MCMC constraints from the parameter posterior. To test the need for this additional component, we compare fits with and without this additional single power-law in our model. Ultimately these tests revealed that the addition of the jet ejecta component not only better describes our data in this epoch, but also results in spectral indices for the compact jet component that are more consistent with those produced via synchrotron emission ($\alpha_{\mathrm{thick}}\approx 0.04$ and $\alpha_{\mathrm{thin}}\approx -0.14$ without the extra power-law). Thus we favor the addition of the single power-law component to model this epoch.
Furthermore, a radio flare detected at 15 GHz, on top of an overall radio emission decline, illustrates the rapid evolution of the jet properties on this epoch \citep[see Extended Data Figure 1 in][]{2020NatAs...4..697B}, while on July 7, observations at 8.4 GHz with VLBA show no radio core, suggesting that the compact jet had already quenched \citep{2021MNRAS.505.3393W}. Thus, we caution that as the jet ejecta component dominates the emission at longer wavelengths, and due to the rapid evolution, the compact jet parameters are not as well constrained in this epoch.

%%%%%%%%%% July 13
\subsubsection{July 13}

In the broad-band spectrum of July 13 (Figure \ref{fig:allseds}, {\it Panel b}) we notice an excess of emission in the IR bands above our best-fit model. Although some of the IR data correspond to $3\sigma$ upper limits, we explore different explanations for this possible excess. Some of the IR data used in this epoch are only quasi-simultaneous (1--2 days prior) with the other multi-wavelength data, thus we suspect that flux variability may be causing this IR excess. Upon comparing the IR data taken in the B10.7 band, we find that the flux density decreases from 4.8 mJy to at least 1.49 mJy over a 24 hour period, confirming our suspicions. Such rapidly variable and fading (over timescales of $\sim$hours) IR emission has been observed recently in another BH XRB, MAXI J1535--571 \citep{Baglio18,2020MNRAS.498.5772R}, where it also affected broad-band spectral modeling efforts. Therefore, given the rapid variability occurring during the time period sampled by the July 13 epoch, where the compact jet is rapidly fading between the days sampled, the data presented here likely only represent an average snapshot of the broad-band spectrum at this stage of the outburst. Alternatively, evidence of a disk wind present in the soft state \citep{2020A&A...640L...3S} can explain the near-IR excess. Using X-shooter data from July 13/15, \citealt{2023MNRAS.521.4190K} modeled the effects of a disk wind/atmosphere in the broad-band spectrum (near-IR/optical, UV, X-ray) of J1820. They found that the wind/atmosphere sitting above the disk reprocesses the disk's thermal emission into a quasi-thermal near-IR/optical bump, although their model somewhat underpredicts the data. Other explanations to this potential excess include synchrotron emission from the recently switched off jet \citep[IR excess due to the onset of the jet has been observed in other sources, e.g.,][]{2001ApJ...554L.181J,2019ApJ...887...21S} and non-thermal emission from a hot inner flow that becomes optically thin in the hard to soft state transition \citep{2014MNRAS.445.3987P}, although we stress that VISIR data on this epoch are not detections, but $3\sigma$ upper limits.

%%%%%%%%%%%%%%%%%%%%%%%%%%%%%%%%% DISCUSSION %%%%%%%%%%%%%%%%%%%%%%%%%%%%%%%%%%
\section{Discussion}\label{sec:discussion}

The results of our broad-band spectral modeling allow us to observe and connect spectral changes of J1820 throughout the course of its full outburst. In this section we first describe the spectral parameter evolution and compare to the observed behavior in other sources. Then, we use these results to connect spectral parameters to jet and accretion flow properties. We include similar analyses of other BH XRBs in the literature and we discuss the implications of our findings.

\begin{figure*}[t]
    \centering
    \includegraphics[width=0.85\textwidth]{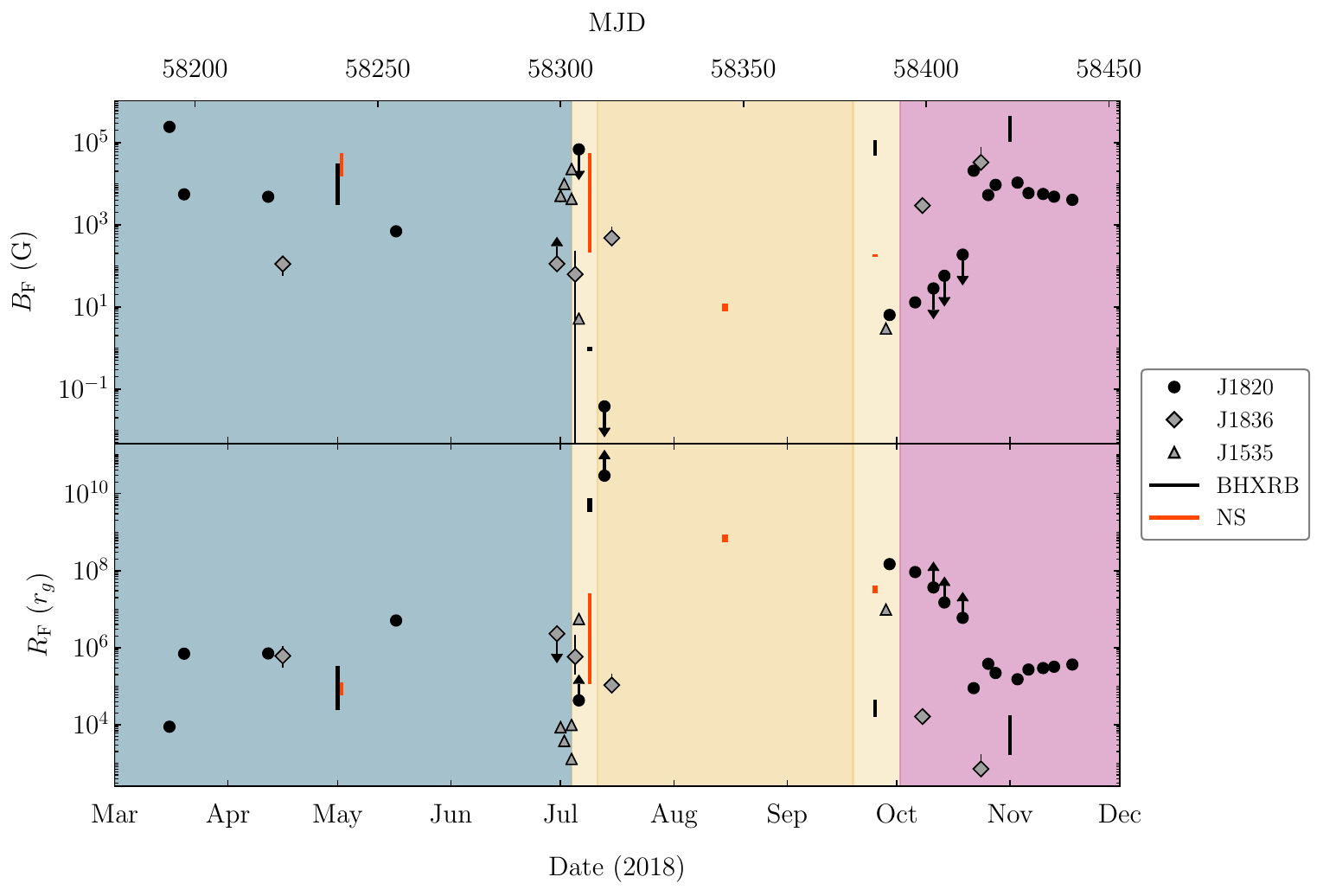}
    \caption{Evolution of the magnetic field ($B_{\mathrm{F}}$, {\it top panel}) and distance ($R_{\mathrm{F}}$, {\it bottom panel}) of the jet base (where particle acceleration begins) in the J1820 compact jet during the 2018/2019 outburst. $R_{\mathrm{F}}$ is measured in units of gravitational radii, where $M_{\mathrm{BH}}=8.48 \, M_{\odot}$ was used \citep{2020ApJ...893L..37T}. Measurements for J1836 \citep{2014MNRAS.439.1390R} and J1535 \citep{2020MNRAS.498.5772R} are included as gray diamonds and triangles, respectively. We include ranges of $B_{\mathrm{F}}$ and $R_{\mathrm{F}}$ values for other BH XRBs and neutron stars (NSs) with spectral break measurements, and represent them with black and red bars, respectively. For visualization purposes, we arbitrarily scaled} the epochs in all sources to match similar phases of J1820's outburst. The background shading represents the accretion states as described in Figure \ref{fig:par_evol}. The overall behavior is consistent with that observed in J1836, J1535, NSs and other BH XRBs in similar phases of the outburst, where the compact jet is quenched towards the state transition and recovered during the outburst decay.
    \label{fig:BandR}
\end{figure*}

\subsection{Source evolution}
\label{subsec:src_ev}
Figure \ref{fig:par_evol} displays the evolution of a selection of parameters over the course of J1820's 2018--2019 outburst. During the rise of the outburst\footnote{While the actual rise was short and took place in March (see Figure \ref{fig:lightcurve}), we classify spectral states following \citealt{2019ApJ...874..183S} and refer to the outburst rise as the period comprising 2018 March-June.}, the synchrotron spectral break moves from the IR into the sub-mm bands\footnote{Due to the sparse data in the IR band, the spectral break evolution can be accounted for with $\alpha_{\mathrm{thin}}$ changes.} ($\approx 10^{14} - 10^{12}$ Hz, see {\it panel a}), accompanied by an increase in the X-ray photon index, suggesting a gradual softening of the X-ray spectrum ({\it panel b}) over a period of 2 months. Over this period, the best-fit disk temperatures tend to be high compared to those inferred from {\it NICER} data \citep[$kT_{\mathrm{in}}\sim0.2$ keV,][]{2020ApJ...896...33W,2021MNRAS.506.2020D}. The disk luminosity is persistently dominated by the corona, with high $L_{\mathrm{c}}/L_{\mathrm{d}}$ values ({\it panel d})\footnote{In the majority of the outburst the upper limit of $L_{\mathrm{c}}/L_{\mathrm{d}}$ is consistent with the hard limit of the parameter in the model.}. During the hard state rise, the X-ray spectrum (up to 100 keV) is well described by two Comptonization zones \citep[e.g.,][]{2021MNRAS.506.2020D,2021ApJ...909L...9Z,2022ApJ...932....7Y,2023MNRAS.519.4434K}, in which an inner and outer corona are responsible for the hard and soft Comptonization components, respectively. However, our \textit{Swift}/XRT data is limited to 1--10 keV, which probes a small portion of the soft zone only. Including a two-component Comptonization model would result in lower disk temperatures, more consistent with the expected ones in a truncated disk in the hard state and explaining why our fits lead to higher temperatures. Another possible bias in the disk temperature is the soft excess due to the reflection component, caused by photons from the corona that are reprocessed by the disk. We do not include this component, since it is not resolved in our \textit{Swift}/XRT spectra.
Lastly, during the hard state rise, a hard-to-hard transition has been reported around MJD 58257 ($\approx$ May 19) which suggests that J1820 underwent a failed outburst \citep{2020ApJ...889..142S,2020ApJ...896...33W,2021NatAs...5...94M}. Our sample includes the epoch right before the transition occurred. Thus, we do not observe its effects, and it does not seem to interfere with the subsequent broad-band evolution.

As the source evolves into the soft state, the jet spectral break moves below the frequencies that our data samples ($\lesssim 10^8$ Hz), while the X-ray photon index reaches its highest value. We observe a cooling trend in the disk temperature, and the disk dominates the source luminosity, as expected for the soft state.

In the outburst decay, the spectral break frequency gradually increases from radio, back to IR bands, while the X-ray spectrum hardens as in the initial phase of the outburst. The disk reaches its coolest temperatures, staying relatively steady at $\approx 0.2$ keV. The source luminosity is again dominated by the corona, but with lower luminosities compared to the rise of the outburst.

The other two sources that have displayed a similar behaviour are MAXI J1836--194 \citep{2014MNRAS.439.1390R,2013ApJ...768L..35R} and MAXI J1535--571 (\citealt{2020MNRAS.498.5772R}; hereafter J1836 and J1535, respectively), although unlike J1820 neither was observed while evolving through all of the typical accretion states in succession. J1836 was observed to evolve from the hard state to the HIMS, after which the X-ray softening stalled,
and the source decayed back to the hard state. This spectral evolution took place over the course of $\approx$ 6 weeks. 
The source was observed once during its outburst rise \citep{2014MNRAS.439.1390R}, where the broad-band spectrum was characterized by an inverted radio spectrum ($\alpha_{\mathrm{thick}}\approx 0.7$), a cool disk ($kT_{\mathrm{disk}}\approx 0.23$ keV) and a hard power-law component in the X-rays ($\Gamma\approx 1.73$). Initially, the spectral break was located at $\sim2.3\times 10^{11}$ Hz. In the following three observational epochs, J1836 was settled in the HIMS, during which it began to soften but never reached the full soft state. During this state, and over a couple of weeks, the radio spectrum flattened ($\alpha_{\mathrm{thick}}\approx 0.2$), and the disk contribution increased as it became hotter ($kT_{\mathrm{disk}}\approx 0.42$ keV). The X-ray power-law steepened, with a maximum value of $\Gamma\approx 2.03$. Due to sparse data, the spectral break was difficult to constrain in these epochs, but remained in the $\approx 10^{11}$ Hz range. As the source outburst started to decay, the radio spectrum became inverted again and was relatively steady, while the disk contribution decreased ($kT_{\mathrm{disk}}\approx 0.1$ keV), the X-ray spectrum hardened ($\Gamma\approx 1.78$), and the spectral break frequency increased by over 2 orders of magnitude (up to $\approx 5 \times 10^{13}$ Hz) in $\sim1$ month. We observe an overall similar behavior in J1820. In particular, we can directly compare the outburst decay phase, in which $\alpha_{\mathrm{thick}}$ is consistent with an inverted spectrum, the disk is colder compared to previous epochs, the X-ray component is hard ($\gtrsim 1.6$), and the spectral break frequency increases by a few orders of magnitude. While both sources follow an overall cooling trend, the disk in J1836 became hotter during its softest state in the HIMS, which, based on our fits, we do not observe in J1820, although it is worth noting that \citealt{2023MNRAS.521.4190K} found a temperature of $\approx 1$ keV on July 13.

% ---- J1535:
For the case of J1535, \citealt{2020MNRAS.498.5772R} focused on the rise of the outburst, in which the broad-band spectrum evolved from the HIMS to the soft intermediate state (SIMS) in a matter of days. During the first 5 of the 6 observational epochs the source was in the HIMS. The broad-band spectrum was characterized by a flat radio spectrum ($\alpha_{\mathrm{thick}}\approx 0.1$), a fairly constant disk temperature ($kT_{\mathrm{disk}}\approx 0.2$ keV), a hard X-ray component ($\Gamma=1.74-1.95$) and a spectral break located at $\approx 10^{13}$ Hz. These properties changed drastically as the source entered in the SIMS: the disk cooled down ($kT_{\mathrm{disk}}\approx 0.09$ keV) and the X-ray spectrum softened ($\Gamma\approx 2.17$). More interestingly, the radio spectrum consisted of a single power-law component with a spectral break below $4.5 \times 10^9$ Hz. Overall, this is the same behavior we observe in J1820 during the rise of its outburst and transition to the soft state, where our model favors a steep radio spectrum. In both sources there is also evidence of jet ejecta launched during the transition to their softer states. Radio monitoring of J1535 in December 2017 tracked the motion of jet knots launched from the system, and constrained the time of ejections to around the HIMS-SIMS transition \citep{2019ApJ...883..198R}. In J1820, radio monitoring in early July links the observed radio flaring to the launching of jet ejecta and constrains the ejection time to MJD $58305.60\pm 0.04$ \citep[July 6;][]{2021MNRAS.505.3393W}, which is during the hard to soft state transition. The ejection time coincides with a broad-band spectrum dominated by a steep optically thin radio component associated with the jet ejecta (see Section \ref{subsec:July6}), suggesting that the compact jet had already quenched. Although the launching of the ejecta and the compact jet quenching cannot be linked directly due to the low cadence of observations, they likely occur around the same time.

%%%%%%%%%% B and R
\subsection{Mapping spectral parameters to jet properties} 

\label{subsec:mapping}
The frequency and flux density of the spectral break are key pieces of information to infer the properties of the first acceleration zone at the base of the jet. As outlined by \citealt{2011A&A...529A...3C} and following the analysis done in J1836 \citep{2014MNRAS.439.1390R} and J1535 \citep{2020MNRAS.498.5772R}, assuming equipartition between particle energy and magnetic field energy density, the frequency and flux density can be used to estimate the radius $R_{\mathrm{F}}$ (or height) and magnetic field $B_{\mathrm{F}}$ of the first acceleration zone\footnote{For the full equations see \citealt{2020MNRAS.498.5772R}, Appendix C.}:

\begin{equation}
    B_{\mathrm{F}} \propto S_{\mathrm{\nu,b}}^{-2/(2p+13)} \nu_{\mathrm{b}},
    \label{eq:bf}
\end{equation}
and
\begin{equation}
    R_{\mathrm{F}} \propto S_{\mathrm{\nu,b}}^{(p+6)/(2p+13)} \nu_{\mathrm{b}}^{-1},
    \label{eq:rf}
\end{equation}
where $\nu_{\mathrm{b}}$ is the spectral break frequency, $S_{\mathrm{\nu,b}}$ is the flux density at the spectral break, and $p$ is the power-law index of the electron energy distribution (where $p=1-2\alpha_{\mathrm{thin}}$). Using our spectral modeling results in Equations \ref{eq:bf} and \ref{eq:rf}, we obtain the evolution of $R_{\mathrm{F}}$ and $B_{\mathrm{F}}$ presented in Figure \ref{fig:BandR}, where errors are calculated using Monte Carlo methods and using the parameter posteriors. The distance of the base of the jet (where particles are first accelerated) above the BH has been previously estimated from J1820's timing properties. For example, on April 12, \citealt{2021MNRAS.504.3862T} estimated a physical distance\footnote{We use $1 \, r_{\mathrm{g}} \sim 10^6$ cm for J1820.} of $\approx 10^{12}$ cm and magnetic field $> 6 \times 10^3$ G. These estimates are comparable to our result on April 12. 
Our May 17 estimate is in good agreement with the limit placed for the size scale of the IR emitting region $\lesssim 10^{12}$ cm \citep{2020MNRAS.495..525M} on May 31. We can now compare our results again to the observed evolution in J1836 and J1535 during similar phases of the outburst, represented by gray diamonds and triangles in Figure \ref{fig:BandR}. When interpreting this figure, note that the time scales of evolution are different for each system, implying a different duration of each state. For the outburst rise in J1535, the radius and magnetic field of the first acceleration zone at the jet base were relatively constant during the HIMS ($R_{\mathrm{F}}\sim 10^3-10^4$ $r_{\mathrm{g}}$, $B_{\mathrm{F}}\sim 10^4$ G). However, in the HIMS-SIMS transition, $R_{\mathrm{F}}$ increased by 3 orders of magnitude while $B_{\mathrm{F}}$ decreased by the same amount over the course of one day. The highest value of $R_{\mathrm{F}}$ and lowest of $B_{\mathrm{F}}$ were found in the SIMS. In the case of J1820, the values at the transition follow the same overall trend. We caution that, on July 6, the flux density and position of the jet break need to be interpreted carefully (see Section \ref{subsec:July6}). In the soft state, we observe the highest $R_{\mathrm{F}}$ and lowest $B_{\mathrm{F}}$, consistent with J1535's evolution. For the reversed transition in J1836, $R_{\mathrm{F}}$ was largest during the HIMS ($\sim 10^5-10^6$ $r_g$) but receded $\sim3$ orders of magnitude during the declining hard state. Combining these results with measurements of the inner radius of the disk ($R_{\mathrm{in}}$) from their spectral fits, \citealt{2014MNRAS.439.1390R,2020MNRAS.498.5772R} inferred that particle acceleration must occur at larger scales than $R_{\mathrm{in}}$. Additionally, the opposite evolution of $\nu_{\mathrm{b}}$ and $R_{\mathrm{F}}$ implies that the jet becomes fainter as particles are accelerated further from the black hole and in larger scales, but it recovers as the acceleration point recedes and becomes smaller. In J1820 we see a similar behavior during the outburst decay. For the first time we observe the compact jet quenching and recovery throughout the full outburst of a BH XRB. Moreover, its evolution is consistent with that of other sources in similar phases of the outburst, as shown in Figure \ref{fig:BandR}, where we have included $B_{\mathrm{F}}$ and $R_{\mathrm{F}}$ for BH XRBs and neutron stars (NSs) with spectral break measurements from the literature: 4U 1728-34 \citep{2017A&A...600A...8D}, 4U 0614+091 \citep{2010ApJ...710..117M}, Aql X-1 \citep{2018A&A...616A..23D}, XTE J1118+480 \citep{2013MNRAS.429..815R}, 4U 1543-47 \citep{2013MNRAS.429..815R}, XTE J1550-564 \citep{2013MNRAS.429..815R}, GX 339-4 \citep{2011ApJ...740L..13G,2013MNRAS.429..815R}, Cyg X-1 \citep{2013MNRAS.429..815R}, and V404 Cyg \citep{2017ApJ...846..111C,2019MNRAS.482.2950T}. From these sources, only Aql X-1 and V404 Cyg have multiple measurements over the course of their respective outbursts. These results motivate multi-wavelength campaigns for future XRBs in outburst (BHs and NSs), to better understand whether this is a universal behavior among jet-launching sources.

%%%%%%%%%% Corona discussion
\subsection{Connecting jet properties to accretion flow properties} 
\label{sub:connecting}
During the rise phase of the outburst, J1535 and J1820 displayed a rapid broad-band spectral evolution. In J1535, the HIMS-SIMS transition was particularly fast, as $\nu_{\mathrm{b}}$ shifted 4 orders of magnitude to lower frequencies over the course of a day, accompanied by a softening X-ray spectrum. The state transition in J1820 spans almost a week, where $\nu_{\mathrm{b}}$ moves at least 3 orders of magnitude in frequency, again with a X-ray softening. 
On the decay phase the jet recovery in J1820 occurs more gradually, where $\nu_{\mathrm{b}}$ shifts back to frequencies comparable to the rise phase in $\sim2$ weeks, and the X-ray spectrum hardens. Even more gradual was the jet recovery in J1836, in which $\nu_{\mathrm{b}}$ shifted to higher frequencies over the course of a month while the X-ray component hardened. A similar time scale was observed in GX 339-4, in which the jet recovered in about a month, during the transition back to the hard state in the decay of its 2010-2011 outburst \citep{2013MNRAS.431L.107C}. This behavior indicates that the jet recovery ($\nu_{\mathrm{b}}$ shifting back to higher frequencies) is closely tied to the accretion flow becoming hot and optically thin or ``accretion flow recovery'' (X-ray hardening). We discuss more direct evidence of these changes and possible implications below. 

\begin{figure}
    \centering
    \includegraphics[width=0.45\textwidth]{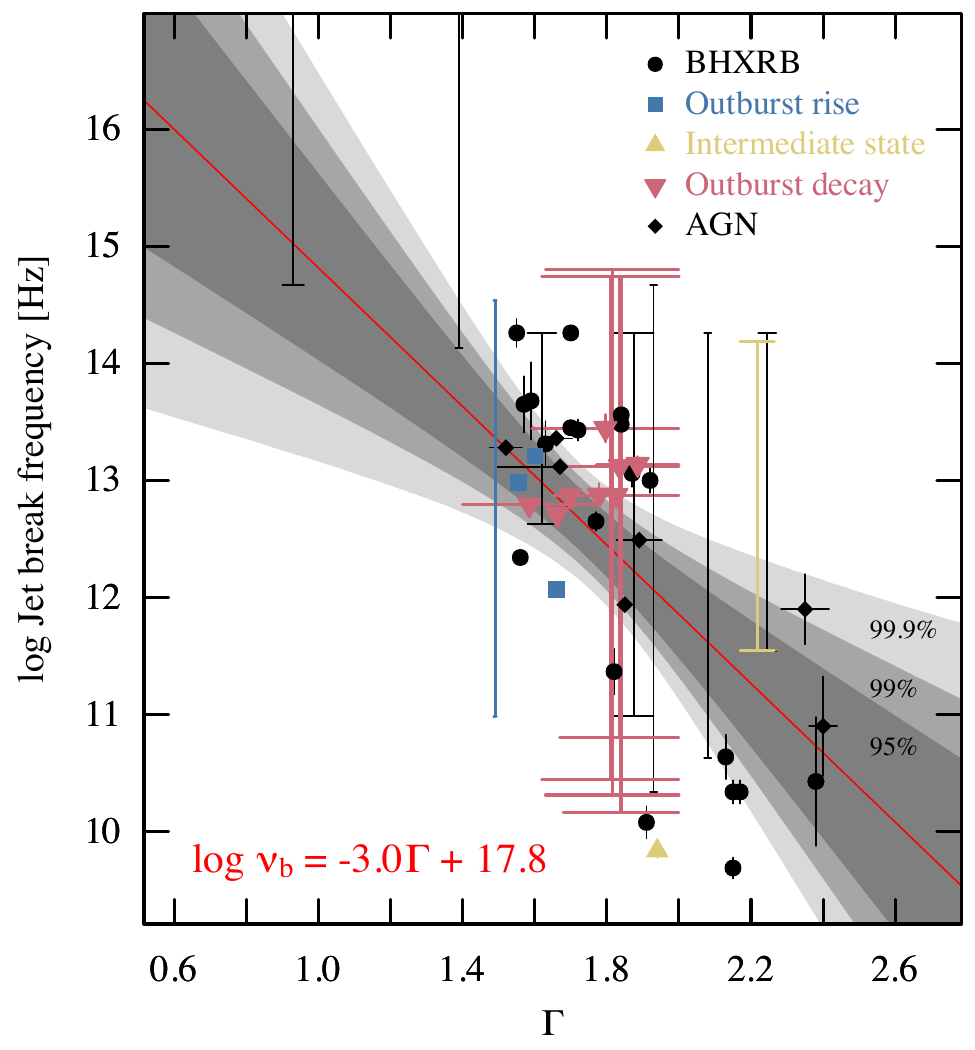}
    \caption{Measurements of the relationship between the jet spectral break frequency ($\nu_b$) and the X-ray power-law photon index ($\Gamma$). Black circles represent measurements from other BH XRB sources from the literature, while black triangles correspond to AGN measurements from the literature \citep[see][]{2015ApJ...814..139K,2020ApJ...893L..37T}. J1820's measurements presented in this work are shown in blue (rising hard state) and pink (declining hard state) squares, where bars indicate data with only limits on $\nu_{\mathrm{b}}$ (vertical) and $\Gamma$ (horizontal). The dark shaded regions represent the 95\%, 99\% and 99.9\% confidence intervals of a linear regression fit to all of the data, including J1820, \citep{2015ApJ...814..139K}. Our measurements of J1820 are in good agreement with the relationship observed in other sources (with the exception of epochs with weak constraints on the jet spectral break), suggesting a similar mechanism governing the accretion flow and jet changes (see Section \ref{sub:connecting}).}
    \label{fig:nu_vs_G}
\end{figure}

\subsubsection{Soft Comptonization Component}

One way to connect the accretion flow properties to the jet properties is  
to search for correlations between the parameters that dominate their emission. For example, one possible correlation is between $\nu_{\mathrm{b}}$ and $\Gamma$, which was studied by \citealt{2015ApJ...814..139K}. The relation was determined from observations of BH XRBs and AGN (see Figure \ref{fig:nu_vs_G}), including hard X-rays. This comparison can be made for AGN with sub-arcsecond resolution multi-wavelength broad-band spectra, which allows to isolate the core emission from the host galaxy. \citealt{2015ApJ...814..139K} observed that these AGN display a broken power-law, self-absorbed synchrotron spectrum, that is flat/inverted in the optically thick portion, just as in XRBs. When comparing BH XRBs and AGNs, it is observed that as $\nu_{\mathrm{b}}$ moves to lower frequencies, $\Gamma$ increases (the spectrum is softer). 

One of the sources included in this analysis is J1836, in which $\Gamma$ follows a clear anti-correlation with jet frequency \citep[see][ Figure 2]{2015ApJ...814..139K}.
During the HIMS, J1535 also followed this correlation \citep{2020MNRAS.498.5772R}. However, as the source softened, $\nu_{\mathrm{b}}$ was lower than the expected value for that $\Gamma$. In a similar phase of the outburst, the source MAXI J1659--152 \citep{2013MNRAS.436.2625V} also showed deviations from this correlation. In Figure \ref{fig:nu_vs_G} we show our results for J1820. Several of our measurements are in good agreement with the correlation observed for the other sources displayed in the figure, suggesting a similarity in the mechanism governing these changes among sources. However, some exceptions are October 6, 11, 14 and 19, where the $\nu_{\mathrm{b}}$  estimates are less constrained in our fits. While in these cases the values do not confirm the correlation, they certainly encompass part of the area predicted by it. We also caution that for both J1535 and J1820, the $\Gamma$ measurements are obtained from X-ray spectra covering up to 10 keV, while \citealt{2015ApJ...814..139K} included much higher energies.

\begin{figure}[t]
    \centering
    \includegraphics[width=0.5\textwidth]{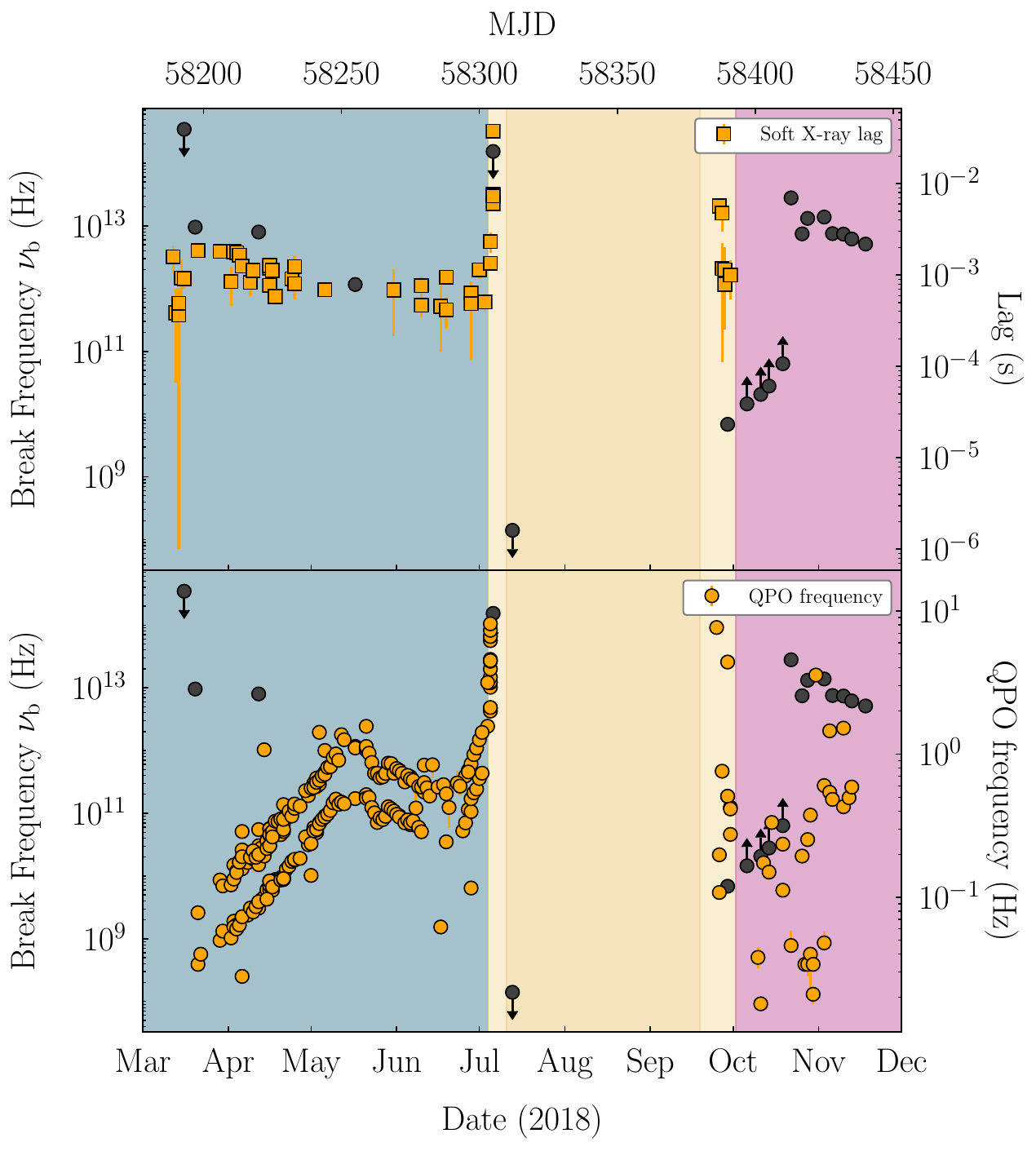}
    \caption{Evolution of soft X-ray lags (orange squares, \textit{top panel}) and soft X-ray lags quasi-periodic oscillation frequencies from \citealt{2020ApJ...889..142S} (orange circles, \textit{bottom panel}) compared to the evolution of the jet spectral break (dark grey markers). The background shading correspond to the accretion states described in Figure \ref{fig:par_evol}. The evolution of QPO frequency and soft X-ray lags appears to be inversely correlated with the jet spectral break evolution, and thus is suggestive of a connection between accretion inflow and jet outflow.}
    \label{fig:qpo_lag}
\end{figure}

\subsubsection{Reverberation Lags}

The origin of the hard X-ray emission in BH XRBs is typically attributed to a corona of hot electrons, although its geometry and physical properties remain unclear. For simplicity, it is sometimes treated as a lamp post, where the corona is situated above the accretion disk, but it could also be an extended region comprising the inner part of the accretion disk. Extensive work has been done in studying J1820's timing properties to understand the structure of the corona. In this section, we discuss these studies and connect them to our findings from broad-band phenomenological modeling.

Regardless of coronal geometry, a fraction of the hard X-ray photons may be intercepted and reprocessed by the disk, producing a soft component in the X-ray spectrum. The disk-corona separation provides key observational evidence to probe the changing coronal geometry, since a larger separation would cause the soft X-ray photons to reach the observer with a delay with respect to the hard ones. This delay is known as reverberation lag, and could allow us to connect the corona to the jet observed in the HIMS.

J1820 was observed by \textit{NICER} beginning the day following its discovery \citep{2018ATel11423....1U}, providing a complete data set to observe reverberation lags as the source evolved through the outburst. Using this technique, \citealt{2019Natur.565..198K} reported reverberation lags corresponding to the rising hard state, and found that soft lags progressively move towards shorter time scales (i.e., probing a progressively smaller emitting region) and constant iron emission line profile, suggesting that the corona contracts over time. They proposed an initially vertically extended corona with a compact core, that becomes more compact as the source evolves through the rising hard state. In addition, they found that the reverberation lag becomes longer during the state transition in J1820 \citep{2021ApJ...910L...3W,2021A&A...654A..14D}, probing an increasingly more extended emitting region. Later, \citealt{2022ApJ...930...18W} systematically searched for reverberation lags from BH XRBs (and candidates) in the \textit{NICER} archive, and found that this is a generic behavior. Their search also included later stages of the outburst, including the transition to the soft state.
Using the Hard X-ray Modulation Telescope (HXMT), \citealt{2020ApJ...896...33W} reported lags in the hard state rise as well. The time lags observed in the same frequency range as in \citealt{2019Natur.565..198K} are harder and larger, implying a disk-corona separation of $\sim 1000 \, R_{\mathrm{g}}$, as opposed to the $\sim 10 \, R_{\mathrm{g}}$ in \citealt{2019Natur.565..198K}. They also found a correlation between high-frequency lags and photon index of hard X-ray emission. They interpreted this result as two regions where the hard X-rays are emitted. One of them is the compact corona suggested in the aforementioned works, and the other is a large-scale jet where energetic electrons up-scatter soft photons from the disk \citep[see also][]{2020ApJ...895L..31E}. Other BH XRBs present evidence of two Comptonization regions based on timing properties: interference of disk and synchrotron Comptonization emission at different radii, applied to GX 339-4 and XTE J1748-288 \citep{2016ApJ...832..181V}; corona and compact jet in GRS 1915+105 \citep{2022MNRAS.513.4196G}; and dual-corona model in GX 339-4 \citep{2023MNRAS.519.1336P}.

As we emphasize in Section \ref{subsec:src_ev}, our limited X-ray coverage hinders the possibility to constrain different Comptonization components. However, because we track the evolution of the spectral break, and consequently, $R_{\mathrm{F}}$, we consider more appropriate to compare to \citealt{2019Natur.565..198K} and subsequent work. Their findings suggest that the corona could correspond to the base of the jet, as proposed originally in \citealt{2005ApJ...635.1203M}, which vertically expands and launches the jet ejecta in the intermediate states, although a potential connection to the compact jet component was not thoroughly explored. Particularly in J1820, the coronal expansion precedes the radio flaring activity \citep{2018ATel11827....1B} linked to the launch of jet ejecta \citep{2021MNRAS.505.3393W}, further supporting this scenario.

In Figure \ref{fig:BandR}, bottom panel, we observe how the location of the first acceleration zone increases during the rising hard state, with its highest value in the soft state. As discussed previously in Section \ref{subsec:mapping}, J1535 displays the same behavior during the HIMS-SIMS transition. For the case of J1836, \citealt{2021MNRAS.501.5910L} studied its behavior during the hard-HIMS transition, finding a strong correlation between the initial jet radius and power-spectral hue, which characterizes the shape of the power-law spectrum (whether it is flat or peaked). The correlation showed that as the jet base radius increased in size, the hard X-ray spectrum softened and became more peaked \citep[this trend has been observed in other sources,][]{2022MNRAS.509.2517C}. Since one way to soften the Comptonization spectrum is by decreasing the optical depth of the emitting region, they interpreted this result as an expanding jet base during the state transition, a consistent scenario with \citealt{2022ApJ...930...18W} and our results. Figure \ref{fig:qpo_lag}, top panel, compares the evolution of the jet spectral break with the evolution of the soft X-ray lags up to the onset of the outburst decay (Wang J., private communication). The details of this data set can be found in \citealt{2022ApJ...930...18W}. The increasing timescale of lags between the rising hard and intermediate state suggests that we are probing an emitting region that is increasing in size. This coincides with $\nu_{\mathrm{b}}$ moving to lower frequencies (base of jet moving away) and an increasing $R_{\mathrm{F}}$ (expanding base of jet). The opposite evolution in the soft to hard state transition indicates a smaller emitting region, consistent to the recovery of the compact jet. This clear trend strongly argues for the scenario where the corona corresponds to the jet base region.

\subsubsection{Quasi-periodic Oscillations}

The X-ray emission in BH XRBs can display variability features at certain frequencies, which are known as quasi-periodic oscillations \citep[QPOs, e.g.,][]{2019NewAR..8501524I}. Depending on their frequency, QPOs can be broadly classified as low-frequency QPOs (LFQPOs, $\sim 10^{-2}-10$ Hz), typically observed in the hard state and HIMS, and high-frequency QPOs ($\sim 10-10^3$ Hz), characteristic of the soft state. The changing frequency of QPOs is closely related to changes in the geometry of the accretion flow, since both frequency and inner radius of the disk change during the course of an outburst. The origin of LFQPOs is still debated, but it is typically attributed to instabilities in the accretion flow, instabilities in the jet \citep{2022A&A...660A..66F}, or
Lense-Thirring precession of either the inner hot accretion flow present in the hard state \citep{1998ApJ...492L..59S,2009MNRAS.397L.101I}, which is directly related to the Compton tail in the X-ray spectrum, or of a small-scale jet \citep{2021NatAs...5...94M}. 
In J1820 only low-frequency QPOs (LFQPOs) have been reported throughout the outburst \citep[e.g.,][from \textit{Swift}-XRT and \textit{NICER}]{2020ApJ...889..142S}, as shown in Figure \ref{fig:qpo_lag}, top panel. In the rising hard state, QPO frequencies remain $\lesssim 1$ Hz suggesting a larger radial extent of the inner accretion flow during this period \citep{2009MNRAS.397L.101I}. In the state transition, the QPO frequency increases remarkably fast (on timescales similar to the jet spectral break), suggesting a small extent to the inner accretion flow during this phase. When J1820 transitions back to the (declining) hard state, QPO frequencies fall again to $\lesssim 2$ Hz, consistent with the inner disk radius receding and being replaced by a radially extended hot inner flow. 

The \textit{Insight}-HXMT has played an important role in the detection of LFQPOs. Particularly in the hard state rise, properties of LFQPOs discovered above 30 keV  can be explained by the precession of a small-scale jet \citep{2021NatAs...5...94M}. In this model, a small-scale jet precesses above the disk, producing LFQPOs at different energies: high-energy LFQPOs come from the jet base while low-energy LFQPOs from the top, where cooling becomes important. This model seems to explain well the LFQPO properties observed in J1820 (as well as the QPOs observed in the optical band, \citealt{2022MNRAS.513L..35T}), particularly if the jet is located a few $r_g$ above the BH (lamppost geometry) and if its height decreases over the rising hard state period (consistent with the picture proposed in \citealt{2019Natur.565..198K}). The magnetic field can accelerate the small scale-jet into a relativistic, large-scale jet, producing broad-band synchrotron emission. Later, \citealt{2023ApJ...948..116M} extended this model to include lower energy ($< 1$ keV) LFQPOs from \textit{NICER} data, and found that the jet and inner disk ring precess together. Our broad-band observations and location of the first acceleration zone at the base of the jet
support the presence of a relativistic jet that dominates the spectrum throughout the hard state, as expected from this model. More recently, \citealt{2023MNRAS.525..854M} focused on a \textit{NICER} observation on July 6, i.e., the state transition, and proposed a dual-corona model to explain the observed QPO evolution. The model consists of a horizontally expanded corona that envelops the inner region of the accretion disk, with a compact corona inside the inner edge. During the transition, the compact corona remains unchanged while the other vertically expands and is associated with the ejecta on this epoch.

While we cannot directly link QPO frequencies to the spectral break frequencies (due to lack of data in the soft state and variability in the declining hard state), they seem to evolve in the opposite way, suggesting that changes in the jet are connected to changes in the geometry of the accretion flow. An analysis of QPO in the source GRS 1915+105 was able to establish such connection, coupling the corona and launching of the jet  \citep{2022NatAs...6..577M}. The study revealed that, as the QPO frequency decreases below $\lesssim 2$ Hz, the corona becomes less radially extended and more vertically extended until the compact jet is launched. This study shows that QPOs are another key piece of information to better understand the corona-jet connection, and highlights the importance of combining this technique with multi-wavelength spectral modeling and reverberation lags to have the most complete picture of the physical processes at play in BH XRBs.

%%%%%%%%%%%%%%%%%%%%%%%%%%%%%%%%% CONCLUSION %%%%%%%%%%%%%%%%%%%%%%%%%%%%%%%%%%
\section{Conclusions}\label{sec:Conclusion}

We present the results of a quasi-simultaneous, multi-wavelength observing campaign on the black hole X-ray binary MAXI J1820+070 over the course of its 2018/2019 outburst. This campaign allowed us to observe, for the first time, the full evolution of a BH XRB as it transitions through the different spectral states, with the potential to provide new insights into the behavior of the compact jet and its connection to the contemporaneous changes in the accretion flow. We collect the most complete data set spanning from radio through X-rays, which we compile into 19 single epochs, encompassing 7 months of the outburst. We fit the broad-band spectrum of each observational epoch with a phenomenological model that accounts for interstellar absorption and extinction, the persistent compact jet, the discrete jet ejecta (when applicable), and the accretion flow and the companion star. Our findings can be summarized as follows:

\begin{enumerate}
    \item The broad-band evolution of J1820 is consistent with that observed in other BH XRBs. In particular, we observe the jet spectral break shift to lower frequencies as the jet emission fades in the soft state, and move back to IR frequencies as the system returns to the hard state, indicating that the compact jet has recovered.
    \item Tracking the jet spectral break and flux density at its location allows us to place estimates on the distance of the first acceleration zone at the base of the jet ($R_{\mathrm{F}}$) and magnetic field strength ($B_{\mathrm{F}}$) in this region. These measurements have been conducted in only two other sources while evolving through spectral states: J1836 and J1535. The behavior of J1820 in corresponding phases of the outburst is consistent with both. When combining these results with other X-ray binaries with jet spectral break measurements, we observe a similar evolution and values of $B_{\mathrm{F}}$ and $R_{\mathrm{F}}$ among all sources. Moreover, the $R_{\mathrm{F}}$ evolution distinctly shows the change in the jet base location during state changes, indicating the quenching and recovery of the compact jet. 
    \item Measurements of the jet spectral break and X-ray photon index in BH XRBs (including J1820) and AGN seem to follow a negative correlation, indicating a connection between inflow-outflow. The fact that several accreting systems at different scales follow this anti-correlation strongly suggests a similarity in the mechanism governing these changes among accreting sources.  
    \item The evolving jet activity observed in J1820 appears to be related with X-ray timing properties (reverberation lags). This trend points towards the scenario where the corona is thought to be the base of the jet.
\end{enumerate}

Our findings further motivate the need for high cadence monitoring of BH XRBs in outburst, to obtain well-sampled broad-band spectra and to characterize the broad-band evolution. This will help to better constrain the jet spectrum, which in turn will allow us better understand the complex connection between the jet and the accretion flow. In this context, next generation telescopes will play a key role in following up transient systems. The James Webb Space telescope instruments now make possible to observe the spectral break when positioned through the mid-IR bands ($1.1 \times 10^{13} - 5 \times 10^{14}$ Hz), which is essential to track the jet base, where particle acceleration begins, and its evolution. The Vera C. Rubin Observatory will provide wide-area coverage of the optical sky, enabling rapid detection of optical transients. Since BH XRBs in outburst tend to brighten first in the optical, the real-time alerts will allow us to improve the multi-wavelength campaigns to follow-up these objects. Finally, timing and polarimetry techniques in the sub-mm region are already providing valuable insight into the jet physics (see results from the PITCH-BLACK Survey\footnote{https://www.eaobservatory.org/jcmt/science/large-programs/pitch-black/}). The combined capabilities of current and next generation telescopes when monitoring future BH XRBs in outburst, together with the outgoing development of relativistic jet simulations, foresees a promising future for the jet physics field.

\section*{Acknowledgements}

First, we would like to express our gratitude to our colleague, Dr. Tomaso Belloni, who passed away in August 2023. He was an inspiration to many of us. It has been hard to normalize his absence, his infectious enthusiasm, and sharp insight.

The authors wish to recognize and acknowledge the very significant cultural role and reverence that the summit of Maunakea has always had within the indigenous Hawaiian community. We are most fortunate to have the opportunity to conduct observations from this mountain. We also offer a special thanks to the NRAO for granting our DDT request for some of the VLA observations presented in this paper. We acknowledge with thanks observations from the AAVSO International Database contributed by observers worldwide. 
 
CET acknowledges support from McGill’s Wolfe Graduate Fellowship and the Trottier Space Institute at McGill. DH acknowledges funding from the Natural Sciences and Engineering Research Council of Canada (NSERC) and the Canada Research Chairs (CRC) program. AJT acknowledges partial support for this work provided by NASA through the NASA Hubble Fellowship grant \#HST--HF2--51494.001 awarded by the Space Telescope Science Institute, which is operated by the Association of Universities for Research in Astronomy, Inc., for NASA, under contract NAS5--26555. TDR acknowledges financial contribution from
the agreement ASI--INAF n.2017--14--H.0.  
TMB acknowledges financial contribution from grant PRIN INAF 2019 n.15.
TS acknowledges  financial support from the Spanish Ministry of Science, Innovation
and Universities (MICIU) under grant PID2020-114822GB-I00. ML, acknowledges support from NASA~ADAP grant 80NSSC17K0515 and  NWO VICI award (Netherlands Organization for Scientific Research) grant Nr. 639.043.513. VT acknowledges support from the Romanian Ministry of Research, Innovation and Digitalization through the Romanian National Core Program LAPLAS VII –
contract no. 30N/2023. JW acknowledges support from the NASA~FINESST Graduate Fellowship, under grant 80NSSC22K1596. VT acknowledges support from the Romanian Ministry of Research, Innovation
and Digitalization through the Romanian National Core Program LAPLAS VII –
contract no. 30N/2023. 

The National Radio Astronomy Observatory is a facility of the National Science Foundation operated under cooperative agreement by Associated Universities, Inc. This paper makes use of the following ALMA data: ADS/JAO.ALMA\#2017.1.01103.T. ALMA is a partnership of ESO (representing its member states), NSF (USA) and NINS (Japan), together with NRC (Canada), MOST and ASIAA (Taiwan), and KASI (Republic of Korea), in cooperation with the Republic of Chile. The Joint ALMA Observatory is operated by ESO, AUI/NRAO and NAOJ. This material is based upon work supported by Tamkeen under the NYU Abu Dhabi Research Institute grant CASS. This work is based on observations carried out under project numbers W17BM \& W17BN  with the IRAM NOEMA Interferometer. IRAM is supported by INSU/CNRS (France), MPG (Germany) and IGN (Spain). The Sub-millimeter Array is a joint project between the Smithsonian Astrophysical Observatory and the Academia Sinica Institute of Astronomy and Astrophysics, and is funded by the Smithsonian Institution and the Academia Sinica. The James Clerk Maxwell Telescope is operated by the East Asian Observatory on behalf of The National Astronomical Observatory of Japan; Academia Sinica Institute of Astronomy and Astrophysics; the Korea Astronomy and Space Science Institute; the National Astronomical Research Institute of Thailand; Center for Astronomical Mega-Science (as well as the National Key R\&D Program of China with No. 2017YFA0402700). Additional funding support is provided by the Science and Technology Facilities Council of the United Kingdom and participating universities and organizations in the United Kingdom and Canada. Additional funds for the construction of SCUBA-2 were provided by the Canada Foundation for Innovation. 

This work is based on observations collected at the European Southern Observatory (ESO) under programmes 0101.D-0634 and 0102.D-0514 (PI D. Russell). This work uses data from the Faulkes Telescope Project, which is an education partner of Las Cumbres Observatory (LCO). The Faulkes Telescopes are maintained and operated by LCO. This work uses observations made with the REM Telescope, INAF Chile. This work uses data collected with the Al Sadeem Observatory, whose Owner and Co-founder is Thabet Al Qaissieh, the Director/Co-founder is Alejandro Palado and the Resident Astronomer is Aldrin B. Gabuya.

\vspace{5mm}
\facilities{AMI-LA, VLA, ALMA, NOEMA, SMA, JCMT, VLT (VISIR, XSHOOTER), REM, LCO, {\it Swift} (XRT and UVOT), AAVSO.}

\software{\textsc{casa} \citep{mc07}, \textsc{gildas} (\url{http://www.iram.fr/IRAMFR/GILDAS}), \textsc{starlink} (\url{http://starlink.eao.hawaii.edu/}), \textsc{esoreflex} \citep{freudling13}, \textsc{aqua} \citep{2004SPIE.5496..729T}, \textsc{daophot} \citep{1987PASP...99..191S}, \textsc{heasoft} (\url{https://heasarc.gsfc.nasa.gov/docs/software/heasoft/}), \textsc{xspec} \citep{1996ASPC..101...17A}, \textsc{ftools} \citep{black95}.}

\bibliography{J1820_main}{}
\bibliographystyle{aasjournal}

\appendix

\section{Markov Chain Monte Carlo Convergence}

\begin{figure*}[ht]
    \centering
    \includegraphics[width=0.99\textwidth]{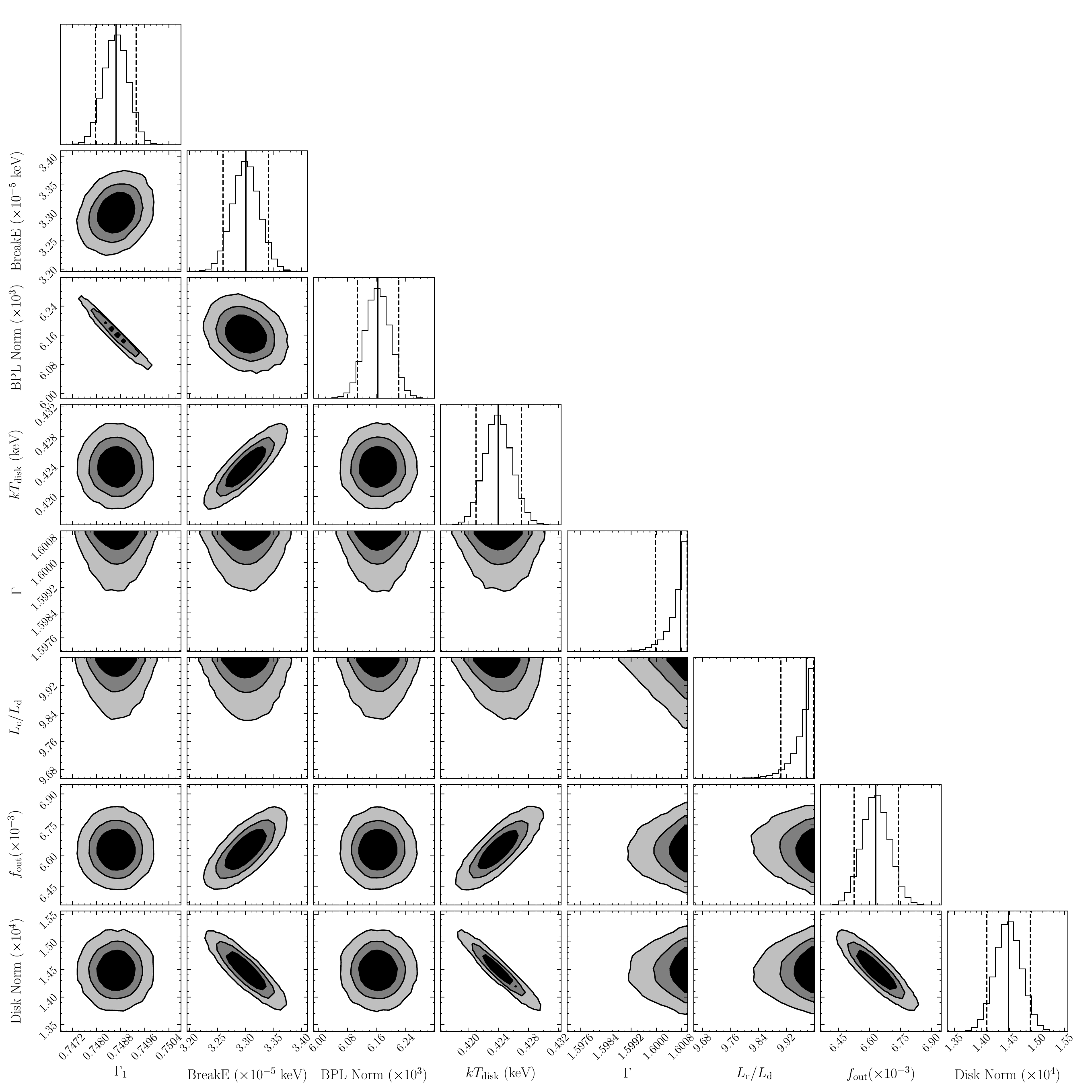}
    \caption{The one- and two-dimensional MCMC posterior distributions for the parameters in the spectral fitting of April 12. The black, dark gray and light gray regions correspond to the 68\%, 90\% and 99\% credible intervals, respectively. The solid line in the middle of the one-dimensional posterior distributions represents the median value, while the left/right dashed lines represent the 16\% and 84\% percentiles, respectively. Note that in the $\Gamma$ and $L_{\mathrm{c}}/L_{\mathrm{d}}$ parameters the upper limit is unconstrained, since it is consistent with the hard limit of the model.}
    \label{fig:April12cornerplot}
\end{figure*}

\clearpage
\section{Spectral Modeling}

\begin{figure*}[ht]
    \centering
        \includegraphics[width=0.8\textwidth]{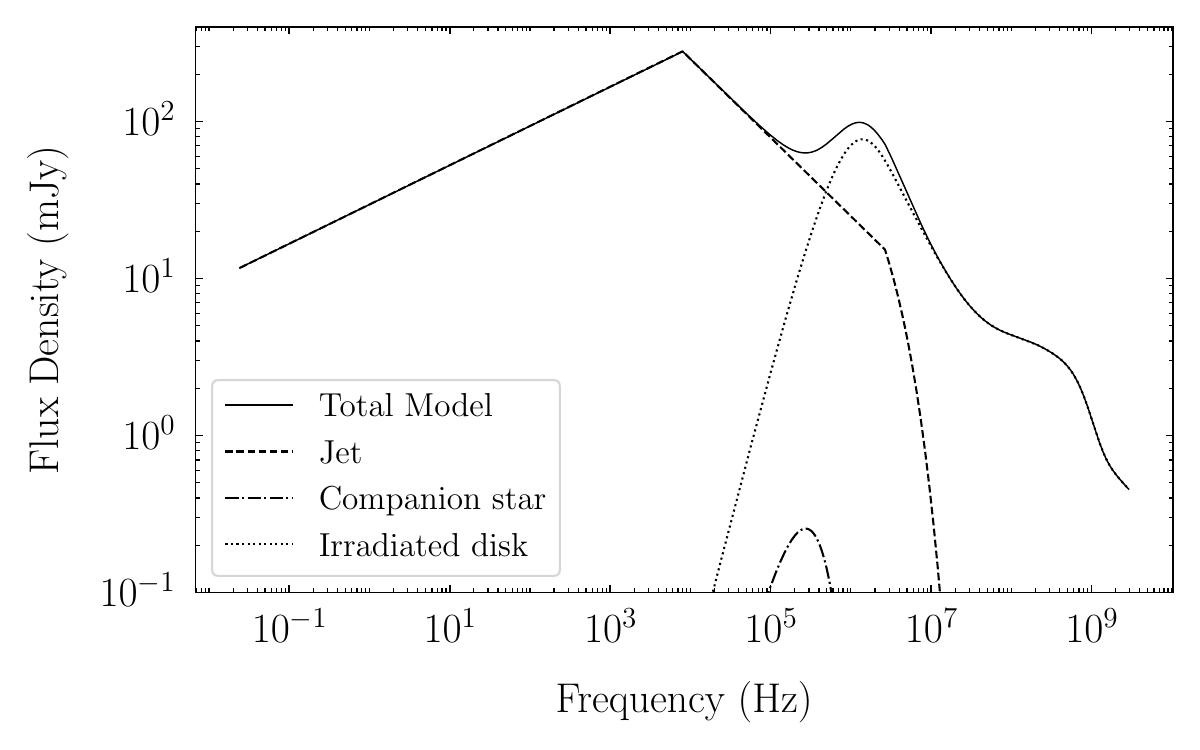}
    \caption{A sample broad-band spectrum corresponding to April 12, displaying the contribution of each component to the emission. }
    \label{fig:model_components}
\end{figure*}

\renewcommand\tabcolsep{3pt}
\begin{deluxetable}{ccc}
\tablecaption{Frequencies ($\nu_{\mathrm{b}}$) and flux densities ($S_{\nu_{\mathrm{b}}}$) at the spectral break, obtained from our spectral modeling (see Section \ref{sec:SpecModel}). \label{tab:snu}}
\tablewidth{0pt}
\tablecolumns{3}
\tablehead{\colhead{Date} & \colhead{$\nu_{\mathrm{b}}$} & \colhead{$S_{\nu_{\mathrm{b}}}$} \\
(2018) & (Hz) & (mJy)}
\def\arraystretch{1.2}
\startdata
March 16 & $\leq 3.45 \times 10^{14}$ & $76.88_{-2.09}^{+2.13}$ \\
March 20 & $9.54^{+2.06}_{-1.11} \times 10^{12}$ & $400.01_{-110.24}^{+113.26}$ \\
April 12 & $7.98^{+0.06}_{-0.06} \times 10^{12}$ & $284.96_{-1.80}^{+1.84}$ \\
May 17 & $1.16^{+0.03}_{-0.03} \times 10^{12}$ & $311.18_{-14.27}^{+14.91}$ \\
July 6 & $\leq 1.53 \times 10^{14}$ & $194.58_{-124.68}^{+394.14}$ \\
July 13 & $\leq 1.40 \times 10^{8}$ & $24.14_{-9.48}^{+15.67}$ \\
Sep 29 & $6.88^{+1.60}_{-0.49} \times 10^{9}$ & $8.05_{-3.64}^{+3.16}$ \\
Oct 6 & $\ge 1.46 \times 10^{10}$ & $16.08_{-0.23}^{+0.23}$ \\
Oct 11  & $\ge 2.06\times 10^{10}$ & $11.52_{-2.23}^{+3.79}$ \\
Oct 14  & $\ge 2.80 \times 10^{10}$ & $9.26_{-2.31}^{+3.71}$ \\
Oct 19  & $\ge 6.35 \times 10^{10}$ & $15.86_{-6.06}^{+10.21}$ \\
Oct 22  & $2.79^{+0.86}_{-0.66} \times 10^{13}$ & $54.83_{-21.17}^{+38.79}$ \\
Oct 26 & $7.46^{+0.91}_{-0.84} \times 10^{12}$ & $66.61_{-9.91}^{+11.79}$ \\
Oct 28  & $1.32^{+0.12}_{-0.11} \times 10^{13}$ & $69.76_{-11.94}^{+14.83}$ \\
Nov 3  & $1.38^{+0.21}_{-0.19} \times 10^{13}$ & $35.20_{-7.36}^{+10.29}$ \\
Nov 6  & $7.52^{+1.80}_{-1.61} \times 10^{12}$ & $36.13_{-12.76}^{+27.82}$ \\
Nov 10  & $7.39^{+1.61}_{-1.59} \times 10^{12}$ & $38.37_{-8.35}^{+10.30}$ \\
Nov 13  & $6.17^{+0.70}_{-0.60} \times 10^{12}$ & $30.74_{-5.70}^{+7.12}$ \\
Nov 18  & $5.13^{+0.83}_{-0.70} \times 10^{12}$ & $27.66_{-7.13}^{+8.59}$ \\
\enddata
\end{deluxetable}
\renewcommand\tabcolsep{6pt}

\begin{figure*}[ht]
    \centering
        \includegraphics[width=0.65\textwidth]{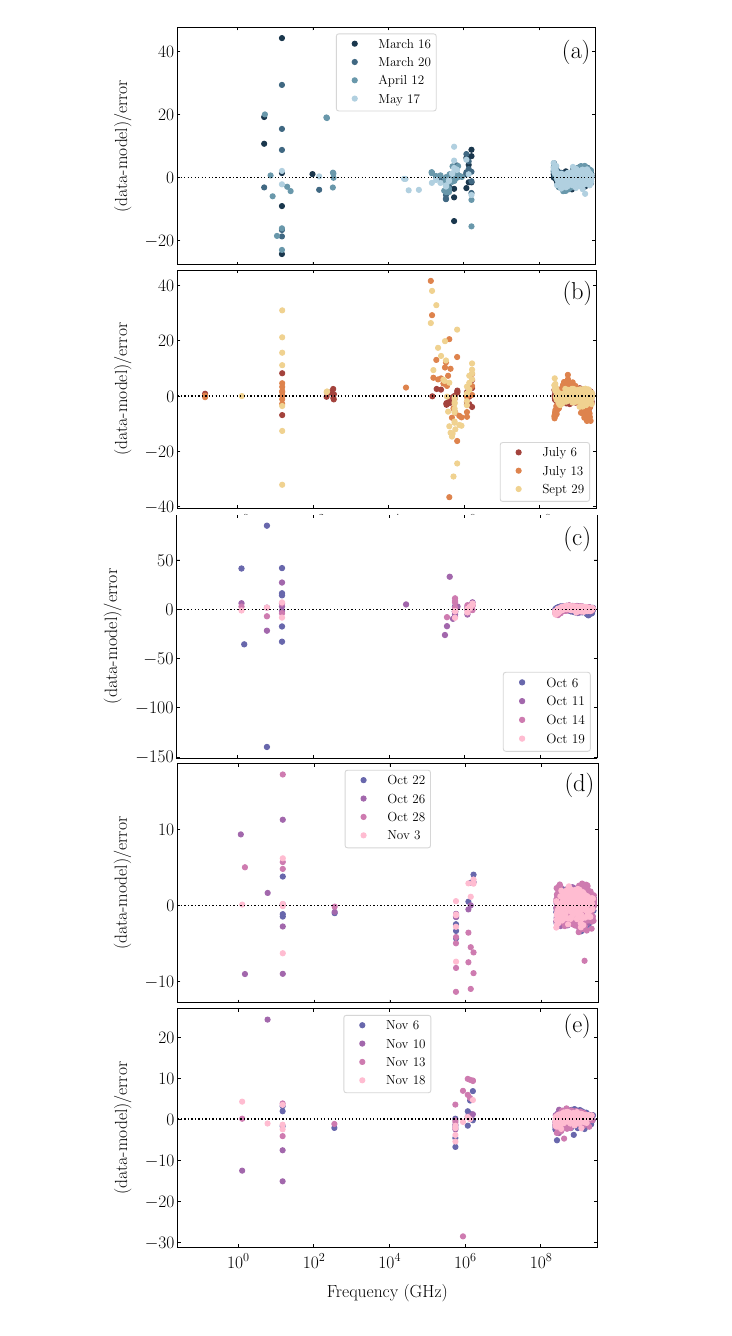}
    \caption{Fit residuals, defined as (data-model)/error, corresponding to the broad-band spectral fits presented in Figure \ref{fig:allseds}. {\it Panel a} displays the model residuals corresponding to the rising hard state (color code blue); {\it Panel b} shows the model residuals of the intermediate (July 6 and September 29) and soft (July 13) states (color code yellow); {\it Panels c, d} and {\it e} show the residuals corresponding to the declining hard state (color code pink). Note that the color code for the accretion states are matched in Figures \ref{fig:lightcurve}, \ref{fig:allseds}, \ref{fig:par_evol}, \ref{fig:BandR} and \ref{fig:qpo_lag}.}
    \label{fig:allseds_res}
\end{figure*}

\clearpage
\section{Observations}
%%%%%% SWIFT-UVOT  SIMULTANEOUS DATA TABLE %%%%%%
\begin{deluxetable}{cccccccc}[b]
 \tablecaption{Summary of \emph{Swift}/UVOT observations of MAXI J1820+070. These observations are simultaneous to the XRT data used in this work, while additional non-simultaneous UV observations can be found in Table \ref{tab:UVOTdata_nonsim}. Exposure times are measured in seconds, while the units of count rates are counts s$^{-1}$. \label{tab:UVOTdata_sim}}
 \tablewidth{0pt}
 \tablehead{
  & & & & & \textit{UVOT Filter} & & \\
 \cmidrule(lr){4-8}
 \colhead{Obs. ID} & \colhead{Date} & \colhead{} & \colhead{U} & \colhead{V} &
 \colhead{UVW1} & \colhead{UVM2} & \colhead{UVW2}
 }
 \scriptsize
 \startdata
       00010627005 & 2018-03-16 03:01:32 & Exposure & - & - & 183.8 & 276.3 & 393.4 \\
       & & Count rate & - & - & 176.9 & 97.7 & 187.4 \\
       00010627009 & 2018-03-20 18:37:18 & Exposure & - & - & 189.7 & 285.2 & 390.6 \\
       & & Count rate & - & - & 254.8 & 148.1 & 267.1 \\
       \midrule 
       00088657004 & 2018-05-17 21:43:10 & Exposure & - & - & 177.9 & 368.2 & 357.1 \\
       & & Count rate & - & - & 137.7 & 73.3 & 132 \\
       \midrule
       00010754001 & 2018-07-06 15:24:07 & Exposure & - & - & 316.7 & 597.9 & 634.6\\
       & & Count rate & - & - & 125.7 & 67.4 & 121.9 \\
       \midrule
       00010754004 & 2018-07-13 08:15:03 & Exposure & - & - & 269.5 & 481.7 & 540.1 \\
       & & Count rate & - & - & 154.3 & 81.6 & 152.3 \\
       \midrule
       00010627102 & 2018-09-29 07:38:22 & Exposure & - & 95.24 & 189.7 & 285.2 & 379.7 \\
       & & Count rate & - & 25.4 & 42.86 & 23.1 & 42.8 \\
       \midrule 
       00010627109 & 2018-10-06 02:15:36 & Exposure & - & - & 176.9 & 265.5 & - \\
       & & Count rate & - & - & 33.1 & 17.6 & - \\
       \midrule 
       00010627111 & 2018-10-11 06:41:03 & Exposure & - & - & 102.1 & 152.3 & 203.5 \\
       & & Count rate & - & - & 46.3 & 23 & 40.8 \\
       \midrule 
       00010627113 & 2018-10-15 17:34:08 & Exposure & - & - & 163.2 & 244.8 & 326.5 \\
       & & Count rate & - & - & 38.7 & 20.1 & 35 \\
       \midrule 
       00010627115 & 2018-10-19 07:12:57 & Exposure & - & 113.9 & 227.1 & 341.4 & 455.5 \\
       & & Count rate & - & 27.9 & 37.2 & 19 & 34.4 \\
       \midrule 
       00010627116 & 2018-10-21 00:42:47 & Exposure & - & - & 100.2 & 364.4 & 561.7 \\
       & & Count rate & - & - & 34.3 & 18.5 & 33.4 \\
       \midrule 
       00010627119 & 2018-10-26 11:27:29 & Exposure & - & 137.6 & 275.4 & 414.1 & 551.9 \\
       & & Count rate & - & 21.2 & 28.1 & 14.4 & 27.2 \\
       \midrule 
       00010627120 & 2018-10-28 14:33:23 & Exposure & - & 48.98 & 97.21 & 146.4 & 195.6 \\
       & & Count rate & - & 22.4 & 28.9 & 14.3 & 25.1 \\
       \midrule 
       00010627123 & 2018-11-03 10:55:07 & Exposure & - & 69.66  & 140.5 & 210.4 & 280.3 \\
       & & Count rate & - & 15 & 22.7 & 10.8 & 19.6 \\
       \midrule 
       00010627125 & 2018-11-06 04:07:05 & Exposure & - & 102.1 & 203.5 & 305.9 & 407.2 \\
       & & Count rate & - & 14.8 & 19 & 10.3 & 18.3 \\
       \midrule 
       00010627127 & 2018-11-10 13:11:47 & Exposure & - & 34.23 & 68.67 & 103.1 & 137.6 \\
       & & Count rate & - & 12.1 & 15.7 & 7.7 & 13.2 \\
       \midrule 
       00010627128 & 2018-11-12 08:29:53 & Exposure & - & 120.8 & 241 & 362 & 483 \\
       & & Count rate & - & 10.3 & 13.8 & 7.1 & 11.5 \\
       \midrule 
       00010627131 & 2018-11-18 04:26:15 & Exposure & 153.7 & 145.4 & 291.1 & 436.8 & 582.5 \\
       & & Count rate & 19.3 & 7.2 & 8.8 & 4.2 & 8.2 \\
\enddata
\end{deluxetable}

%%%%%% SWIFT-UVOT NON-SIMULTANEOUS DATA TABLE %%%%%%
\begin{deluxetable}{cccccccc}
\tablecaption{Summary of \emph{Swift}/UVOT observations of MAXI J1820+070. These are not simultaneous to the XRT exposures used in this  work. Exposure times are measured in seconds, while the units of count rates are counts s$^{-1}$. \label{tab:UVOTdata_nonsim}}
\tablewidth{0pt}
 \tablehead{
  & & & & & \textit{UVOT Filter} & & \\
 \cmidrule(lr){4-8}
 \colhead{Obs. ID} & \colhead{Date} & \colhead{} & \colhead{U} & \colhead{V} &
 \colhead{UVW1} & \colhead{UVM2} & \colhead{UVW2}
 }
 \scriptsize
 \startdata
       00010627001 & 2018-03-14 20:54:55 & Exposure & \dots & \dots & 219.3 & 329.5 & 345.4 \\
       & & Count rate & \dots & \dots & 161.5 & 85.8 & 156.1 \\
       \midrule 
       00010627006 & 2018-03-17 02:52:57 & Exposure & \dots & \dots & 203.5 & 305.9 & 408.6 \\
       & & Count rate & \dots & \dots & 199.9 & 104.2 & 196.9 \\
       \midrule 
       00010627008 & 2018-03-19 18:40:57 & Exposure & \dots & \dots & 195.6 & 294 & 359.1 \\
       & & Count rate & \dots & \dots & 257.3 & 145.3 & 248.2 \\
       \midrule 
       00010627010 & 2018-03-21 18:45:57 & Exposure & \dots & \dots & 210.4 & 315.7 & 426.7 \\
       & & Count rate & \dots & \dots & 295.8 & 157.2 & 284.4 \\
       \midrule 
       00010627030 & 2018-04-11 05:25:57 & Exposure & \dots & \dots & \dots & \dots & 1134 \\
       & & Count rate & \dots & \dots & \dots & \dots & 261.2 \\
       \midrule 
       00010627037 & 2018-04-14 06:45:57 & Exposure & \dots & \dots & \dots & \dots & 959 \\
       & & Count rate & \dots & \dots & \dots & \dots & 227.4 \\
       \midrule
       00010627076 & 2018-07-08 05:36:57 & Exposure & \dots & \dots & 112 & 183.2 & 224.1 \\
       & & Count rate & \dots & \dots & 114.0 & 62 & 114.7 \\
       \midrule 
       00010627079 & 2018-07-08 23:19:57 & Exposure & \dots & \dots & 92.3 & 151.6 & 184.8 \\
       & & Count rate & \dots & \dots & 116.1 & 58.1 & 98.8 \\
       \midrule 
       00010627083 & 2018-07-11 21:35:44 & Exposure & \dots & \dots & 175 & 297.7 & 285.2 \\
       & & Count rate & \dots & \dots & 135.2 & 77.5 & 138.1 \\
       \midrule 
       00010754002 & 2018-07-11 03:34:09 & Exposure & \dots & \dots & 417.1 & \dots & 158.2 \\
       & & Count rate & \dots & \dots & 132.1 & \dots & 128.6 \\
       \midrule 
       00088657008 & 2018-07-15 19:41:48 & Exposure & \dots & \dots & 218.4 & \dots & \dots \\
       & & Count rate & \dots & \dots & 151.6 & \dots & \dots \\
       \midrule 
       00010627104 & 2018-09-30 10:43:57 & Exposure & \dots & 90.3 & 180.9 & 355.7 & 362 \\
       & & Count rate & \dots & 23 & 40.1 & 21.4 & 41.3 \\
       \midrule 
       00010627105 & 2018-10-01 13:49:57 & Exposure & \dots & 87.4 & 175 & 370.7 & 349.2 \\
       & & Count rate & \dots & 22.7 & 38.8 & 21.5 & 39.1 \\
       \midrule 
       00088657010 & 2018-09-27 22:00:57 & Exposure & \dots & 83.4 & 167.1 & 249.8 & 333.4 \\
       & & Count rate & \dots & 28.1 & 45.7 & 25 & 45.8 \\
       \midrule 
       00010627110 & 2018-10-09 00:13:57 & Exposure & \dots & 50 & 100.2 & 150.4 & 200.5 \\
       & & Count rate & \dots & 28.7 & 37.9 & 19.8 & 37.5 \\
       \midrule 
       00010627112 & 2018-10-13 06:13:57 & Exposure & \dots & \dots & 186.8 & 280.3 & 373.8 \\
       & & Count rate & \dots & \dots & 43.7 & 22.3 & 38.9 \\
       \midrule 
       00010627114 & 2018-10-17 05:48:56 & Exposure & \dots & \dots & 170.1 & 341.4 & 339.3 \\
       & & Count rate & \dots & \dots & 40.4 & 19 & 35.7 \\
       \midrule 
       00088657011 & 2018-10-30 00:04:57 & Exposure & \dots & 102.1 & 205.5 & 450.7 & 411.2 \\
       & & Count rate & \dots & 19.5 & 26.9 & 13.2 & 24.2 \\
       \midrule 
       00010627124 & 2018-11-04 15:25:57 & Exposure & \dots & 110.1 & \dots & \dots & 582.5 \\
       & & Count rate & \dots & 17.2 & \dots & \dots & 19.7 \\
       \midrule 
       00010627126 & 2018-11-08 16:41:57 & Exposure & \dots & 69.7 & 139.5 & 208.4 & 278.3 \\
       & & Count rate & \dots & 12.1 & 17.2 & 9.3 & 15.2 \\
       \midrule 
       00010627129 & 2018-11-14 12:42:34 & Exposure & 90.1 & 114.9 & 231.1 & 346.2 & 461.4 \\
       & & Count rate & 28 & 9.591 & 12.1 & 6 & 11.5 \\
    \enddata
\end{deluxetable}

%%%%%% SWIFT-XRT SIMULTANEOUS DATA TABLE %%%%%%
\begin{deluxetable}{ccccc}
  \caption{Summary of \emph{Swift}/XRT observations of MAXI J1820+070.\label{tab:XRTdata}}
  \tablewidth{0pt}
  \tablehead{
  \colhead{Obs. ID} & \colhead{Date} & \colhead{Exposure} & \colhead{Count rate} & \colhead{Pile-up corrected} \\
  & & (s) & (s$^{-1}$) & }
  \scriptsize
  \startdata
       00010627005 & 2018-03-16 03:01:32 & 978.1 & 48.15 & Yes  \\
       \noalign{\vskip 1.2mm}
       00010627009 & 2018-03-20 18:37:18 & 991.5 & 131.9 & Yes  \\
       \noalign{\vskip 1.2mm}
       00010627034 &  2018-04-12 06:58:53 & 1018 & 100.4 & Yes  \\
       \noalign{\vskip 1.2mm}
       00010627035 &  2018-04-12 10:11:03 & 888.3 & 102.5 & Yes  \\
       \noalign{\vskip 1.2mm}
       00088657004 & 2018-05-17 21:43:10 & 777.4 & 123.7 & Yes  \\
       \noalign{\vskip 1.2mm}
       00010754001 & 2018-07-06 15:24:07 & 1994.9 & 110.7 & Yes  \\
       \noalign{\vskip 1.2mm}
       00010754004	& 2018-07-13 08:15:03 & 1607 & 119 & Yes  \\
       \noalign{\vskip 1.2mm}
       00010627102	& 2018-09-29 07:38:22 & 1108 & 115.8 & Yes  \\
       \noalign{\vskip 1.2mm}
       00010627109	& 2018-10-06 02:15:36 & 1037  & 86.6 & No  \\
       \noalign{\vskip 1.2mm}
       00010627111	& 2018-10-11 06:41:03 & 588.8  & 25.4  & No \\
       \noalign{\vskip 1.2mm}
       00010627113	& 2018-10-15 17:34:08 & 943 & 13.9 & No  \\
       \noalign{\vskip 1.2mm}
       00010627115	& 2018-10-19 07:17:24 &  1348  & 10.3 & No  \\
       \noalign{\vskip 1.2mm}
       00010627116	& 2018-10-21 00:42:47 &  1732  &  7.7  & No  \\
       \noalign{\vskip 1.2mm}
       00010627119	& 2018-10-26 11:27:29 &  448.1  &  5.1  & No  \\
       \noalign{\vskip 1.2mm}
       00010627120	& 2018-10-28 14:33:23 & 12508.1 & 3.9 & No \\
       \noalign{\vskip 1.2mm}
       00010627123	& 2018-11-03 10:55:07 &  1386  & 2.7 & No \\
       \noalign{\vskip 1.2mm}
       00010627125	& 2018-11-06 04:07:05 &  2197 &  2.2  & No  \\
       \noalign{\vskip 1.2mm}
       00010627127	& 2018-11-10 13:11:47 &  399.1  & 1.4 & No  \\
       \noalign{\vskip 1.2mm}
       00010627128	& 2018-11-12 08:29:53 & 2014 & 1.1 & No  \\
       \noalign{\vskip 1.2mm}
       00010627131	& 2018-11-18 04:26:15 & 1657  & 0.6 & No  \\
       \enddata
\end{deluxetable}

\clearpage
\section{Summary of flux densities}
%%%%%% RADIO FLUXES TABLE %%%%%%
\startlongtable
\begin{deluxetable}{cccccc}
\tablecaption{Flux densities of MAXI J1820+070 at radio/sub-mm frequencies. \label{tab:radio_fluxes}}
\tablehead{
\colhead{Telescope} & \colhead{Date} & \colhead{MJD} & \colhead{Frequency}  & \colhead{Flux Density} & \colhead{Reference}\\
&  (2018)   &     &  (GHz)   &  (mJy) &\\}
\scriptsize
\startdata
% March 16 
AMI-LA  & March 14 & 58191 & 15 & $17.32 \pm 0.15$ & \dots\\
AMI-LA  & March 15 & 58192 & 15 & $21.53 \pm 0.42$ &\dots\\
eMERLIN  & March 16 & 58193 & 5.07 & $23.2 \pm 0.4$ & 2\\
AMI-LA  & March 16 & 58193 & 15 & $32.18 \pm 0.25$ &\dots\\
VLBA  & March 16 & 58193 & 15 & $20.01 \pm 0.10$ & 9 \\
NOEMA  & March 16  & 58193 & 97 & $30 \pm 3$ & 8\\
eMERLIN  & March 17  & 58194 & 5.07 & $26.6 \pm 0.4$ & 2\\
\midrule
% March 20 
AMI-LA  & March 18 & 58195 & 15 & $50.65 \pm 0.30$ & \dots\\
AMI-LA  & March 19 & 58196 & 15 & $52.08 \pm 0.19$ &\dots\\
AMI-LA  & March 20 & 58197 & 15 & $58.44 \pm 0.31$ &\dots\\
NOEMA  & March 20 & 58197 & 146 & $80.8 \pm 8.0$ & 8\\
AMI-LA  & March 21 & 58198 & 15 & $60.46 \pm 0.31$ &\dots\\
eMERLIN  & March 22 & 58199 & 5.07 & $38 \pm 1$ & 2\\
AMI-LA  & March 22 & 58199  & 15 & $66.96 \pm 0.38$ &\dots\\
\midrule
% April 12 
AMI-LA  & April 11 & 58219  & 15 & $47.48 \pm 0.43$ & \dots \\
VLA  & April 12 & 58220 & 5.3 & $46.0 \pm 0.1$ & 5\\
VLA  & April 12  & 58220  & 7.5 & $48.1 \pm 0.2$ & 5\\
VLA  & April 12  & 58220 & 8.5 & $48.3 \pm 0.2$ & 5\\
VLA  & April 12  & 58220 & 11.1 & $49.2 \pm 0.2$ & 5\\
VLA  & April 12  & 58220 & 20.7 & $58.7 \pm 1.1$ & 5\\
VLA  & April 12  & 58220 & 25.5 & $60.5 \pm 1.1$ & 5\\
SMA  & April 12 & 58220 & 226.6 & $151.1 \pm 2.0$ & 6\\
SMA  & April 12  & 58220  & 234.6 & $166.9 \pm 2.8$ & 6\\
ALMA  & April 12  & 58220 & 336.6 & $124.4 \pm 0.07$ & 4\\
ALMA  & April 12  & 58220 & 338.6 & $124.9 \pm 0.06$ & 4\\
ALMA  & April 12  & 58220 & 348.6 & $125.8 \pm 0.06$ & 4\\
ALMA  & April 12  & 58220 & 350.4 & $125.9 \pm 0.06$ & 4\\
AMI-LA  & April 13  & 58221 & 15 & $48.65 \pm 0.54$ &\dots\\
\midrule
% May 17 
AMI-LA & May 16 & 58254 & 15 & $52.21 \pm 0.46$ &\dots \\
AMI-LA & May 17 & 58255 & 15 & $49.98 \pm 0.65$ &\dots\\
NOEMA & May 18 & 58256 & 146 & $133.57 \pm 0.46$ & 8\\
\midrule
% July 6 
AMI-LA  & July 5  & 58304 & 15 & $10.72 \pm 0.17$ & \dots\\
AMI-LA  & July 6  & 58305 & 15 & $8.32 \pm 0.14$ & \dots\\
SMA  & July 6  & 58305 & 224.6 & $5.54 \pm 2.4$ & 6\\
ALMA  & July 6  & 58305 & 336.4 & $5.91 \pm 0.02$& 4\\
ALMA  & July 6  & 58305 & 338.4  & $5.84 \pm 0.02$ & 4 \\
ALMA  & July 6  & 58305 & 348.4  & $5.81 \pm 0.02$ & 4\\
ALMA  & July 6  & 58305 & 350.4  & $5.85 \pm 0.03$ & 4\\
LOFAR  & July 7 & 58305 & 0.1365  & $27.1 \pm 5.6$ & 1\\
SMA  & July 7 & 58305 & 276.1  & $8.8 \pm 2.2$ & 6\\
\midrule
% July 13 
LOFAR  & July 11 & 58310 & 0.1365 & $11.5 \pm 2.7$ & 1\\
AMI-LA  & July 11 & 58310 & 15 & $1.08 \pm 0.16$ &\dots\\
AMI-LA  & July 11 & 58310 & 15 & $1.10 \pm 0.05$ &\dots\\
AMI-LA  & July 12 & 58311 & 15 & $1.29 \pm 0.16$ &\dots\\
LOFAR  & July 13 & 58312 & 0.1365 & $14.7 \pm 5.1$ & 1\\
AMI-LA  & July 13 & 58312 & 15 & $1.39 \pm 0.09$ &\dots\\
AMI-LA  & July 13 & 58312 & 15 & $1.61 \pm 0.20$ &\dots\\
AMI-LA  & July 14 & 58313 & 15 & $1.78 \pm 0.11$ &\dots\\
AMI-LA  & July 14 & 58313 & 15 & $1.80 \pm 0.16$ &\dots\\
\midrule
% Sept 29 
MeerKAT  & September 28 & 58389 & 1.28  & $3.47 \pm 0.05$ & 2\\
AMI-LA  & September 28 & 58389 & 15 & $3.70 \pm 0.04$ &\dots\\
SMA  & September 29 & 58390 & 225.1 & $2.68 \pm 1.0$ & 6\\
SMA  & September 29 & 58390 & 231.1 & $3.36 \pm 1.28$ & 6\\
AMI-LA  & September 29 & 58390 & 15 & $4.17 \pm 0.06$ &\dots\\
AMI-LA  & September 29 & 58390 & 15 & $4.54 \pm 0.11$ &\dots\\
AMI-LA  & September 30 & 58391 & 15 & $7.05 \pm 0.07$ &\dots\\
AMI-LA  & September 30 & 58391 & 15 & $7.37 \pm 0.22$ &\dots\\
AMI-LA  & October 1 & 58392 & 15 & $6.33 \pm 0.07$ &\dots\\
AMI-LA  & October 1 & 58392 & 15 & $6.22 \pm 0.08$ &\dots\\
\midrule
% Oct 6 
AMI-LA  & October 4 & 58395 & 15 & $10.36 \pm 0.10$ &\dots\\
AMI-LA  & October 5 & 58396 & 15 & $15.37 \pm 0.11$ &\dots\\
MeerKAT  & October 5 & 58396 & 1.28 & $11.8 \pm 0.1$ & 2\\
AMI-LA  & October 6 & 58397 & 15 & $19.88 \pm 0.15$ &\dots\\
eMERLIN  & October 7 & 58398 & 1.51 & $5.26 \pm 0.08$ & 2\\
VLA   & October 7 & 58398 & 6 & $16.99 \pm 0.03$ & 3\\
AMI-LA  & October 7 & 58398 & 15 & $14.92 \pm 0.09$ &\dots\\
VLA  & October 8 & 58399 & 6 & $7.46 \pm 0.05$ & 3\\
AMI-LA  & October 8 & 58399 & 15 & $12.33 \pm 0.07$ &\dots\\
\midrule
% Oct 11 
AMI-LA  & October 9 & 58400 & 15 & $10.61 \pm 0.10$ &\dots\\
AMI-LA  & October 10 & 58401 & 15 & $9.05 \pm 0.37$ &\dots\\
VLA  & October 11 & 58402 & 6 & $5.12 \pm 0.03$ & 3\\
AMI-LA  & October 11 & 58402 & 15 & $7.89 \pm 0.08$ &\dots\\
MeerKAT  & October 12 & 58403 & 1.28 & $2.62 \pm 0.04$ & 2\\
VLA  & October 12 & 58403 & 6 & $4.20 \pm 0.04$ & 3\\
\midrule
% Oct 14 
AMI-LA  & October 13 & 58404 & 15 & $5.84 \pm 0.06$ &\dots\\
MeerKAT  & October 14 & 58405 & 1.28 & $2.41 \pm 0.03$  & 2\\
VLA  & October 14 & 58405 & 6 & $3.59 \pm 0.05$ & 3\\
AMI-LA  & October 16 & 58407 & 15 & $4.80 \pm 0.17$ &\dots\\
\midrule
% Oct 19 
VLA  & October 18 & 58409 & 6 & $3.19 \pm 0.05$ & 3\\
AMI-LA  & October 18 & 58409 & 15 & $5.0 \pm 0.05$ &\dots\\
MeerKAT  & October 19 & 58410 & 1.28 & $1.52 \pm 0.06$ & 2\\
AMI-LA  & October 19 & 58410 & 15 & $4.22 \pm 0.05$ &\dots\\
\midrule
% Oct 22 
AMI-LA  & October 20 & 58411 & 15 & $4.53 \pm 0.07$ &\dots\\
AMI-LA  & October 21 & 58412 & 15 & $4.06 \pm 0.15$ &\dots\\
AMI-LA  & October 22 & 58413 & 15 & $4.18 \pm 0.04$ &\dots\\
JCMT  & October 22 & 58413 & 350 & $10.4 \pm 1.9$ & 7\\
\midrule
% Oct 26 
AMI-LA  & October 24 & 58415 & 15 & $4.18 \pm 0.04$ &\dots\\
JCMT  & October 24 & 58415 & 350 & $14.3 \pm 2.1$ & 7\\
AMI-LA  & October 25 & 58416 & 15 & $3.75 \pm 0.049$ &\dots\\
eMERLIN  & October 26 & 58417 & 1.51 & $0.93 \pm 0.04$ & 2\\
AMI-LA  & October 26 & 58417 & 15 & $3.44 \pm 0.03$ &\dots\\
MeerKAT  & October 27 & 58418 & 1.28 & $1.61 \pm 0.05$ & 2\\
VLA  & October 27 & 58418 & 6 & $2.49 \pm 0.03$ & 3\\
AMI-LA  & October 27 & 58418 & 15 & $3.56 \pm 0.07$ &\dots\\
\midrule
% Oct 28 
eMERLIN  & October 28 & 58419 & 1.51 & $1.15 \pm 0.03$ & 2\\
AMI-LA  & October 28 & 58419 & 15 & $3.35 \pm 0.10$ &\dots\\
JCMT  & October 28 & 58419 & 350 & $11.8 \pm 2.5$ & 7\\
AMI-LA  & October 29 & 58420 & 15 &$4.01 \pm 0.07$  &\dots\\
AMI-LA  & October 30 & 58421 & 15 & $3.06 \pm 0.03$ &\dots\\ 
\midrule
% Nov 3 
AMI-LA  & November 1 & 58423 & 15 & $2.80 \pm 0.09$ &\dots\\
AMI-LA  & November 2 & 58424 & 15 & $3.03 \pm 0.04$ &\dots\\
MeerKAT  & November 3 & 58425 & 1.28 & $1.15 \pm 0.04$ & 2\\
AMI-LA  & November 3 & 58425 & 15 & $2.58 \pm 0.04$ &\dots\\
AMI-LA  & November 4 & 58426 & 15 & $2.82 \pm 0.06$ &\dots\\
\midrule
% Nov 6 
AMI-LA  & November 5 & 58427 & 15 & $2.86 \pm 0.11$ &\dots\\
JCMT  & November 6 & 58428 & 350 & $5.4 \pm 2.0$ & 7\\
AMI-LA  & November 7 & 58429 & 15 & $2.75 \pm 0.11$ &\dots\\
AMI-LA  & November 8 & 58430 & 15 & $2.46 \pm 0.04$ &\dots\\
\midrule
% Nov 10 
VLA  & November 9 & 58431 & 6 & $3.59 \pm 0.05$ & 3\\
MeerKAT  & November 10 & 58432 & 1.28 & $0.82 \pm 0.04$ & 2\\
AMI-LA  & November 10 & 58432 & 15 & $2.01 \pm 0.09$ &\dots\\
AMI-LA  & November 11 & 58433 & 15 & $2.32 \pm 0.14$ &\dots\\
\midrule
% Nov 13 
AMI-LA  & November 12 & 58434 & 15 & $2.39 \pm 0.05$ &\dots\\
MeerKAT  & November 13 & 58435 & 1.28 & $0.75 \pm 0.02$ & 2\\
JCMT  & November 13 & 58435 & 350 & $5.5 \pm 2.7$ & 7\\
AMI-LA  & November 14 & 58436 & 15 & $1.91 \pm 0.07$ &\dots\\
\midrule
% Nov 18 
AMI-LA  & November 16 & 58438 & 15 & $1.93 \pm 0.04$ &\dots\\
MeerKAT  & November 17 & 58439 & 1.28 & $0.79 \pm 0.05$ & 2\\
VLA  & November 18 & 58440 & 6 & $1.16 \pm 0.01$ & 3\\ 
AMI-LA  & November 18 & 58440 & 15 & $1.62 \pm 0.07$ &\dots\\
AMI-LA  & November 20 & 58442 & 15 & $1.67 \pm 0.10$ &\dots\\
\enddata
\tablerefs{1:\citealt{2018ATel11887....1B}; 2: \citealt{2020NatAs...4..697B}; 3: Project Code 18A-277 \citep{2021ApJ...907...34S}; 4: Project Code 2017.1.01103.T; 5: Project Code 18A-470; 6: Project Codes 2018A-S011 and 2017B-S010 ; 7: Project Code M18BP025; 8: Project Codes W17BM and W17BN; 9: \citep{2020MNRAS.493L..81A}.}
\end{deluxetable}

\clearpage
%%%%%% IR/OPT FLUXES TABLE %%%%%%
\startlongtable
\begin{deluxetable}{cccccccc}
\tablecaption{Flux densities of MAXI J1820+070 at infrared and optical frequencies.\label{tab:iropt_fluxes}}
\tablehead{
\colhead{Facility} & \colhead{Band} & \colhead{Date} & \colhead{MJD} & \colhead{Wavelength} & \colhead{Frequency}  & \colhead{Flux Density} & \colhead{Reference}\\
& &  (2018)   & &  ($\mu$m)  &  (GHz)   &  (mJy) & \\}
\scriptsize
\startdata
% March 16 
AAVSO & V & March 14 & 58191 & 0.55 & $5.5\times 10^5$ & $24.27 \pm 0.10$  & 1\\
AAVSO & V & March 15 & 58192 & 0.55 & $5.5\times 10^5$ & $23.98 \pm 2.25$ & 1\\
AAVSO & V & March 16 & 58193 & 0.55 & $5.5\times 10^5$ & $28.50 \pm 2.69$ & 1\\
AAVSO & V & March 17 & 58194 & 0.55 & $5.5\times 10^5$ & $38.60 \pm 3.80$ & 1\\
\midrule
% March 20 
AAVSO & V & March 18 & 58195 & 0.55 & $5.5\times 10^5$ & $36.05 \pm 3.81$ & 1\\
AAVSO & V & March 19 & 58196 & 0.55 & $5.5\times 10^5$ & $34.80 \pm 3.59$ & 1\\
AAVSO & V & March 20 & 58197 & 0.55 & $5.5\times 10^5$ & $36.77 \pm 2.89$ & 1\\
AAVSO & V & March 21 & 58198 & 0.55 & $5.5\times 10^5$ & $41.18 \pm 3.85$ & 1\\
REM & K & March 22 & 58199 & 2.13 & $1.41\times 10^5$ & $119.23 \pm 12.34$ & 2\\
REM & H & March 22 & 58199 & 1.64 & $1.83\times 10^5$ & $91.51 \pm 26.10$ & 2\\
REM & J & March 22 & 58199 & 1.25 & $2.40\times 10^5$ & $70.36 \pm 5.72$ & 2\\
REM & z & March 22 & 58199 & 0.89 & $3.36\times 10^5$ & $32.22 \pm 3.61$ & 2\\
REM & z & March 22 & 58199 & 0.89 & $3.36\times 10^5$ & $32.21 \pm 3.53$ & 2\\
REM & z & March 22 & 58199 & 0.89 & $3.36\times 10^5$ & $33.12 \pm 4.21$ & 2\\
REM & i & March 22 & 58199 & 0.75 & $4.02\times 10^5$ & $34.68 \pm 4.64$ & 2\\
REM & i & March 22 & 58199 & 0.75 & $4.02\times 10^5$ & $34.81 \pm 4.45$ & 2\\
REM & i  & March 22 & 58199 & 0.75 & $4.02\times 10^5$ & $35.71 \pm 4.37$ & 2\\
REM & r & March 22 & 58199 & 0.61 & $4.88\times 10^5$ & $40.29 \pm 4.552$ & 2\\
REM & r & March 22 & 58199 & 0.61 & $4.88\times 10^5$ & $39.79 \pm 5.07$ & 2\\
REM & r & March 22 & 58199 & 0.61 & $4.88\times 10^5$ & $40.78 \pm 4.99$ & 2\\
REM & g & March 22 & 58199 & 0.47 & $6.42\times 10^5$ & $47.98 \pm 3.61$ & 2\\
REM & g & March 22 & 58199 & 0.47 & $6.42\times 10^5$ & $46.44 \pm 4.72$ & 2\\
REM & g  & March 22 & 58199 & 0.47 & $6.42\times 10^5$ & $48.10 \pm 4.43$ & 2\\
\midrule
% April 12 
AAVSO & V & April 11 & 58219 & 0.55 & $5.5\times 10^5$ & $48.72 \pm 4.32$ & 1\\
AAVSO & I & April 12 & 58220 & 0.80 & $3.76\times 10^5$ & $51.40 \pm 6.05$ & 1\\
AAVSO & V & April 12 & 58220 & 0.55 & $5.5\times 10^5$ & $40.90 \pm 5.93$ & 1\\
AAVSO & I & April 13 & 58221 & 0.80 & $3.76\times 10^5$ & $47.38 \pm 6.78$ & 1\\
AAVSO & V & April 13 & 58221 & 0.55 & $5.5\times 10^5$ & $42.21 \pm 7.0$ & 1\\
AAVSO & B & April 13 & 58221 & 0.44 & $6.85\times 10^5$ & $42.65 \pm 5.07$ & 1\\
REM & K & April 13 & 58221 & 2.13 & $1.41\times 10^5$ & $91.83 \pm 12.56$ & 2\\
REM & H & April 13 & 58221 & 1.64 & $1.83\times 10^5$ & $75.70 \pm 21.88$ & 2\\
REM & J & April 13 & 58221 & 1.25 & $2.40\times 10^5$ & $60.50 \pm 5.76$ & 2\\
REM & z & April 13 & 58221 & 0.89 & $3.36\times 10^5$ & $28.94 \pm 4.68$ & 2\\
REM & z & April 13 & 58221 & 0.89 & $3.36\times 10^5$ & $28.88 \pm 18.45$ & 2\\
REM & z & April 13 & 58221 & 0.89 & $3.36\times 10^5$ & $28.53 \pm 18.52$ & 2\\
REM & i & April 13 & 58221 & 0.75 & $4.02\times 10^5$ & $33.31 \pm 4.04$ & 2\\
REM & i & April 13 & 58221 & 0.75 & $4.02\times 10^5$ & $33.90 \pm 3.12$ & 2\\
REM & i  & April 13 & 58221 & 0.75 & $4.02\times 10^5$ & $33.28 \pm 3.13$ & 2\\
REM & r & April 13 & 58221 & 0.61 & $4.88\times 10^5$ & $39.45 \pm 4.93$ & 2\\
REM & r & April 13 & 58221 & 0.61 & $4.88\times 10^5$ & $40.33 \pm 5.18$ & 2\\
REM & r & April 13 & 58221 & 0.61 & $4.88\times 10^5$ & $39.34 \pm 5.64$ & 2\\
REM & g & April 13 & 58221 & 0.47 & $6.42\times 10^5$ & $47.85 \pm 4.45$ & 2\\
REM & g & April 13 & 58221 & 0.47 & $6.42\times 10^5$ & $48.27 \pm 4.36$ & 2\\
REM & g  & April 13 & 58221 & 0.47 & $6.42\times 10^5$ & $45.89 \pm 4.57$ & 2\\
\midrule
% May 17 
AAVSO & V & May 16 & 58254 & 0.55 & $5.5\times 10^5$ & $25.60 \pm 1.23$ & 1\\
AAVSO & V & May 17 & 58255 & 0.55 & $5.5\times 10^5$ & $23.0 \pm 1.74$ & 1\\
AAVSO & V & May 18 & 58256 & 0.55 & $5.5\times 10^5$ & $20.47 \pm 2.23$ & 1\\
REM & K & May 18 & 58256 & 2.13 & $1.41\times 10^5$ & $22.19 \pm 3.86$ & 2\\
REM & H & May 18 & 58256 & 1.64 & $1.83\times 10^5$ & $21.49 \pm 4.62$ & 2\\
REM & J & May 18 & 58256 & 1.25 & $2.40\times 10^5$ & $18.85 \pm 1.48$ & 2\\
REM & z & May 18 & 58256 & 0.89 & $3.36\times 10^5$ & $12.45 \pm 2.01$ & 2\\
REM & z & May 18 & 58256 & 0.89 & $3.36\times 10^5$ & $12.48 \pm 2.06$ & 2\\
REM & z & May 18 & 58256 & 0.89 & $3.36\times 10^5$ & $12.41 \pm 1.83$ & 2\\
REM & i & May 18 & 58256 & 0.75 & $4.02\times 10^5$ & $14.05 \pm 1.85$ & 2\\
REM & i & May 18 & 58256 & 0.75 & $4.02\times 10^5$ & $13.97 \pm 1.84$ & 2\\
REM & i  & May 18 & 58256 & 0.75 & $4.02\times 10^5$ & $13.92 \pm 1.67$ & 2\\
REM & r & May 18 & 58256 & 0.61 & $4.88\times 10^5$ & $16.70 \pm 2.0$ & 2\\
REM & r & May 18 & 58256 & 0.61 & $4.88\times 10^5$ & $16.54 \pm 1.77$ & 2\\
REM & r & May 18 & 58256 & 0.61 & $4.88\times 10^5$ & $16.50 \pm 1.70$ & 2\\
REM & g  & May 18 & 58256 & 0.47 & $6.42\times 10^5$ & $20.16 \pm 2.85$ & 2\\
REM & g  & May 18 & 58256 & 0.47 & $6.42\times 10^5$ & $20.01 \pm 2.77$ & 2\\
REM & g  & May 18 & 58256 & 0.47 & $6.42\times 10^5$ & $19.72 \pm 2.76$ & 2\\
VISIR & B11.7 & May 20 & 58258 & 11.56 & $2.59\times 10^4$ & $66.27 \pm 3.30$ & \dots\\
VISIR & B10.7 & May 20 & 58258 & 10.67 & $2.81\times 10^4$ & $63.18 \pm 3.53$ & \dots\\
VISIR & J8.9 & May 20 & 58258 & 8.74 & $3.43\times 10^4$ & $46.60 \pm 3.01$ & \dots\\
VISIR & M & May 20 & 58258 & 4.89 & $6.16\times 10^4$  & $23.12 \pm 5.16$ & \dots\\
\midrule
% July 6 
AAVSO & V & July 4 & 58303 & 0.55 & $5.5\times 10^5$ & $14.71 \pm 2.08$ & 1\\
AAVSO & V & July 5 & 58304 & 0.55 & $5.5\times 10^5$ & $15.90 \pm 3.16$ & 1\\
AAVSO & V & July 6 & 58305 & 0.55 & $5.5\times 10^5$ & $15.99 \pm 2.61$ & 1\\
AAVSO & V & July 7 & 58306 & 0.55 & $5.5\times 10^5$ & $17.06 \pm 2.98$ & 1\\
REM & K & July 8 & 58307 & 2.13 & $1.41\times 10^5$ & $7.20 \pm 0.79$ & 2\\
REM & H & July 8 & 58307 & 1.64 & $1.83\times 10^5$ & $10.55 \pm 0.70$ & 2\\
REM & J & July 8 & 58307 & 1.25 & $2.40\times 10^5$ & $13.20 \pm 1.05$ & 2\\
REM & z & July 8 & 58307 & 0.89 & $3.36\times 10^5$ & $10.24 \pm 1.32$ & 2\\
REM & z & July 8 & 58307 & 0.89 & $3.36\times 10^5$ & $10.21 \pm 1.17$ & 2\\
REM & z & July 8 & 58307 & 0.89 & $3.36\times 10^5$ & $10.19 \pm 1.45$ & 2\\
REM & i & July 8 & 58307 & 0.75 & $4.02\times 10^5$ & $12.33 \pm 1.13$ & 2\\
REM & i & July 8 & 58307 & 0.75 & $4.02\times 10^5$ & $12.42 \pm 1.45$ & 2\\
REM & r & July 8 & 58307 & 0.61 & $4.88\times 10^5$ & $15.22 \pm 2.0$ & 2\\
REM & r & July 8 & 58307 & 0.61 & $4.88\times 10^5$ & $15.47 \pm 2.0$ & 2\\
REM & r & July 8 & 58307 & 0.61 & $4.88\times 10^5$ & $15.47 \pm 1.94$ & 2\\
REM & g  & July 8 & 58307 & 0.47 & $6.42\times 10^5$ & $19.24 \pm 1.57$ & 2\\
REM & g  & July 8 & 58307 & 0.47 & $6.42\times 10^5$ & $19.55 \pm 2.75$ & 2\\
REM & g  & July 8 & 58307 & 0.47 & $6.42\times 10^5$ & $19.33 \pm 2.03$ & 2\\
\midrule
% July 13 
LCO & i' & July 11 & 58310 & 0.76 & $3.97\times 10^5$ & $20.23 \pm 0.05$ & \dots\\
LCO & g' & July 11 & 58310 & 0.48 & $6.29\times 10^5$ & $22.22 \pm 0.04$ & \dots\\
VISIR & B10.7 & July 12 & 58311 & 10.67 & $2.81\times 10^4$ & $4.84 \pm 1.44$ & \dots\\
VISIR & M & July 13 & 58312 & 4.89 & $6.16\times 10^4$  & $<5.41$ & \dots\\
VISIR & B10.7 & July 13 & 58312 & 10.67 & $2.81\times 10^4$ & $<1.56$ & \dots\\
VISIR & B11.7 & July 13 & 58312 & 11.56 & $2.59\times 10^4$ & $<3.78$ & \dots\\
VISIR &J 8.9 & July 13 & 58312 & 8.74 & $3.43\times 10^4$ & $<3.48$ & \dots\\
\midrule
% Sept 29 
AAVSO & V & September 27 & 58388 & 0.55 & $5.5\times 10^5$ & $6.86 \pm 1.0$ & \dots\\
VISIR & B10.7 & September 28 & 58389 & 10.67 & $2.81\times 10^4$ & $<1.04$ & \dots\\
VISIR & B10.7 & September 28 & 58389 & 10.67 & $2.81\times 10^4$ & $<1.24$& \dots\\
LCO & Y & September 28 & 58389 & 1.0 & $2.99\times 10^5$ & $6.67 \pm 0.10$ & \dots\\
LCO & i' & September 28 & 58389 & 0.76 & $3.97\times 10^5$ & $7.30 \pm 0.03$ & \dots\\
LCO & r' & September 28 & 58389 & 0.62 & $4.82\times 10^5$ & $7.02 \pm 0.02$ & \dots\\
AAVSO & V & September 28 & 58389 & 0.55 & $5.5\times 10^5$ & $6.44 \pm 0.95$ & 1\\
LCO & g' & September 28 & 58389 & 0.48 & $6.29\times 10^5$ & $7.0 \pm 0.02$ & \dots\\
AAVSO & V & September 29 & 58390 & 0.55 & $5.5\times 10^5$ & $5.50 \pm 0.70$ & 1\\
AAVSO & V & September 30 & 58391 & 0.55 & $5.5\times 10^5$ & $5.78 \pm 0.29$ & 1\\
AAVSO & V & October 1 & 58392 & 0.55 & $5.5\times 10^5$ & $4.98 \pm 0.53$ & 1\\
\midrule
% October 6 
AAVSO & V & October 4 & 58395 & 0.55 & $5.5\times 10^5$ & $4.96 \pm 0.30$ & 1\\
AAVSO & V & October 5 & 58396 & 0.55 & $5.5\times 10^5$ & $4.59 \pm 0.48$ & 1\\
AAVSO & V & October 6 & 58397 & 0.55 & $5.5\times 10^5$ & $5.44 \pm 0.52$ & 1\\
AAVSO & V & October 7 & 58398 & 0.55 & $5.5\times 10^5$ & $6.59 \pm 0.38$ & 1\\
AAVSO & V & October 8 & 58399 & 0.55 & $5.5\times 10^5$ & $6.93 \pm 0.45$ & 1\\
\midrule
% October 11 
AAVSO & V & October 9 & 58400 & 0.55 & $5.5\times 10^5$ & $7.61 \pm 0.57$ & 1\\
AAVSO & I & October 10 & 58401 & 0.80 & $3.76\times 10^5$ & $2.65 \pm 0.35$ & 1\\
AAVSO & V & October 10 & 58401 & 0.55 & $5.5\times 10^5$ & $7.01 \pm 0.93$ & 1\\
AAVSO & V & October 11 & 58402 & 0.55 & $5.5\times 10^5$ & $8.16 \pm 0.23$ & 1\\
LCO & Y & October 12 & 58403 & 1.0 & $2.99\times 10^5$ & $11.78 \pm 0.09$ & \dots\\
LCO & i' & October 12 & 58403 & 0.76 & $3.97\times 10^5$ & $10.21 \pm 0.03$ & \dots\\
LCO & r' & October 12 & 58403 & 0.62 & $4.82\times 10^5$ & $8.88 \pm 0.02$ & \dots\\
AAVSO & V & October 12 & 58403 & 0.55 & $5.5\times 10^5$ & $7.36 \pm 0.31$ & 1\\
LCO & g' & October 12 & 58403 & 0.48 & $6.29\times 10^5$ & $7.93 \pm 0.20$ & \dots\\
VISIR & B10.7 & October 12 & 58404 & 10.67 & $2.81\times 10^4$ & $21.34 \pm 4.33$ & \dots\\
\midrule
% October 14 
AAVSO & V & October 13 & 58404 & 0.55 & $5.5\times 10^5$ & $8.19 \pm 0.49$ & 1\\
AAVSO & I & October 14 & 58405 & 0.80 & $3.76\times 10^5$ & $3.13 \pm 0.07$ & 1\\
AAVSO & V & October 14 & 58405 & 0.55 & $5.5\times 10^5$ & $6.97 \pm 1.24$ & 1\\
AAVSO & V & October 15 & 58406 & 0.55 & $5.5\times 10^5$ & $7.73 \pm 0.26$ & 1\\
AAVSO & V & October 16 & 58407 & 0.55 & $5.5\times 10^5$ & $7.50 \pm 0.32$ & 1\\
\midrule
% October 19 
AAVSO & V & October 17 & 58408 & 0.55 & $5.5\times 10^5$ & $7.40 \pm 0.48$ & 1\\
AAVSO & V & October 18 & 58409 & 0.55 & $5.5\times 10^5$ & $7.14 \pm 0.79$ & 1\\
\midrule
% October 22 
AAVSO & V & October 20 & 58411 & 0.55 & $5.5\times 10^5$ & $6.39 \pm 0.92$ & 1\\
AAVSO & V & October 21 & 58412 & 0.55 & $5.5\times 10^5$ & $6.48 \pm 0.73$ & 1\\
AAVSO & V & October 22 & 58413 & 0.55 & $5.5\times 10^5$ & $6.16 \pm 0.43$ & 1\\
AAVSO & V & October 23 & 58414 & 0.55 & $5.5\times 10^5$ & $5.82 \pm 0.41$ & 1\\
\midrule
% October 26 
AAVSO & V & October 26 & 58417 & 0.55 & $5.5\times 10^5$ & $5.35 \pm 0.25$ & 1\\
AAVSO & V & October 27 & 58418 & 0.55 & $5.5\times 10^5$ & $4.76 \pm 0.56$ & 1\\
\midrule
% October 28 
AAVSO & V & October 28 & 58419 & 0.55 & $5.5\times 10^5$ & $4.74 \pm 0.52$ & 1\\
AAVSO & V & October 29 & 58420 & 0.55 & $5.5\times 10^5$ & $4.74 \pm 0.26$ & 1\\
\midrule
% Nov 3 
AAVSO & V & November 1 & 58423 & 0.55 & $5.5\times 10^5$ & $3.76 \pm 0.43$ & 1\\
AAVSO & V & November 2 & 58424 & 0.55 & $5.5\times 10^5$ & $3.61 \pm 0.52$ & 1\\
AAVSO & V & November 4 & 58426 & 0.55 & $5.5\times 10^5$ & $3.14 \pm 0.15$ & 1\\
\midrule
% Nov 6 
AAVSO & V & November 7 & 58429 & 0.55 & $5.5\times 10^5$ & $2.74 \pm 0.14$ & 1\\
AAVSO & V & November 8 & 58430 & 0.55 & $5.5\times 10^5$ & $3.48 \pm 0.15$ & 1\\
\midrule
% Nov 10 
AAVSO & V & November 9 & 58431 & 0.55 & $5.5\times 10^5$ & $3.16 \pm 0.25$ & 1\\
AAVSO & V & November 10 & 58432 & 0.55 & $5.5\times 10^5$ & $3.05 \pm 0.18$ & 1\\
\midrule
% Nov 13 
AAVSO & V & November 13 & 58435 & 0.55 & $5.5\times 10^5$ & $2.16 \pm 0.26$ & 1\\
AAVSO & V & November 14 & 58436 & 0.55 & $5.5\times 10^5$ & $2.01 \pm 0.30$ & 1\\
\midrule
% Nov 18
AAVSO & V & November 17 & 58439 & 0.55 & $5.5\times 10^5$ & $1.34 \pm 0.12$ & 1\\
AAVSO & V & November 18 & 58440 & 0.55 & $5.5\times 10^5$ & $1.42 \pm 0.16$ & 1\\
AAVSO & V & November 19 & 58441 & 0.55 & $5.5\times 10^5$ & $1.50 \pm 0.32$ & 1\\
\enddata
\tablerefs{1: \citealt{AAVSO_CITE}; 2: Proposal code 37025}
\end{deluxetable}

%%%%%% XSHOOTER BINNED FLUXES TABLE %%%%%%
\startlongtable
\begin{deluxetable}{cccccc}
\tablecaption{Flux densities of MAXI J1820+070 at optical frequencies from the X-shooter instrument (Project code: 0101.D-0356). The original date of the observation included in the representative epoch July 13 is 2018 July 14. \label{tab:xshooter_fluxes}}
\tablehead{
\colhead{Date} & \colhead{Wavelength} & \colhead{Frequency}  & \colhead{Flux Density}\\
(2018)   &  ($\mu$m)   &  (GHz)   &  (mJy) \\}
\scriptsize
\startdata
% April 12
April 12 & 2.14 & 1.40$\times10^{5}$ & 83.19 $\pm$ 5.25 \\
& 1.61 & 1.87$\times10^{5}$ & 64.18 $\pm$ 4.61\\
& 0.99 & 3.01$\times10^{5}$ & 50.83 $\pm$ 0.13\\
& 0.94 & 3.19$\times10^{5}$ & 50.40 $\pm$ 0.38\\
& 0.87 & 3.38$\times10^{5}$ & 48.66 $\pm$ 0.45\\
& 0.83 & 3.60$\times10^{5}$ & 48.27 $\pm$ 0.14\\
& 0.78 & 3.85$\times10^{5}$ & 47.38 $\pm$ 0.26\\
& 0.72 & 4.14$\times10^{5}$ & 46.80 $\pm$ 0.37\\
& 0.67 & 4.47$\times10^{5}$ & 46.0 $\pm$ 0.29\\
& 0.62 & 4.87$\times10^{5}$ & 45.49 $\pm$ 0.04\\
& 0.56 & 5.34$\times10^{5}$ & 44.44 $\pm$ 0.38\\
& 0.50 & 5.98$\times10^{5}$ & 43.64 $\pm$ 0.16\\
& 0.45 & 6.70$\times10^{5}$ & 43.20 $\pm$ 0.36\\
& 0.39 & 7.62$\times10^{5}$ & 41.26 $\pm$ 0.65\\
& 0.35 & 8.55$\times10^{5}$ & 40.63 $\pm$ 0.02\\
\midrule
% July 13 
July 13 & 2.32 & 1.29$\times10^{5}$ & 10.31 $\pm$ 0.12\\
& 2.14 & 1.40$\times10^{5}$ & 10.83 $\pm$ 0.15\\
& 1.99 & 1.51$\times10^{5}$ & 11.86 $\pm$ 0.68 \\
& 1.66 & 1.81$\times10^{5}$ & 13.54 $\pm$ 0.31\\
& 1.49 & 2.01$\times10^{5}$ &  15.51 $\pm$ 0.72\\
& 1.24 & 2.42$\times10^{5}$ & 17.29 $\pm$ 0.50\\
& 1.07 & 2.79$\times10^{5}$ & 19.15 $\pm$ 0.55\\
& 0.97 & 3.08$\times10^{5}$ & 20.32 $\pm$ 0.19\\
& 0.92 & 3.26$\times10^{5}$ & 21.02 $\pm$ 0.14\\
& 0.86 & 3.47$\times10^{5}$ & 21.61 $\pm$ 0.35\\
& 0.81 & 3.72$\times10^{5}$ & 22.72 $\pm$ 0.21\\
& 0.75 & 3.99$\times10^{5}$ & 23.13 $\pm$ 0.05\\
& 0.70 & 4.31$\times10^{5}$ & 23.27 $\pm$ 0.05\\
& 0.64 & 4.69$\times10^{5}$ & 23.39 $\pm$ 0.03\\
& 0.58 & 5.16$\times10^{5}$ & 23.15 $\pm$ 0.28\\
& 0.52 & 5.77$\times10^{5}$ & 22.56 $\pm$ 0.37\\
& 0.46 & 6.47$\times10^{5}$ & 22.01 $\pm$ 0.37\\
& 0.41 & 7.36$\times10^{5}$ & 21.15 $\pm$ 0.26\\
& 0.36 & 8.39$\times10^{5}$ & 20.67 $\pm$ 0.19\\
\midrule
% September 29 
September 29 & 2.32 & 1.29$\times10^{5}$ & 43.41 $\pm$ 0.08\\
& 2.14 & 1.40$\times10^{5}$ & 46.23 $\pm$ 0.06\\
& 1.99 & 1.51$\times10^{5}$ & 49.39 $\pm$ 0.24 \\
& 1.66 & 1.81$\times10^{5}$ & 54.32 $\pm$ 0.06\\
& 1.49 & 2.01$\times10^{5}$ &  58.43 $\pm$ 0.11\\
& 1.24 & 2.42$\times10^{5}$ & 62.04 $\pm$ 0.09\\
& 1.07 & 2.79$\times10^{5}$ & 65.11 $\pm$ 0.14\\
& 0.97 & 3.08$\times10^{5}$ & 67.43 $\pm$ 0.03\\
& 0.92 & 3.26$\times10^{5}$ & 68.34 $\pm$ 0.03\\
& 0.86 & 3.47$\times10^{5}$ & 67.79 $\pm$ 0.05\\
& 0.81 & 3.72$\times10^{5}$ & 69.22 $\pm$ 0.02\\
& 0.75 & 3.99$\times10^{5}$ & 68.41 $\pm$ 0.03\\
& 0.70 & 4.31$\times10^{5}$ & 67.03 $\pm$ 0.05\\
& 0.64 & 4.69$\times10^{5}$ & 65.24 $\pm$ 0.07\\
& 0.58 & 5.16$\times10^{5}$ & 62.98 $\pm$ 0.05\\
& 0.52 & 5.77$\times10^{5}$ & 60.53 $\pm$ 0.14\\
& 0.46 & 6.47$\times10^{5}$ & 57.13 $\pm$ 0.07\\
& 0.41 & 7.36$\times10^{5}$ & 53.23 $\pm$ 0.15\\
& 0.36 & 8.39$\times10^{5}$ & 48.50 $\pm$ 0.15\\
\enddata
\end{deluxetable}

%%%%%% UVOT FLUXES TABLE %%%%%%
\startlongtable
\begin{deluxetable}{cccccc}
\tablecaption{Flux densities of MAXI J1820+070 at UV frequencies. \label{tab:uv_fluxes}}
\tablehead{
\colhead{Band} & \colhead{Date} & \colhead{MJD} & \colhead{Wavelength} & \colhead{Frequency}  & \colhead{Flux Density} \\
&  (2018)   &     & ($\mu$m) &  (GHz)   &  (mJy) \\}
\scriptsize
\startdata
% March 16 
UW1 & March 14 & 58191 & 0.26 & $1.16 \times 10^6$ & $15.2 \pm 0.6$ \\
UM2 & March 14 & 58191 & 0.22 & $1.34 \times 10^6$ & $12.0 \pm 0.3$  \\
UW2 & March 14 & 58191 & 0.20 & $1.48 \times 10^6$ & $12.9 \pm 0.4$ \\
UW1 & March 16 & 58193 & 0.26 & $1.16 \times 10^6$ & $16.7 \pm 0.6$  \\
UM2 & March 16 & 58193 & 0.22 & $1.34 \times 10^6$ & $13.7 \pm 0.3$  \\
UW2 & March 16 & 58193 & 0.20 & $1.48 \times 10^6$ & $15.4 \pm 0.4$  \\
UW1 & March 17 & 58194 & 0.26 & $1.16 \times 10^6$ & $18.8 \pm 0.7$  \\
UM2 & March 17 & 58194 & 0.22 & $1.34 \times 10^6$ & $14.6 \pm 0.3$  \\
UW2 & March 17 & 58194 & 0.20 & $1.48 \times 10^6$ & $16.2 \pm 0.5$  \\
\midrule
% March 20 
UW1 & March 19 & 58196 & 0.26 & $1.16 \times 10^6$ & $24.3 \pm 0.9$  \\
UM2 & March 19 & 58196 & 0.22 & $1.34 \times 10^6$ & $20.3 \pm 0.5$  \\
UW2 & March 19 & 58196 & 0.20 & $1.48 \times 10^6$ & $20.4 \pm 0.6$  \\
UW1 & March 20 & 58197 & 0.26 & $1.16 \times 10^6$ & $24 \pm 0.9$  \\
UM2 & March 20 & 58197 & 0.22 & $1.34 \times 10^6$ & $20.7 \pm 0.5$  \\
UW2 & March 20 & 58197 & 0.20 & $1.48 \times 10^6$ & $22 \pm 0.7$ \\
UW1 & March 21 & 58198 & 0.26 & $1.16 \times 10^6$ & $27.9 \pm 1.1$  \\
UM2 & March 21 & 58198 & 0.22 & $1.34 \times 10^6$ & $22 \pm 0.5$  \\
UW2 & March 21 & 58198 & 0.20 & $1.48 \times 10^6$ & $23.4 \pm 0.7$  \\
\midrule
% April 12
UM2 & April 11 & 58219 & 0.22 & $1.34 \times 10^6$ & $21.5 \pm 0.6$  \\
UM2 & April 14 & 58222 & 0.22 & $1.34 \times 10^6$ & $18.7 \pm 0.6$  \\
\midrule
% May 17
UW1 & May 17 & 58253 & 0.26 & $1.16 \times 10^6$ & $13 \pm 0.5$  \\
UM2 & May 17 & 58253 & 0.22 & $1.34 \times 10^6$ & $10.3 \pm 0.2$  \\
UW2 & May 17 & 58253 & 0.20 & $1.48 \times 10^6$ & $10.9 \pm 0.3$  \\
\midrule
% July 6
UW1 & July 6 & 58305 & 0.26 & $1.16 \times 10^6$ & $11.9 \pm 0.4$  \\
UM2 & July 6 & 58305 & 0.22 & $1.34 \times 10^6$ & $9.3 \pm 0.2$  \\
UW2 & July 6 & 58305 & 0.20 & $1.48 \times 10^6$ & $10 \pm 0.3$  \\
UW1 & July 8 & 58307 & 0.26 & $1.16 \times 10^6$ & $10.7 \pm 0.4$  \\
UM2 & July 8 & 58307 & 0.22 & $1.34 \times 10^6$ & $8.7 \pm 0.2$  \\
UW2 & July 8 & 58307 & 0.20 & $1.48 \times 10^6$ & $9.5 \pm 0.3$  \\
UW1 & July 8 & 58307 & 0.26 & $1.16 \times 10^6$ & $10.9 \pm 0.4$  \\
UM2 & July 8 & 58307 & 0.22 & $1.34 \times 10^6$ & $8.2 \pm 0.2$  \\
UW2 & July 8 & 58307 & 0.20 & $1.48 \times 10^6$ & $8.1 \pm 0.3$ \\
\midrule
% July 13
UW1 & July 11 & 58310 & 0.26 & $1.16 \times 10^6$ & $12.4 \pm 0.5$  \\
UW2 & July 11 & 58310 & 0.20 & $1.48 \times 10^6$ & $10.6 \pm 0.3$  \\
UW1 & July 11 & 58310 & 0.26 & $1.16 \times 10^6$ & $12.7 \pm 0.5$  \\
UM2 & July 11 & 58310 & 0.22 & $1.34 \times 10^6$ & $10.8 \pm 0.2$  \\
UW2 & July 11 & 58310 & 0.20 & $1.48 \times 10^6$ & $11.4 \pm 0.3$  \\
UW1 & July 13 & 58312 & 0.26 & $1.16 \times 10^6$ & $14.5 \pm 0.5$  \\
UM2 & July 13 & 58312 & 0.22 & $1.34 \times 10^6$ & $11.4 \pm 0.3$  \\
UW2 & July 13 & 58312 & 0.20 & $1.48 \times 10^6$ & $12.5 \pm 0.4$  \\
UW1 & July 15 & 58314 & 0.26 & $1.16 \times 10^6$ & $14.3 \pm 0.5$  \\
\midrule
% September 29
V & September 27 & 58389 & 0.54 & $5.6 \times 10^5$ & $7.2 \pm 0.2$  \\
UW1 & September 27 & 58389 & 0.26 & $1.16 \times 10^6$ & $4.3 \pm 0.2$  \\
UM2 & September 27 & 583892 & 0.22 & $1.34 \times 10^6$ & $3.5 \pm 0.1$  \\
UW2 & September 27 & 58389 & 0.20 & $1.48 \times 10^6$ & $3.8 \pm 0.1$  \\
V & September 27 & 58389 & 0.54 & $5.6 \times 10^5$ & $6.5 \pm 0.2$  \\
UW1 & September 29 & 58390 & 0.26 & $1.16 \times 10^6$ & $4.0 \pm 0.2$ \\
UM2 & September 29 & 58390 & 0.22 & $1.34 \times 10^6$ & $3.2 \pm 0.1$  \\
UW2 & September 29 & 58390 & 0.20 & $1.48 \times 10^6$ & $3.5 \pm 0.1$  \\
V & September 30 & 58391 & 0.54 & $5.6 \times 10^5$ & $5.9 \pm 0.2$  \\
UW1 & September 30 & 58391 & 0.26 & $1.16 \times 10^6$ & $3.8 \pm 0.2$  \\
UM2 & September 30 & 58391 & 0.22 & $1.34 \times 10^6$ & $3.0 \pm 0.1$  \\
UW2 & September 30 & 58391 & 0.20 & $1.48 \times 10^6$ & $3.4 \pm 0.1$  \\
V & October 1 & 58392 & 0.54 & $5.6 \times 10^5$ & $5.8 \pm 0.2$  \\
UW1 & October 1 & 58392 & 0.26 & $1.16 \times 10^6$ & $3.7 \pm 0.1$  \\
UM2 & October 1 & 58392 & 0.22 & $1.34 \times 10^6$ & $3.0 \pm 0.1$  \\
UW2 & October 1 & 58392 & 0.20 & $1.48 \times 10^6$ & $3.2 \pm 0.1$  \\
\midrule
% October 6
UW1 & October 6 & 58397 & 0.26 & $1.16 \times 10^6$ & $3.1 \pm 0.1$  \\
UM2 & October 6 & 58397 & 0.22 & $1.34 \times 10^6$ & $2.5 \pm 0.1$  \\
\midrule
% October 11
V & October 9 & 58400 & 0.54 & $5.6 \times 10^5$ & $7.3 \pm 0.3$  \\
UW1 & October 9 & 58400 & 0.26 & $1.16 \times 10^6$ & $3.6 \pm 0.1$  \\
UM2 & October 9 & 58400 & 0.22 & $1.34 \times 10^6$ & $2.8 \pm 0.1$  \\
UW2 & October 9 & 58400 & 0.20 & $1.48 \times 10^6$ & $3.1 \pm 0.1$  \\
UW1 & October 11 & 58402 & 0.26 & $1.16 \times 10^6$ & $4.4 \pm 0.2$  \\
UM2 & October 11 & 58402 & 0.22 & $1.34 \times 10^6$ & $3.2 \pm 0.1$  \\
UW2 & October 11 & 58402 & 0.20 & $1.48 \times 10^6$ & $3.4 \pm 0.1$  \\
\midrule
% October 14
UW1 & October 13 & 58404 & 0.26 & $1.16 \times 10^6$ & $4.1 \pm 0.2$  \\
UM2 & October 13 & 58404 & 0.22 & $1.34 \times 10^6$ & $3.1 \pm 0.1$  \\
UW2 & October 13 & 58404 & 0.20 & $1.48 \times 10^6$ & $3.2 \pm 0.1$  \\
V & October 15 & 58406 & 0.54 & $5.6 \times 10^5$ & $7.6 \pm 0.3$  \\
UW1 & October 15 & 58406 & 0.26 & $1.16 \times 10^6$ & $3.7 \pm 0.1$  \\
UM2 & October 15 & 58406 & 0.22 & $1.34 \times 10^6$ & $2.8 \pm 0.1$  \\
UW2 & October 15 & 58406 & 0.20 & $1.48 \times 10^6$ & $2.9 \pm 0.1$  \\
\midrule
% October 19
UW1 & October 17 & 58408 & 0.26 & $1.16 \times 10^6$ & $3.8 \pm 0.2$  \\
UM2 & October 17 & 58408 & 0.22 & $1.34 \times 10^6$ & $2.8 \pm 0.1$  \\
UW2 & October 17 & 58408 & 0.20 & $1.48 \times 10^6$ & $3.0 \pm 0.1$  \\
V & October 19 & 58410 & 0.54 & $5.6 \times 10^5$ & $7.1 \pm 0.2$  \\
UW1 & October 19 & 58410 & 0.26 & $1.16 \times 10^6$ & $3.5 \pm 0.1$  \\
UM2 & October 19 & 58410 & 0.22 & $1.34 \times 10^6$ & $2.7 \pm 0.1$  \\
UW2 & October 19 & 58410 & 0.20 & $1.48 \times 10^6$ & $2.8 \pm 0.1$  \\
\midrule
% October 22
V & October 21 & 58412 & 0.54 & $5.6 \times 10^5$ & $7.1 \pm 0.2$  \\
UW1 & October 21 & 58412 & 0.26 & $1.16 \times 10^6$ & $3.2 \pm 0.1$  \\
UM2 & October 21 & 58412 & 0.22 & $1.34 \times 10^6$ & $2.6 \pm 0.1$  \\
UW2 & October 21 & 58412 & 0.20 & $1.48 \times 10^6$ & $2.8 \pm 0.1$  \\
\midrule
% October 26
V & October 26 & 58417 & 0.54 & $5.6 \times 10^5$ & $5.4 \pm 0.2$  \\
UW1 & October 26 & 58417 & 0.26 & $1.16 \times 10^6$ & $2.6 \pm 0.1$  \\
UM2 & October 26 & 58417 & 0.22 & $1.34 \times 10^6$ & $2.0 \pm 0.1$  \\
UW2 & October 26 & 58417 & 0.20 & $1.48 \times 10^6$ & $2.2 \pm 0.1$  \\
\midrule
% October 28
V & October 28 & 58419 & 0.54 & $5.6 \times 10^5$ & $5.7 \pm 0.2$  \\
UW1 & October 28 & 58419 & 0.26 & $1.16 \times 10^6$ & $2.7 \pm 0.1$  \\
UM2 & October 28 & 58419 & 0.22 & $1.34 \times 10^6$ & $2.0 \pm 0.1$  \\
UW2 & October 28 & 58419 & 0.20 & $1.48 \times 10^6$ & $2.1 \pm 0.1$  \\
V & October 30 & 58421 & 0.54 & $5.6 \times 10^5$ & $5.0 \pm 0.2$  \\
UW1 & October 30 & 58421 & 0.26 & $1.16 \times 10^6$ & $2.5 \pm 0.1$  \\
UM2 & October 30 & 58421 & 0.22 & $1.34 \times 10^6$ & $1.9 \pm 0.1$  \\
UW2 & October 30 & 58421 & 0.20 & $1.48 \times 10^6$ & $2.0 \pm 0.1$  \\
\midrule
% November 3
V & November 3 & 58425 & 0.54 & $5.6 \times 10^5$ & $3.8 \pm 0.2$  \\
UW1 & November 3 & 58425 & 0.26 & $1.16 \times 10^6$ & $2.1 \pm 0.1$  \\
UM2 & November 3 & 58425 & 0.22 & $1.34 \times 10^6$ & $1.5 \pm 0.1$  \\
UW2 & November 3 & 58425 & 0.20 & $1.48 \times 10^6$ & $1.6 \pm 0.1$  \\
V & November 4 & 58426 & 0.54 & $5.6 \times 10^5$ & $4.4 \pm 0.2$  \\
UW2 & November 4 & 58426 & 0.20 & $1.48 \times 10^6$ & $1.6 \pm 0.1$  \\
\midrule
% November 6
V & November 6 & 58428 & 0.54 & $5.6 \times 10^5$ & $3.8 \pm 0.1$  \\
UW1 & November 6 & 58428 & 0.26 & $1.16 \times 10^6$ & $1.8 \pm 0.1$  \\
UM2 & November 6 & 58428 & 0.22 & $1.34 \times 10^6$ & $1.44 \pm 0.04$  \\
UW2 & November 6 & 58428 & 0.20 & $1.48 \times 10^6$ & $1.5 \pm 0.1$  \\
V & November 8 & 58430 & 0.54 & $5.6 \times 10^5$ & $3.1 \pm 0.2$  \\
UW1 & November 8 & 58430 & 0.26 & $1.16 \times 10^6$ & $1.6 \pm 0.1$  \\
UM2 & November 8 & 58430 & 0.22 & $1.34 \times 10^6$ & $1.30 \pm 0.04$  \\
UW2 & November 8 & 58430 & 0.20 & $1.48 \times 10^6$ & $1.25 \pm 0.04$ \\
\midrule
% November 10
V & November 10 & 58432 & 0.54 & $5.6 \times 10^5$ & $3.1 \pm 0.2$  \\
UW1 & November 10 & 58432 & 0.26 & $1.16 \times 10^6$ & $1.5 \pm 0.1$  \\
UM2 & November 10 & 58432 & 0.22 & $1.34 \times 10^6$ & $1.1 \pm 0.1$  \\
UW2 & November 10 & 58432 & 0.20 & $1.48 \times 10^6$ & $1.09 \pm 0.04$  \\
\midrule
% November 13
V & November 12 & 58434 & 0.54 & $5.6 \times 10^5$ & $2.6 \pm 0.1$  \\
UW1 & November 12 & 58434 & 0.26 & $1.16 \times 10^6$ & $1.3 \pm 0.1$  \\
UM2 & November 12 & 58434 & 0.22 & $1.34 \times 10^6$ & $0.99 \pm 0.03$  \\
UW2 & November 12 & 58434 & 0.20 & $1.48 \times 10^6$ & $0.95 \pm 0.03$  \\
V & November 14 & 58436 & 0.54 & $5.6 \times 10^5$ & $2.4 \pm 0.1$  \\
U & November 14 & 58436 & 0.35 & $8.6 \times 10^5$ & $1.9 \pm 0.1$  \\
UW1 & November 14 & 58436 & 0.26 & $1.16 \times 10^6$ & $1.1 \pm 0.1$  \\
UM2 & November 14 & 58436 & 0.22 & $1.34 \times 10^6$ & $0.84 \pm 0.03$  \\
UW2 & November 14 & 58436 & 0.20 & $1.48 \times 10^6$ & $0.95 \pm 0.03$  \\
\midrule
% November 18
V & November 18 & 58440 & 0.54 & $5.6 \times 10^5$ & $1.84 \pm 0.1$  \\
U & November 18 & 58440 & 0.35 & $8.6 \times 10^5$ & $1.29 \pm 0.04$  \\
UW1 & November 18 & 58440 & 0.26 & $1.16 \times 10^6$ & $0.83 \pm 0.04$  \\
UM2 & November 18 & 58440 & 0.22 & $1.34 \times 10^6$ & $0.59 \pm 0.02$  \\
UW2 & November 18 & 58440 & 0.20 & $1.48 \times 10^6$ & $0.68 \pm 0.02$  \\
\enddata
\end{deluxetable}

%%%%%%%%%%%%%%%%%%%

\end{document}